\RequirePackage{fix-cm} 
\documentclass[a4paper, twoside, reqno, dvips, 12pt]{amsart}
\usepackage{fixltx2e}   

\usepackage{etex}

\usepackage[latin1]{inputenc}
\usepackage[T1]{fontenc}

\usepackage{algorithm}
\usepackage[noend]{algpseudocode}

\usepackage[titletoc,title]{appendix}

\usepackage{esint}
\usepackage{dsfont}
\usepackage{xspace}
\usepackage{amsgen}
\usepackage{amsthm}
\usepackage{amssymb}
\usepackage{amsmath}
\usepackage{wasysym}
\usepackage{upgreek}
\usepackage{amsfonts}
\usepackage{stmaryrd}
\usepackage{mathtools}

\usepackage{relsize}
\usepackage{textcomp}
\usepackage{textgreek}
\usepackage[mathcal, mathscr]{euscript}

\usepackage{mathrsfs}
\DeclareMathAlphabet{\mathscrbf}{OMS}{mdugm}{b}{n}

\usepackage{a4wide}

\usepackage[dvipsnames, table]{xcolor}
\definecolor{bckg}{RGB}{20.8, 20.8, 20.8}
\definecolor{oneblue}{rgb}{0.0, 0.0, 0.85}
\definecolor{Lightblue}{RGB}{214, 214, 214}
\definecolor{bluepigment}{rgb}{0.2, 0.2, 0.6}
\definecolor{charcoal}{rgb}{0.21, 0.27, 0.31}
\definecolor{denimblue}{rgb}{0.08, 0.38, 0.74}
\definecolor{Lightgray}{rgb}{0.89, 0.89, 0.89}
\definecolor{darkgrey}{rgb}{0.273, 0.281, 0.30}
\definecolor{darkelectricblue}{rgb}{0.33, 0.41, 0.47}

\usepackage{multirow}

\usepackage[sort&compress, comma, square, numbers]{natbib}

\usepackage{psfrag}
\usepackage{graphicx}
\usepackage{adjustbox}
\usepackage[tight]{subfigure}
\usepackage{morefloats}
\usepackage{indentfirst}

\usepackage[usenames, dvipsnames, pdf]{pstricks}
\usepackage{epsfig}
\usepackage{pst-grad} 
\usepackage{pst-plot} 

\usepackage{rotating}
\usepackage{pdflscape}

\usepackage{acronym}
\usepackage{microtype}
\usepackage[labelsep=period,%
            labelfont={bf,sf,color=bluepigment},%
            justification=raggedright]{caption}

\usepackage[perpage, symbol]{footmisc}

\usepackage[colorlinks,
           urlcolor=oneblue,
           linkcolor=denimblue,
           citecolor=NavyBlue,
           bookmarksopen=false,
           pdfpagemode=UseNone,
           pagebackref]{hyperref}

\usepackage[explicit]{titlesec}

\titleformat{\section}[block]
  {\color{NavyBlue}\Large\sffamily\bfseries}
  {}
  {0.0em}
  {\colorbox{bckg!5}{\strut\parbox{\dimexpr\linewidth-2\fboxsep\relax}{\thesection. #1}}}
  [\vspace*{0.33em}]

\titleformat{name=\section,numberless}[block]
  {\color{NavyBlue}\Large\sffamily\bfseries}
  {}
  {0.0em}
  {\colorbox{bckg!5}{\strut\parbox{\dimexpr\linewidth-2\fboxsep\relax}{#1}}}
  [\vspace*{0.33em}]

\titleformat{\subsection}
  {\color{NavyBlue}\large\sffamily\bfseries}
  {}
  {0.0em}
  {\colorbox{bckg!5}{\parbox{\dimexpr\linewidth-2\fboxsep\relax}{\thesubsection. #1}}}
  [\vspace*{0.33em}]

\titleformat{name=\subsection,numberless}
  {\color{NavyBlue}\Large\sffamily\bfseries}
  {}
  {0em}
  {\colorbox{bckg!5}{\parbox{\dimexpr\linewidth-2\fboxsep\relax}{#1}}}
  [\vspace*{0.33em}]

\titleformat{\subsubsection}
  {\color{bluepigment}\sffamily\normalsize\bfseries}
  {\thesubsubsection}
  {0.5em}
  {#1}
  [\vspace*{0.33em}]

\titleformat{\paragraph}[runin]
  {\color{bluepigment}\sffamily\small\bfseries}
  {}
  {0em}
  {#1}

\titlespacing{\section}{0.0em}{1.5em plus 2pt minus 2pt}%
{1.0em plus 2pt minus 2pt}[0em]
\titlespacing{\subsection}{0.5em}{1.5em plus 2pt minus 2pt}%
{1.0em}[0em]
\titlespacing{\subsubsection}{0.5em}{1.5em plus 2pt minus 2pt}%
{1.0em plus 2pt minus 2pt}[0em]

\usepackage{titletoc}

\contentsmargin{0.5em}
\newlength{\tocsep} 
\titlecontents{section}[\tocsep]
  {\addvspace{10pt}\bfseries\sffamily}
  {\contentslabel[\thecontentslabel]{\tocsep}}
  {}
  {\ \titlerule*[0.75pc]{.}\ \thecontentspage}
  []
\titlecontents{subsection}[\tocsep]
  {\addvspace{8pt}\sffamily}
  {\contentslabel[\thecontentslabel]{\tocsep}}
  {}
  {\ \titlerule*[0.5pc]{.}\ \thecontentspage}
  []
\titlecontents*{subsubsection}[\tocsep]
  {\addvspace{2pt}\footnotesize\sffamily}
  {}
  {}
  {\ \titlerule*[0.35pc]{.}\ \thecontentspage}
  [\\*]

\makeatletter
\def\@setauthors{%
  \begingroup
  \def\thanks{\protect\thanks@warning}%
  \trivlist
  \centering\footnotesize \@topsep30\p@\relax
  \advance\@topsep by -\baselineskip
  \item\relax
  \author@andify\authors
  \def\\{\protect\linebreak}%
  \textsc{\normalsize\textcolor{darkelectricblue}{\authors}}%
  \ifx\@empty\contribs
  \else
    ,\penalty-3 \space \@setcontribs
    \@closetoccontribs
  \fi
  \endtrivlist
  \endgroup
}
\def\@settitle{\begin{center}%
  \baselineskip14\p@\relax
    \bfseries
    \textsc{\Large\textcolor{charcoal}{\@title}}
  \end{center}%
}
\makeatother

\usepackage{enumitem}
\setlist[description]{%
  topsep=30pt,               
  itemsep=5pt,               
  font={\bfseries\sffamily\color{NavyBlue}}, 
}

\usepackage{fancyhdr}
\usepackage{lastpage}

\newcommand*\Title{\textcolor{bluepigment}{Adaptive simulation of heat and moisture transfer}}
\newcommand*\Authors{\textcolor{bluepigment}{S.~Gasparin, J.~Berger, D.~Dutykh \& N.~Mendes}}
\newcommand*{\plogo}{\textcolor{gray}{{\texttt{arXiv.org} / \textsc{hal}}}} 

\numberwithin{equation}{section}

\newcommand{\ie}{\emph{i.e.\xspace}}

\newcommand{\etc}{\emph{etc.\xspace}}
\newcommand{\etal}{\emph{et al.\xspace}}

\newcommand*\egal{\ = \ }
\newcommand*\plus{\ + \ }
\newcommand*\moins{\ - \ }
\newcommand*\egalb{\, = \, }

\newcommand*{\Ox}{\Omega_{\, x}}
\newcommand*{\Ot}{\Omega_{\, t}}

\newcommand{\cM}{c_{\,M}}
\newcommand{\cMref}{c_{\,M,\,0}}
\newcommand{\cTM}{c_{\,TM}}
\newcommand{\cTMref}{c_{\,TM,\,0}}
\newcommand{\cT}{c_{\,T}}
\newcommand{\cTref}{c_{\,T,\,0}}
\newcommand{\cw}{c_{\,w}}
\newcommand{\cz}{c_{\,0}}
\newcommand{\hM}{h_{\,M}}
\newcommand{\hT}{h_{\,T}}
\newcommand{\kl}{k_{\,l}}
\newcommand{\kM}{k_{\,M}}
\newcommand{\kMref}{k_{\,M,\,0}}
\newcommand{\kTM}{k_{\,TM}}
\newcommand{\kTMref}{k_{\,TM,\,0}}
\newcommand{\kT}{k_{\,T}}
\newcommand{\kTref}{k_{\,T,\,0}}
\newcommand{\Lv}{L_{\,v}}
\newcommand{\Pc}{P_{\,c}}
\newcommand{\Ps}{P_{\,s}}
\newcommand{\Pv}{P_{\,v}}
\newcommand{\Pvi}{P_{\,v, \,i}}
\newcommand{\Pvref}{P_{\,v,\,0}}
\newcommand{\Pvinf}{P_{\,v, \, \infty}}
\newcommand{\Rv}{R_{\,v}}
\newcommand{\Ti}{T_{\,i}}
\newcommand{\Tinf}{T_{\, \infty}}
\newcommand{\Tref}{T_{\,0}}
\newcommand{\tref}{t_{\,0}}

\newcommand{\rhol}{\rho_{\,l}}
\newcommand{\rhow}{\rho_{\,w}}
\newcommand{\rhoz}{\rho_{\,0}}

\newcommand{\BiM}{\mathrm{Bi}_{\,M}}
\newcommand{\BiML}{\mathrm{Bi}_{\,M,\,L}}
\newcommand{\BiMR}{\mathrm{Bi}_{\,M,\,R}}
\newcommand{\BiT}{\mathrm{Bi}_{\,T}}
\newcommand{\BiTL}{\mathrm{Bi}_{\,T,\,L}}
\newcommand{\BiTR}{\mathrm{Bi}_{\,T,\,R}}
\newcommand{\BiTM}{\mathrm{Bi}_{\,TM}}
\newcommand{\BiTML}{\mathrm{Bi}_{\,TM,\,L}}
\newcommand{\BiTMR}{\mathrm{Bi}_{\,TM,\,R}}
\newcommand{\cMs}{c_{\,M}^{\,\star}}
\newcommand{\cTs}{c_{\,T}^{\,\star}}
\newcommand{\cTMs}{c_{\,TM}^{\,\star}}
\newcommand{\dt}{\Delta t}
\newcommand{\dx}{\Delta x}
\newcommand{\FoT}{\mathrm{Fo}_{\,T}}
\newcommand{\FoM}{\mathrm{Fo}_{\,M}}
\newcommand{\kMs}{k_{\,M}^{\,\star}}
\newcommand{\kTs}{k_{\,T}^{\,\star}}
\newcommand{\kTMs}{k_{\,TM}^{\,\star}}
\newcommand{\gsinf}{g^{\,\star}_{\,\infty}}
\newcommand{\gsinfL}{g^{\,\star}_{\,\infty, \,L}}
\newcommand{\qsinf}{q^{\,\star}_{\,\infty}}
\newcommand{\qsinfL}{q^{\,\star}_{\,\infty, \,L}}
\newcommand{\ts}{t^{\,\star}}
\newcommand{\uinf}{u_{\,\infty}}
\newcommand{\uinfL}{u_{\,\infty ,\,L}}
\newcommand{\uinfR}{u_{\,\infty ,\,R}}
\newcommand{\vinf}{v_{\,\infty}}
\newcommand{\vinfL}{v_{\,\infty ,\,L}}
\newcommand{\vinfR}{v_{\,\infty ,\,R}}
\newcommand{\xs}{x^{\,\star}}

\newcommand*\pd[2]{\dfrac{\partial #1}{\partial #2}}
\newcommand{\pdd}[2]{\dfrac{\partial\/ #1}{\partial\/ #2}}
\newcommand*\od[2]{\frac{\mathrm{d} #1}{\mathrm{d} #2}}
\newcommand{\eqdef}{\mathop{\stackrel{\,\mathrm{def}}{:=}\,}}

\newcommand{\dix}[1]{ \cdot 10^{\,#1}}

\newcommand{\unitfrac}[2]{{\mathsf{#1}}/{\mathsf{#2}}}
\newcommand{\gC}{^{\circ}\ \mathsf{C}}

\newcommand{\R}{\mathds{R}}

\newcommand{\ab}{\bar{a}}
\newcommand{\db}{\bar{d}}
\newcommand{\N}{\mathds{N}}
\newcommand{\D}{\mathscr{D}}
\newcommand{\F}{\mathscr{F}}

\newcommand{\I}{\mathcal{I}}
\renewcommand{\O}{\mathcal{O}\,}
\renewcommand{\geq}{\geqslant}
\newcommand{\uv}{\vartheta}
\renewcommand{\d}{\mathrm{d}\,}

\begin{document}

\title[\Title]{An adaptive simulation of nonlinear heat and moisture transfer as a boundary value problem}

\author[S.~Gasparin]{Suelen Gasparin$^*$}
\address{\textbf{S.~Gasparin:} LAMA, UMR 5127 CNRS, Universit\'e Savoie Mont Blanc, Campus Scientifique, F-73376 Le Bourget-du-Lac Cedex, France and Thermal Systems Laboratory, Mechanical Engineering Graduate Program, Pontifical Catholic University of Paran\'a, Rua Imaculada Concei\c{c}\~{a}o, 1155, CEP: 80215-901, Curitiba -- Paran\'a, Brazil}
\email{suelengasparin@hotmail.com}
\urladdr{https://www.researchgate.net/profile/Suelen\_Gasparin/}
\thanks{$^*$ Corresponding author}

\author[J.~Berger]{Julien Berger}
\address{\textbf{J.~Berger:} LOCIE, UMR 5271 CNRS, Universit\'e Savoie Mont Blanc, Campus Scientifique, F-73376 Le Bourget-du-Lac Cedex, France}
\email{Berger.Julien@univ-smb.fr}
\urladdr{https://www.researchgate.net/profile/Julien\_Berger3/}

\author[D.~Dutykh]{Denys Dutykh}
\address{\textbf{D.~Dutykh:} Univ. Grenoble Alpes, Univ. Savoie Mont Blanc, CNRS, LAMA, 73000 Chamb\'ery, France and LAMA, UMR 5127 CNRS, Universit\'e Savoie Mont Blanc, Campus Scientifique, F-73376 Le Bourget-du-Lac Cedex, France}
\email{Denys.Dutykh@univ-smb.fr}
\urladdr{http://www.denys-dutykh.com/}

\author[N.~Mendes]{Nathan Mendes}
\address{\textbf{N.~Mendes:} Thermal Systems Laboratory, Mechanical Engineering Graduate Program, Pontifical Catholic University of Paran\'a, Rua Imaculada Concei\c{c}\~{a}o, 1155, CEP: 80215-901, Curitiba -- Paran\'a, Brazil}
\email{Nathan.Mendes@pucpr.edu.br}
\urladdr{https://www.researchgate.net/profile/Nathan\_Mendes/}


\begin{titlepage}
\thispagestyle{empty} 
\noindent
{\Large Suelen \textsc{Gasparin}}\\
{\it\textcolor{gray}{Pontifical Catholic University of Paran\'a, Brazil}}\\
{\it\textcolor{gray}{LAMA--CNRS, Universit\'e Savoie Mont Blanc, France}}
\\[0.02\textheight]
{\Large Julien \textsc{Berger}}\\
{\it\textcolor{gray}{LOCIE--CNRS, Universit\'e Savoie Mont Blanc, France}}
\\[0.02\textheight]
{\Large Denys \textsc{Dutykh}}\\
{\it\textcolor{gray}{LAMA--CNRS, Universit\'e Savoie Mont Blanc, France}}
\\[0.02\textheight]
{\Large Nathan \textsc{Mendes}}\\
{\it\textcolor{gray}{Pontifical Catholic University of Paran\'a, Brazil}}
\\[0.10\textheight]

\colorbox{Lightblue}{
  \parbox[t]{1.0\textwidth}{
    \centering\huge\sc
    \vspace*{0.7cm}
    
    \textcolor{bluepigment}{An adaptive simulation of nonlinear heat and moisture transfer as a boundary value problem}

    \vspace*{0.7cm}
  }
}

\vfill 

\raggedleft     
{\large \plogo} 
\end{titlepage}


\newpage
\thispagestyle{empty} 
\par\vspace*{\fill}   
\begin{flushright} 
{\textcolor{denimblue}{\textsc{Last modified:}} \today}
\end{flushright}


\newpage
\maketitle
\thispagestyle{empty}


\begin{abstract}

This work presents an alternative view on the numerical simulation of diffusion processes applied to the heat and moisture transfer through porous building materials. Traditionally, by using the finite-difference approach, the discretization follows the Method Of Lines (MOL), when the problem is first discretized in space to obtain a large system of coupled Ordinary Differential Equations (ODEs). Thus, this paper proposes to change this viewpoint. First, we discretize in time to obtain a small system of coupled ODEs, which means instead of having a \textsc{Cauchy} (Initial Value) Problem (IVP), we have a Boundary Value Problem (BVP). Fortunately, BVPs can be solved efficiently today using adaptive collocation methods of high order. To demonstrate the benefits of this new approach, three case studies are presented, in which one of them is compared with experimental data. The first one considers nonlinear heat and moisture transfer through one material layer while the second one considers two material layers. Results show how the nonlinearities and the interface between materials are easily treated, by reasonably using a fourth-order adaptive method. Finally, the last case study compares numerical results with experimental measurements, showing a good agreement.


\bigskip\bigskip
\noindent \textbf{\keywordsname:} coupled heat and moisture diffusive transfer; boundary value problems; semi-discretization in time; semi-discretization in space; \texttt{bvp4c} \\


\end{abstract}


\newpage
\tableofcontents
\thispagestyle{empty}


\newpage
\section{Introduction}

The hygrothermal transfer through porous structures is a matter of concern in many areas such as building physics, geophysics, environmental engineering and energy systems whose the transient evolution of heat and moisture migration plays an important role. Particularly, in the area of building physics, the heat and moisture transfer process through the porous envelope, roofing systems and the ground can strongly affect the energy efficiency, the thermal comfort of the occupants and the durability of the components \cite{Woloszyn2008, Tariku2010, Berger2015a}. Therefore, reliable assessment of hygrothermal transfer in building materials is a major issue, requiring efficient numerical tools for heat and moisture transfer in building materials \cite{Clarke2013}.

As building material properties are temperature- and moisture-dependent and the boundary conditions are driven by weather variables, the models included in those tools are based on numerical approaches using discrete representations of the continuous equations. To compute the solution, standard discretization and incremental techniques are applied, such as the \textsc{Euler} implicit scheme in \cite{Mendes2005, Janssen2014} to solve large systems of equations. Furthermore, when dealing with nonlinearities, hygrothermal properties of porous materials have to be updated as a function of the temperature and moisture content fields at each iteration \cite{Ferguson1995}. The difficulties to compute the solution increase, particularly when using implicit schemes that require sub-iterations to treat those issues. In the literature \cite{DosSantos2006, Dalgliesh2005, Abuku2009}, the important numerical costs of simulation tools are also mentioned and it is a matter of concern due to the substantial scale of buildings, where heat and moisture transfer phenomena have to be simulated.

In addition, in the models proposed in literature, the problem previously described is generally solved by traditional approaches such as the finite-difference method \cite{Gasparin2017}, the finite-volume method \cite{DosSantos2009, Mendes2005, Manz2003} and the finite-element method  \cite{Thomas1980, Janssen2007, Rouchier2013}, which are well established in the fields of thermal sciences and building physics. In these classical approaches, the higher accuracy obtained for the space discretization of the numerical schemes is to order $\O\,(\Delta x^{\,2})\,$. For a space standard discretization $\Delta x \egalb 10^{\,-2}\,$, it implies that the error $\varepsilon$ on the solution of the equations cannot be lower than $\O(10^{\,-4})\,$. Within the issue of comparing the model numerical predictions with experimental observations as carried out for instance in \cite{Lelievre2014, Qin2009, Olek2016}, it is of major importance to control the accuracy of the solution.

For sure, the accuracy of the computed solution can be increased by reducing the space and time discretization parameters. However, as mentioned before, the standard approaches proposed in literature has a high degree of freedom. Therefore, increasing the number of spatial and temporal grid points will inevitably increase the computational time of the numerical model. With high-order numerical schemes, it is possible to have the same precision of low-order numerical schemes but with a lower computational cost, as shown in \cite{Muller2012}. For this reason, these traditional methods have to be improved or even replaced by innovative and efficient ways of numerical simulation \cite{Pallin2013}, particularly with the issue of comparing the model predictions with experimental observations.

Therefore, this article aims at contributing to the numerical development of hygrothermal transfer, by proposing a new approach to simulate the one-dimensional heat and moisture diffusive transfer trough single and multilayered building  porous materials. The Method of Horizontal Lines is here proposed to solve the nonlinear heat and moisture transfer to increase significantly the accuracy in space with a low computational time of the numerical model. Usually, when using finite-differences, the discretization follows the Method of Lines (MOL). It means that the problem is first discretized in space to obtain a large system of coupled ODEs. Here, a different point of view is proposed based on discretizing first in the time domain to obtain a Boundary Value Problem (BVP). Such problems can be easily solved using adaptive collocation methods of high order. This approach is investigated in this paper to compute with high accuracy combined heat and mass transfer problems in porous materials.

The manuscript is organized as follows. Section~\ref{sec:HAM_transfer} details the physical model of heat and moisture transfer while fundamentals of the proposed method are shown in Section~\ref{sec:MHL}. Numerical results are discussed in Section~\ref{sec:numeric_app} and simulation are compared with experimental data in Section~\ref{sec:validation}. Finally, in Section~\ref{sec:conl}, the main conclusions are outlined with future perspectives.


\section{Physical model}
\label{sec:HAM_transfer}

The physical problem considers one-dimensional heat and moisture transfer through a porous material defined by the spatial domain $\Ox \egalb [\,0,\,L\,]$ and time domain $\Ot \egalb [\, 0, \, \tau \,]\,$. The following convention is adopted: $x \egalb 0$ corresponds to the surface in contact with the inside room and, $x \egal L\,$, corresponds to the outside surface. The moisture transfer occurs due to capillary migration and vapour diffusion. The heat transfer is governed by diffusion and latent mechanisms. The physical problem can be formulated as \cite{Luikov1966, Remki2012}:
\begin{subequations}\label{eq:HAM_equation}
\begin{align}
  \pd{\rhow}{t} &\egal \pd{}{x} \Biggl(\, \kl\, \pd{\Pc}{x} \plus \delta_{\,v}\, \pd{\Pv}{x}\, \Biggr)\,,\label{eq:M_equation}\\[3pt]
  \bigl(\, \rhoz \ \cz \plus \rhow \ \cw \, \bigr) \ \pd{T}{t} \plus \cw \ T \ \pd{\rhow}{t} &\egal \pd{}{x} \Biggl(\, \lambda \ \pd{T}{x} \plus \Lv \ \delta_{\,v} \ \pd{\Pv}{x} \, \Biggr) \,, \label{eq:H_equation}
\end{align}
\end{subequations}
where $\rhow$ is the volumetric moisture content of the material, $\delta_{\,v}$ and $\kl\,$, the vapour and liquid permeabilities, $\Pv\,$, the vapour pressure, $T\,$, the temperature, $\Rv\,$, the water vapour gas constant, $\Pc\,$, the capillary pressure, $\cz\,$, the material heat capacity, $\rhoz\,$, the material density, $\cw\,$, the water heat capacity, $\lambda\,$, the thermal conductivity, and, $\Lv\,$, the latent heat of evaporation. Equation~\eqref{eq:M_equation} can be written using the vapour pressure $\Pv$ as the driving potential. For this, we consider the physical relation, known as the \textsc{Kelvin} equation, between $\Pv$ and $\Pc$, and the \textsc{Clausius}--\textsc{Clapeyron} equation:
\begin{align*}
  \Pc & \egal \rhol \, \Rv \, T \, \ln \Biggl(\, \frac{\Pv}{\Ps\,(T)}\, \Biggr) \,,\\
  \pd{\Pc}{\Pv} & \egal \frac{\rhol\, R_{\,v} \, T}{\Pv} \,.
\end{align*}
where $\Pv$ comes from the relation $\phi \egalb \dfrac{\Pv}{\Ps\,(\,T\,)}\,$, in which $\phi$ is the relative humidity. Thus, by neglecting the variation of the capillary pressure and the mass content with temperature \cite{Rouchier2013}, the partial derivative of $\Pc$ can be written as:
\begin{align*}
\pd{\Pc}{x} \egal \pd{\Pc}{\Pv} \, \pd{\Pv}{x} \plus \pd{\Pc}{T} \, \pd{T}{x} 
\ \simeq \ \frac{\rhol\, \Rv \, T }{\Pv} \, \pd{\Pv}{x}\,. 
\end{align*}
In addition, we have:
\begin{align*}
& \pd{\rhow}{t} \egal \pd{\rhow}{\phi} \, \pd{\phi}{\Pv} \, \pd{\Pv}{t} \plus \pd{\rhow}{T} \, \pd{T}{t}\ \simeq\ \pd{\rhow}{\phi} \, \pd{\phi}{\Pv} \, \pd{\Pv}{t}\,.
\end{align*}
Considering the relation $\rhow \egalb w\,(\phi)\,$, obtained from the sorption isotherm, and from the relation between the vapour pressure $\Pv$ and the relative humidity $\phi\,$, we get: 
\begin{align*}
& \pd{\rhow}{t} \egal \frac{w^{\,\prime}\,(\phi)}{\Ps\,(T)} \ \pd{\Pv}{t} \,.
\end{align*}

We denote by
\begin{align*}
\kM \ &\eqdef \ \kl \, \dfrac{\rhol\, \Rv \, T}{\Pv} \plus \delta_{\,v}\,: && \text{the total moisture transfer coefficient} \\ & && \text{under vapour pressure gradient} ,\\
\kTM \ &\eqdef \ \Lv \ \delta_{\,v}\,: && \text{the heat coefficient due to a  vapour pressure gradient} ,\\
\kT \ &\eqdef \ \lambda\,: && \text{the heat transfer coefficient under temperature gradient} , \\
\cM \ &\eqdef \ \frac{w^{\,\prime}\,(\phi)}{\Ps\,(T)}\,: && \text{the moisture storage coeficient}  ,\\
\cT \ &\eqdef \ \rhoz \, \cz \plus w\,(\phi) \, \cw\,: && \text{the energy storage coeficient} ,\\
\cTM \ &\eqdef \ \cw \ T\ \frac{w^{\,\prime}\,(\phi)}{\Ps\,(T)}\,: && \text{the coupling storage coefficient}.
\end{align*}

Considering the previous notation, Equation~\eqref{eq:HAM_equation} can be rewritten as:
\begin{subequations}
\label{eq:HM_equation2}
\begin{align}
\cM\,(T,\,\Pv)\, \pd{\Pv}{t} &\egal \pd{}{x} \Biggl( \, \kM\,(T,\,\Pv)  \, \pd{\Pv}{x} \,  \Biggr)  \,, \\[3pt]
\cT\,(T,\,\Pv)\, \pd{T}{t} \plus \cTM\,(T,\,\Pv)\, \pd{\Pv}{t} &\egal \pd{}{x} \Biggl(\, \kT\,(T,\,\Pv)\, \pd{T}{x} \plus 
\kTM\,(T,\,\Pv) \, \pd{\Pv}{x} \,\Biggr)\,.
\end{align}
\end{subequations}

Finally, the problem of interest is a coupled system of two nonlinear parabolic partial differential equations, with vapour pressure $\Pv$ and temperature $T$ gradients as driving potentials. Their boundary conditions are expressed as:
\begin{align*}
\mathbf{n}\cdot \Biggl(\,\kM\,(T,\,\Pv)\, \pd{\Pv}{x}\, \Biggr) \egal &\hM \, \Bigl(\, \Pv - \Pvinf\,(t) \,\Bigr) \moins g_{\,\infty}\,(t) \,, \\
\mathbf{n}\cdot \Biggl(\, \kT\,(T,\,\Pv) \, \pd{T}{x} \plus \kTM\,(T,\,\Pv) \, \pd{\Pv}{x}\, \Biggr) \egal &\hT \, \Bigl(\, T\moins \Tinf\,(t) \,\Bigr) \moins q_{\,\infty}\,(t) \nonumber \\
& \plus \Lv \, \hM \, \Bigl(\, \Pv - \Pvinf\,(t) \,\Bigr) \,,
\end{align*}
where $\Pvinf$ and $\Tinf$ stand for the vapour pressure and temperature of the air, $\hM$ and $\hT$ are the convective transfer coefficients and $\mathbf{n}$ is the normal that assumes $+\,1$ or $-\,1$ at the left or right boundary sides. If the bounding surface is in contact with the \textit{outside} air, $g_{\,\infty}$ is the liquid flow from wind driven rain and $q_{\,\infty}$ is the sensible heat from the rain:
\begin{align*}
  q_{\,\infty} \egal g_{\,\infty} \cdot H_{\,l}  \,,
\end{align*} 
where, $H_{\,l}$ is the water enthalpy. If the bounding surface is in contact with the \textit{inside} building air then $g_{\,\infty} \egalb 0 $ and the variable $q_{\,\infty}$ is the enclosure and long-wave radiative heat exchanged among the room surfaces:
\begin{align*}
  q_{\,\infty} \egal \sum_{j=1}^{m}\, \mathrm{s} \, \xi \, \sigma \,\biggl[\, \Bigl(\, T_{\,j}\, (x \egalb 0) \, \Bigr)^{\,4} \moins \Bigl(\, T\, (x \egalb 0) \, \Bigr)^{\,4} \, \biggr] \,,
\end{align*} 
where $\mathrm{s}$ is the view factor between two surfaces, $\sigma$ is the \textsc{Stefan}--\textsc{Boltzmann} constant, $\xi$ is the emissivity of the wall surface, $j$ represents the $m$ bounding walls. 

We consider a uniform vapour pressure and temperature distributions as initial conditions:
\begin{align*}
 \Pv\,(x,\,t\egalb 0) \egal \Pvi && \text{and} && T\,(x,\,t\egalb 0)  \egal \Ti \,.
\end{align*}

One of the interesting outputs in the building physics framework are the sensible $q_{\,s}$ and the latent $q_{\,l}$ heat fluxes \cite{Traore2011}, which are respectively defined as:
\begin{align*}
q_{\,s}\,(t)\ \eqdef\ \moins \kT \,(T,\,\Pv) \,\left. \pd{T}{x}  \, \right\vert_{x_{\,0}}  & & \text{and} & &
q_{\,l}\,(t)\ \eqdef\ \moins \kTM\,(T,\,\Pv) \,\left. \pd{\Pv}{x}\, \right\vert_{x_{\,0}} \,. 
\end{align*}
The moisture flow $g$ is similarly computed:
\begin{align*}
g\,(t)\ \eqdef\ \moins \kM\, (T,\,\Pv) \,\left. \pd{\Pv}{x}\, \right\vert_{x_{\,0}} \,, 
\end{align*}
where $x_{\,0}\, \in\, [\,0,\,L\,]\,$.


\subsection{Dimensionless representation}

Before solving this problem directly, it is of great importance to get a dimensionless formulation. It enables us to determine important scaling parameters such as \textsc{Biot} and \textsc{Fourier} numbers. It allows also to estimate the relative magnitude of various terms in governing equations, and thus, eventually to simplify the problem using asymptotic methods \cite{Nayfeh2000}. In addition, the dimensionless form enables to manipulate numerically the quantities at the order of $\O\,(1)$ where the floating point arithmetics are designed to have minimal rounding errors \citep{Kahan1979}. Considering the temperature range of interest in building applications, temperature dependence was neglected when compared to their dependence with moisture content, with all transport coefficients considered as moisture content dependent. Therefore, the governing Equation~\eqref{eq:HM_equation2} can be written in a dimensionless form as:
\begin{subequations}\label{eq:HAM_equation_dimless}
\begin{align}
  \cMs\,(v) \, \pd{v}{\ts} & \egal \FoM \, \pd{}{\xs} \Biggl( \, \kMs\,(v)  \, \pd{v}{\xs} \,  \Biggr)  \,, \label{eq:moisture_equation_dimless}\\[3pt]
  \cTs\,(v) \, \pd{u}{\ts} \plus \gamma_{\,1}\, \cTMs\,(v)  \, \pd{v}{\ts} & \egal \FoT \, \pd{}{\xs} \Biggl(\, \kTs\,(v)\, \pd{u}{\xs} \plus \gamma_{\,2}\, \kTMs\,(v)\, \pd{v}{\xs}\, \Biggr)\,,  \label{eq:heat_equation_dimless}
\end{align}
\end{subequations}
which can be expressed by the following equation in a matrix form:
\begin{align} \label{eq:HAM_dimlss_vecto}
  \left[ \begin{array}{cc}
    \dfrac{\cMs}{\FoM} & 0 \\[10pt] \dfrac{\gamma_{\,1}\, \cTMs}{\FoT } & \dfrac{\cTs}{\FoT } 
  \end{array} \right]
  \cdot \left[ 
  \begin{array}{c}
    \pd{v}{t} \\[10pt] \pd{u}{t} 
  \end{array} \right] \egal \pd{}{x} \left( \left[
  \begin{array}{cc}
    \kMs & 0 \\[10pt] \gamma_{\,2} \, \kTMs & \kTs
  \end{array} \right] \cdot \left[ 
  \begin{array}{c}
    \pd{v}{x} \\[10pt] \pd{u}{x} 
  \end{array} \right] \right)\,.
\end{align}

The dimensionless formulation of the boundary conditions is:
\begin{subequations}
\label{eq:HAM_BC_dimless}
\begin{align}
\mathbf{n}\cdot \biggl(\,\kMs\,(v) \, \pd{v}{\xs}\, \biggr) \egal & \BiM \, \Bigl(\, v \moins \vinf\,(\ts) \,\Bigr) \moins \gsinf\,(\ts) \,, \\[3pt]
\mathbf{n}\cdot \biggl(\,\kTs\,(v) \, \pd{u}{\xs} \plus \gamma_{\,2} \, \kTMs\,(v) \, \pd{v}{\xs}\, \biggr) \egal & 
\BiT \, \Bigl(\, u\moins \uinf\,(\ts) \,\Bigr)\moins \qsinf\,(\ts) \nonumber\\
&\plus \gamma_{\,2} \,\BiTM\,  \Bigl(\, v \moins \vinf\,(\ts) \,\Bigr)  \,,
\end{align}
\end{subequations}
where the dimensionless quantities are defined as: 
\begin{align*}
 u \ &\eqdef \ \frac{T}{\Tref} \,,
& v \ &\eqdef \ \frac{\Pv}{\Pvref} \,,
& \xs \ &\eqdef \ \frac{x}{L} \,, \\[3pt]
\uinf \ &\eqdef \ \frac{\Tinf}{\Tref} \,, 
& \vinf \ &\eqdef \ \frac{\Pvinf}{\Pvref} \,, 
& \ts \ &\eqdef \ \frac{t}{\tref} \,, \\[3pt]
 \cMs \ &\eqdef \ \frac{\cM }{\cMref} \,,
& \cTs \ &\eqdef \ \frac{\cT }{\cTref} \,, 
& \cTMs \ &\eqdef \ \frac{\cTM }{\cTMref} \,, \\[3pt]
 \kMs \ &\eqdef \ \frac{\kM }{\kMref} \,, 
& \kTs \ &\eqdef \ \frac{\kT }{\kTref} \,,
& \kTMs \ &\eqdef \ \frac{\kTM }{\kTMref} \,,  \\[3pt]
 \BiM \ &\eqdef \ \frac{\hM \cdot L}{\kMref} \,, 
& \BiT \ &\eqdef \ \frac{\hT \cdot L}{\kTref} \,, 
& \BiTM \ &\eqdef \ \frac{\hM \cdot L \cdot \Lv}{\kTMref} \,, \\[3pt]
\FoM \ &\eqdef \ \frac{\tref \cdot \kMref}{L^{\,2} \cdot \cMref} \,,
& \FoT \ &\eqdef \ \frac{\tref \cdot \kTref}{L^{\,2} \cdot \cTref} \,,
& \gsinf \ &\eqdef \ \frac{ g_{\,\infty}\cdot L}{\Pvref \cdot \kMref} \,, \\[3pt]
 \gamma_{\,1} \ &\eqdef \ \frac{\cTMref \cdot \Pvref}{\cTref \cdot \Tref} \,, 
& \gamma_{\,2} \ &\eqdef \ \frac{\kTMref \cdot \Pvref}{\kTref \cdot \Tref} \,, 
& \qsinf \ &\eqdef \ \frac{q_{\,\infty}\cdot L}{\Tref \cdot \kTref}  \,,
\end{align*}
where the subscript $0$ represents a reference value, chosen according to the application problem and the superscript $\star$ represents a dimensionless quantity of the same variable. The dimensionless values of each numerical application can be found in the Appendix~\ref{annexe:dimensionless}.


\section{The Method of Horizontal Lines}
\label{sec:MHL}

The perception of objects might change depending on the viewpoint. 
Keeping that in mind, in this study we apply the same idea to Partial Differential Equations (PDEs). Since our main applications include the heat and/or moisture transfer, we explain the idea on the simplest parabolic (or \textsc{Fourier}) equation, which is traditionally written as \cite{Fourier1822, Evans2010}:
\begin{equation}\label{eq:heat}
  \pd{u}{t} \egal \nu\, \pd{^{\,2}\,u}{x^{\,2}}\,, \qquad x\ \in\ \I\ \equiv\ [\,0,\,\ell\,]\,, \qquad t\ \geq\ 0\,,
\end{equation}
where $\nu\, >\, 0$ is the diffusion coefficient. The last equation has to be supplemented with \emph{right} and \emph{compatible} initial and boundary conditions to obtain a well-posed problem \cite{Hadamard1902}.

The way how Equation~\eqref{eq:heat} is written conditions the tools and, thus, the way how it is solved numerically (or study it theoretically \cite{Evans2010}). Indeed, the most popular approach nowadays is the so-called Method Of Lines (MOL) \cite{Shampine1994, Reddy1992, Kreiss1992, Schiesser1994, Hamdi2007}. It consists in semi-discretization in space as the first step. With this goal in mind, let us consider a uniform (for the sake of simplicity) discretization in space:
\begin{equation*}
  \I_{\,\Delta x}\ \eqdef\ \bigl\{x_{\,j}\ \equiv\ j\cdot\Delta x\ \vert\ j\egal 0,\,1,\,2,\,\ldots \,N,\,N+1\bigr\}\,, \qquad \Delta x\ \eqdef\ \frac{\ell}{N + 1}\,.
\end{equation*}
The classical second-order central finite difference scheme yields the following system of coupled ordinary differential equations (ODEs) \cite{Richtmyer1967}:
\begin{equation*}
  \od{u_{\,j}}{t}\egal \frac{\nu}{\Delta x^{\,2}}\, \bigl(\,u_{\,j-1}\moins 2\,u_{\,j}\plus u_{\,j+1}\,\bigr)\,, \qquad j\egal 1,\,2,\,\ldots,\,N\,,
\end{equation*}
with $u_{\,j}\,(t)\, \eqdef\, u\,(x_{\,j},\,t)$ is the solution nodal value. For simplicity of the discussion, we write here evolution equations only for interior nodes. There are still two remaining equations dependent on the boundary conditions which will be discussed in a further Section.

\begin{figure}
\begin{center}
  \subfigure[][\label{fig:MOL}]{\includegraphics[width=.32\textwidth]{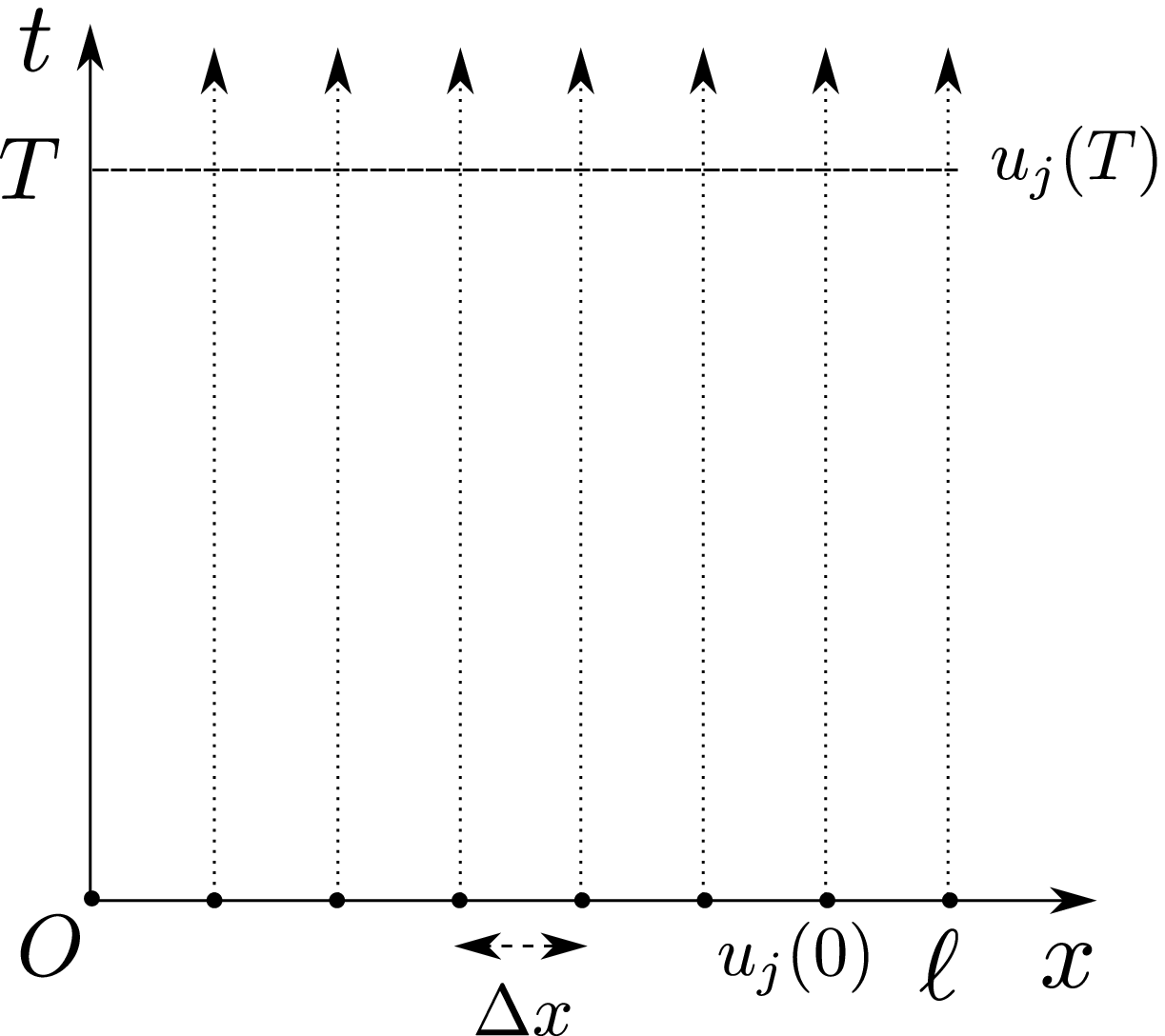}} \hspace{0.5cm}
  \subfigure[][\label{fig:BVP}]{\includegraphics[width=.38\textwidth]{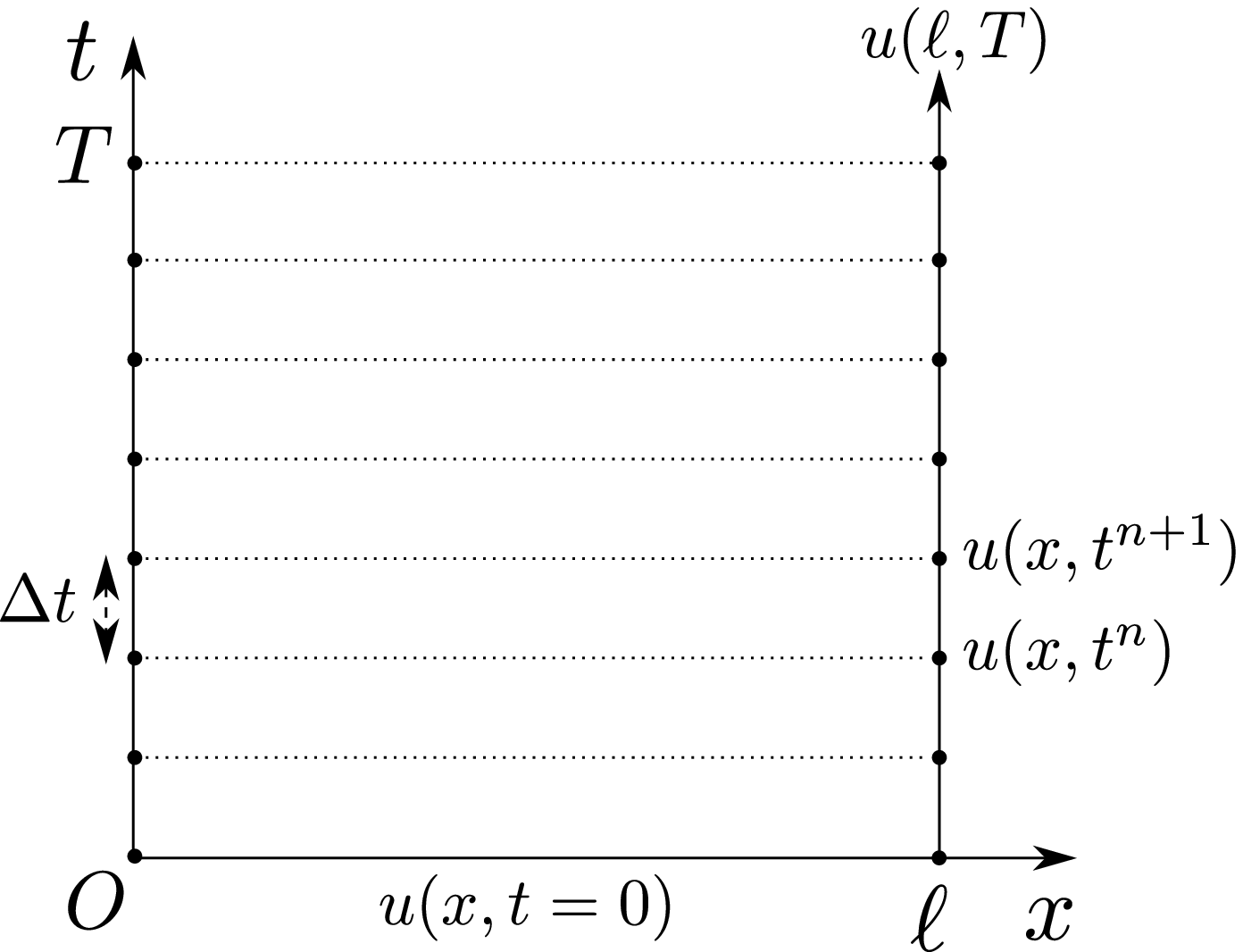}} 
  \caption{\small\em The classical numerical solution strategy based on the MOL (a) and the proposed approach view based on BVPs (b).}
\label{fig:MOLvsBVP}
\end{center}
\end{figure}

Hence, the Partial Differential Equation~\eqref{eq:heat} solved with MOL yields approximatively to a large system of coupled differential equations. In typical accurate numerical simulations $N\, \gtrsim\, 10^{\,2}\,\ldots\,10^{\,3}\,$. Then, the last system of equations can be discretized using various explicit or implicit time-marching schemes \cite{Butcher2016}. This philosophy is shown schematically in Figure~\ref{fig:MOL}. To make the language more accurate, we can refer to this approach as the Method of \emph{Vertical} Lines (MOVL). From a brief literature survey we can conclude that the dominant majority of numerical simulations in the field of HAM transfer follow this philosophy. However, in this study we would like to propose an alternative view on diffusion equations. Namely, we claim that Equation \eqref{eq:heat} can be solved by a totally different numerical technique, if it is rewritten in \textsc{Cauchy--Kovalevskaya} form as:
\begin{equation}\label{eq:heat2}
  \nu\,\pd{^{\,2}\,u}{x^{\,2}}\egal \pd{u}{t}\,, \qquad x\ \in\ \I\ \equiv\ [\,0,\,\ell\,]\,, \qquad t\ \geq\ 0\,.
\end{equation}
By defining a uniform time discretization:
\begin{equation*}
  \I_{\,\Delta t}\ \eqdef\ \bigl\{t^{\,n}\ \equiv\ n\cdot\Delta t\ \vert\ n\egal 0,\,1,\,2,\,\ldots \,M,\,M+1\,\,\bigr\}\,, \qquad \Delta t\ \eqdef\ \frac{\tau}{M+1}\,.
\end{equation*}
Equation~\eqref{eq:heat2} yields to the following system:
\begin{equation*}\label{eq:heat2_dif_view}
  \nu\,\pd{^{\,2}\,u\,(x,\,t^{\,n})}{x^{\,2}}\egal \pd{u}{t}\biggr|_{\,t \egalb t^{\,n}}\,, \qquad x\ \in\ \I\ \equiv\ [\,0,\,\ell\,]\,, \qquad t\ \geq\ 0\,.
\end{equation*}
The main term now is the second-order elliptic differential operator in space, while the time derivative can be seen (and treated) as a source term of secondary importance. In other words, instead of having an evolution problem, we have a Boundary Value Problem (BVP), which can be tackled by appropriate methods. This approach is schematically depicted in Figure~\ref{fig:BVP}, which will be called in this work as the \emph{Method of Horizontal Lines} (MOHL).
In the following Section~\ref{sec:scal}, we propose a reformulation for scalar nonlinear equations.

\subsection{Scalar nonlinear equations}
\label{sec:scal}

Consider the following dimensionless nonlinear diffusion equation, which can describe the heat or the moisture transfer in a 1D section\footnote{In higher dimensional problems we would require the domain $\I\ \subset\ \R^{\,m}\,$ $\bigl(m\ \in\ \bigl\{2,\,3\bigr\}\bigr)$ to be bounded with a \textsc{Lipschitz}-continuous boundary.} of a porous material:
\begin{equation}\label{eq:prob}
  c\,(u)\,u_{\,t}\moins \bigl[\,\D\,(u)\,u_{\,x}\,\bigr]_{\,x}\egal\F\,(x,\,t)\,, \qquad x\ \in\ \I\ \equiv\ [\,0,\,\ell\,]\,,
\end{equation}
where the right-hand side $\F\,(x,\,t)\, \in\, L^{\,2}\bigl((0,\,T),\,L^{\,2}\,(\I)\bigr)$ and describes eventual sources/sinks inside the material. The time horizon $T\ >\ 0$ is fixed. We assume that coefficients $c\,(u)$ and $\D\,(u)$ are positive definite: 
\begin{equation}\label{eq:ass}
  c\,(u)\ \geq\ c_{\,\min}\ >\ 0\,, \qquad
  \D\,(u)\ \geq\ d_{\,\min}\ >\ 0\,, \qquad
  \forall\, u \,.
\end{equation}
For the moisture diffusion, the coefficient $c\,(u)$ is the moisture storage coefficient $\cMs\,(u)$  and the coefficient $\D\,(u)$ is the diffusion coefficient $\FoM \cdot \kMs\, (u)\,$. For the heat transfer, $\F\,(x,\,t)$ incorporates the other terms. The subscripts denote partial derivatives, \ie, $u_{\,t}\ \eqdef\ \pdd{u}{t}\,$, $u_{\,x}\ \eqdef\ \pdd{u}{x}\,$, \etc Equation \eqref{eq:prob} has to be completed by initial and boundary conditions. The boundary conditions are described by Equation~\eqref{eq:HAM_BC_dimless}, which are of (mixed) \textsc{Robin}-type and the initial condition as $u_{\,0}\,(x)\, \equiv\, u\,(x,\,0)\, \in\, H^{\,1}(\I)\,$. Then, the Boundary-Value Problem (BVP) \eqref{eq:prob} has a unique solution $u\,(x,\,t)\, \in\, L^{\,2}\bigl((0,\,T),\,H^{\,1}\,(\I)\bigr)\,$.


\subsection{Problem reformulation as a BVP}

For our purposes it will be more advantageous to recast Equation~\eqref{eq:prob} in a non-conservative form. By assuming that function $\D\,(u)$ is continuously differentiable, we introduce its derivative:
\begin{equation*}
  d\,(u)\ \eqdef\ \od{\D\,(u)}{u}\ \equiv\ \D^{\,\prime}\,(u)\,.
\end{equation*}
After taking one derivative, Equation~\eqref{eq:prob} becomes:
\begin{equation}\label{eq:ncons}
  u_{\,t}\moins \ab\,(u,\,u_{\,x})\,u_{\,x}\moins \db\,(u)\,u_{\,x\,x}\egal f\,(x,\,t)\,,
\end{equation}
where we introduce the \emph{effective} coefficients:
\begin{equation*}
  \ab\,(u,\,u_{\,x})\ \eqdef\ \frac{d\,(u)\,u_{\,x}}{c\,(u)}\,, \qquad
  \db\,(u)\ \eqdef\ \frac{\D\,(u)}{c\,(u)}\,,
\end{equation*}
and also the effective source term:
\begin{equation*}
  f\,(u,\,x,\,t)\ \eqdef\ \frac{\F\,(x,\,t)}{c\,(u)}\,.
\end{equation*}
Equation~\eqref{eq:ncons} is equivalent to Equation~\eqref{eq:prob} for smooth solutions. Under assumptions \eqref{eq:ass} the solutions are indeed smooth.

Equation~\eqref{eq:ncons} is an evolution equation written in time--space: first we write the derivatives with respect to time, then the spatial ones. According to the philosophy of the approach described in Section~\ref{sec:MHL}, let us rewrite it in space--time:
\begin{equation*}
  \db\,(u)\,u_{\,x\,x}  \plus \ab\,(u,\,u_{\,x})\,u_{\,x} \egal f\,(u,\,x,\,t)\plus u_{\,t}\,.
\end{equation*}
In other words, we see the time derivative $u_{\,t}$ as a source term for our convenience. Finally, we rewrite the last equation as a system of first-order differential equations in the spatial variable $x\,$:
\begin{align*}
\left\{ \begin{array}{l}
  u_{\,x}\egal \uv \,, \\
  \uv_{\,x}\egal \dfrac{f \plus u_{\,t}}{\db\,(u)}\moins \dfrac{\ab\,(u,\,\uv)}{\db\,(u)}\; \uv \,. 
  \end{array} \right.
\end{align*}
The last system can be rewritten with original coefficients as well:
\begin{align*}
\left\{ \begin{array}{l}
  u_{\,x} \egal \uv\,,\\ 
  \uv_{\,x} \egal \dfrac{\F \plus c\,(u)\,u_{\,t}}{\D\,(u)} \moins \dfrac{d\,(u)\,\uv}{\D\,(u)}\;\uv\,.
  \end{array} \right.
\end{align*}

In order to have a true BVP, we have to get rid of the solution dependence on time. To do it, we (semi-)discretize the solution $u$ in time, \ie, we replace the surface $u\,(x,\,t)$ by a sequence of lines:
\begin{equation*}
  u^{\,n}\,(x)\ \eqdef\ u\,(x,\,t^{\,n})\,, \qquad t^{\,n}\egal n\,\Delta\,t\,, \qquad n\ \in\ \N_{\,0}\,,
\end{equation*}
where $\Delta t\, >\, 0$ is a chosen time step. See Figure~\ref{fig:BVP} for an illustration. This method is called the Method Of Horizontal Lines (MOHL). At each time layer $t^{\,n}$ we have to solve a \emph{true} BVP:
\begin{align*}
\left\{ \begin{array}{l}
  u_{\,x}^{\,n}\egal \uv^{\,n}\,, \\
  \uv_{\,x}^{\,n}\egal \dfrac{\F^{\,n}\plus c\,(u^{\,n})\,\bigl(u_{\,t}^{\,\star}\bigr)^{\,n}}{\D\,(u^{\,n})}\moins \dfrac{d\,(u^{\,n})\,\uv^{\,n}}{\D\,(u^{\,n})}\;\uv^{\,n}\,,
    \end{array} \right.
\end{align*}
where $\bigl(u_{\,t}^{\,\star}\bigr)^{\,n}$ is an approximation of the time derivative. Depending on the accuracy we would like to achieve, we can take the following backward finite-difference formulas:
\begin{align*}
  \bigl(u_{\,t}^{\,\star}\bigr)^{\,n} &\egal \dfrac{u^{\,n}\moins u^{\,n-1}}{\Delta t}\plus \O\,\bigl(\Delta t\bigr)\,, \\
  \bigl(u_{\,t}^{\,\star}\bigr)^{\,n} &\egal \dfrac{3\,u^{\,n}\moins 4\,u^{\,n-1}\plus 2\,u^{\,n-2}}{2\, \Delta t}\plus \O\,\bigl(\Delta t^{\,2}\bigr)\,.
\end{align*}
In this case, the second-order accuracy in time is implemented in the calculations in order to have a more precise solution, so that we can use $\Delta \ts \, \geq\, 10^{\,-1}\,$. The final scheme is unconditionally stable according to the construction of the method.


\subsection{Numerical methods}
\label{sec:num}

All numerical results in this paper were computed using \texttt{Matlab\texttrademark 2017b}.The solver \texttt{bvp4c} \cite{Shampine2000} is employed to solve the BVP problem at every time step, other solvers also can be employed such as the \texttt{bvp5c} \cite{Shampine2000} and even \texttt{bvp6c} \cite{Hale2008}. The codes for \texttt{bvp4c} \& \texttt{bvp5c} are available within any standard \texttt{Matlab\texttrademark} distribution, while the code \texttt{bvp6c} was developed by Dr. Nick~\textsc{Hale} (Stellenbosh University, South Africa) and is freely available. All these methods are based on finite-difference approximations that implement various orders of \textsc{Lobatto}~IIIA formula. This is a collocation method and the corresponding collocation polynomial provides a $C^{\,1}\,(\I)$ approximation of the uniform fourth, fifth or sixth orders of accuracy in $\Delta x$, respectively. Special attention should be given to the fact that other implementation details are quite different among the solvers, the order is not the only difference between them. In our numerical simulations we use adaptive methods of the fourth order which in most practical applications are enough.

Let us consider a system of ordinary differential equations of the form $u^{\,\prime}\,(x) \egalb f\,(x,\,u)\,$, within the interval $[\,a,\,b\,]$ subjected to two boundary conditions $\psi\,(u\,(a),\,u\,(b)) \egalb 0$. To use any of the BVP solvers, three inputs are required: the initial guess, a function with the boundary conditions and another function with the system of ordinary equations. They will return essentially three outputs: the spatial grid mesh, the solution approximation of $u\,(x)$ at the mesh points and the approximation to $u^{\,\prime}\,(x)$. Note that the solvers produce a solution that is continuous in the considered interval $[\,a,\,b\,]$ and with a continuous first derivative. For more details on the methods and their implementations refer to \cite{Shampine2000, Hale2008}.

The essential feature of these algorithms is the adaptive distribution of collocation nodes. The grid adaptation and error control are based on the residuals (there are two: one for the equation and one for the boundary conditions) of the $C^{\,1}\,(\I)$ continuous solution. The convergence speed towards the BVP solution depends essentially on the quality of the initial guess. Here we deal with an IVP-BVP problem. Thus, the BVP-solution value on the previous time step can be a good approximation. However, some physics-based or vector-based extrapolation techniques might also be used to further reduce internal iterations on every time step. The boundary conditions are directly provide as one of the inputs of the solver. No special treatment need to be carried out before supplying it.


\subsection{Extension to the coupled transfer}

There are two ways of using the BVP approach in the heat and mass transfer System~\eqref{eq:HAM_equation_dimless}. The first one is to take advantage of the solver \texttt{bvp4c} that provides directly the derivative of the field as output. First, it is computed the solution for $v$ and then, it is computed the solution for $u\,$, which is the approach used in the application Section. This approach works because System~\eqref{eq:HAM_equation_dimless} is weakly coupled. The procedure to implement the MOHL is described in Algorithm~\ref{alg:mohl}.

Therefore, using this approach, the system of first-order differential equations for the mass balance equation Equation~\eqref{eq:moisture_equation_dimless} is written as:
\begin{align}
\left\{ \begin{array}{l}
  v_{\,x}^{\,n}   \egal \uv^{\,n}\,, \\[4pt]
  \uv_{\,x}^{\,n} \egal \dfrac{(v_{\,t}^{\,\star})^{\,n}}{\bar{d}\,(v^{\,n})} \moins  \dfrac{\bar{a}\,(v^{\,n}\,,\uv^{\,n})}{\bar{d}\,(v^{\,n})}\;\uv^{\,n}\,,
    \end{array} \right. \label{eq:bvp_moist}
\end{align}
while the system of the energy equation Equation~\eqref{eq:heat_equation_dimless} as: 
\begin{align}
\left\{ \begin{array}{l}
  u_{\,x}^{\,n}   \egal \mu^{\,n}\,, \\[4pt]
  \mu_{\,x}^{\,n} \egal \Biggl[\dfrac{(u_{\,t}^{\,\star})^{\,n}}{\bar{b}\,(v^{\,n})} \plus \dfrac{\bar{c}\,(v^{\,n}) - \bar{f}\,(v^{\,n})}{\bar{b}\,(v^{\,n})} \; \uv_{\,x}^{\,n} \plus \dfrac{\bar{h}\,(v^{\,n},\,\uv^{\,n}) - \bar{g}\,(v^{\,n},\,\uv^{\,n})}{\bar{b}\,(v^{\,n})} \; \uv^{\,n}  \Biggr] \moins \dfrac{\bar{e}\,(v^{\,n},\,\uv^{\,n})}{\bar{b}\,(v^{\,n})}\;\mu^{\,n}\,,
    \end{array} \right.\label{eq:bvp_heat}
\end{align}
where,
\begin{align*}
\bar{a}\,(v^{\,n},\,\uv^{\,n}) &\egal \dfrac{\FoM}{\cMs\,(v^{\,n})}\cdot \dfrac{\d \kMs\,(v^{\,n})}{\d v}\; \uv^{\,n}\,, & 
\bar{b}\,(v^{\,n}) &\egal \dfrac{\FoT \cdot \kTs\,(v^{\,n})}{\cTs\,(v^{\,n})}\,, \\
\bar{e}\,(v^{\,n},\,\uv^{\,n}) &\egal \dfrac{\FoT}{\cTs\,(v^{\,n})}\cdot \dfrac{\d \kTs\,(v^{\,n})}{\d v}\; \uv^{\,n}\,,  & 
\bar{c}\,(v^{\,n}) &\egal \dfrac{\FoM \cdot \gamma_{\,1}\cdot \kMs\,(v^{\,n})\cdot \cTMs\,(v^{\,n})}{\cTs\,(v^{\,n})\cdot \cMs\,(v^{\,n})}\,, \\
\bar{g}\,(v^{\,n},\,\uv^{\,n}) &\egal \dfrac{\FoT \cdot \gamma_{\,2}}{\cTs\,(v^{\,n})}\cdot \dfrac{\d \kTMs\,(v^{\,n})}{\d v}\; \uv^{\,n}\,, & 
\bar{d}\,(v^{\,n}) &\egal \dfrac{\FoM \cdot \kMs\,(v^{\,n})}{\cMs\,(v^{\,n})}\,, \\
\bar{h}\,(v^{\,n},\,\uv^{\,n}) &\egal \dfrac{\FoM \cdot \gamma_{\,1}\cdot \cTMs\,(v^{\,n}) }{\cTs\,(v^{\,n})\cdot \cMs\,(v^{\,n})} \cdot \dfrac{\d \kMs\,(v^{\,n})}{\d v}   \; \uv^{\,n}\,,   & 
\bar{f}\,(v^{\,n}) &\egal \dfrac{\FoT \cdot \gamma_{\,2}\cdot \kTMs\,(v^{\,n})}{\cTs\,(v^{\,n})}\,. 
\end{align*}
In addition, the boundary conditions that complement Equation~\eqref{eq:bvp_moist} are:
\begin{align}
\left\{ \begin{array}{l}
  \mathrm{R}_{\,v,L}  \egal \kMs\, (v^{\,n}) \cdot  \uv^{\,n} \moins \BiML \cdot v^{\,n} \plus \BiML \cdot \vinfL \plus g_{\,\infty}\,, \\[4pt]
  \mathrm{R}_{\,v,R}  \egal \kMs\, (v^{\,n}) \cdot  \uv^{\,n} \plus \BiMR \cdot v^{\,n} \moins \BiMR \cdot \vinfR  \,,
    \end{array} \right.  \label{eq:bvp_moist_bc}
\end{align}
and, the boundary conditions that complement Equation~\eqref{eq:bvp_heat} are:
\begin{align}
  \small
  \left\{\begin{array}{ll}
  \mathrm{R}_{\,u,L}  \egal& \FoT\,\Bigl( \kTs\, (v^{\,n})\, \mu^{\,n} + \kTMs\,(v^{\,n})\,\gamma_{\,2}\, \uv^{\,n}\Bigr) \\
  & - \BiTL ( u^{\,n} - \uinfL) - \BiTML\,\gamma_{\,2} ( v^{\,n} - \vinfL) + q_{\,\infty}\,, \\[4pt]
  \mathrm{R}_{\,u,R}  \egal& \FoT\,\Bigl( \kTs\, (v^{\,n})\, \mu^{\,n} + \kTMs\,(v^{\,n})\,\gamma_{\,2}\, \uv^{\,n}\Bigr)\\
  & + \BiTR ( u^{\,n} - \uinfR)  + \BiTMR\,\gamma_{\,2} ( v^{\,n} - \vinfR)\,,
  \end{array}\right.\label{eq:bvp_heat_bc}
\end{align}
where, $\mathrm{R}$ represents the residual and the subscripts $L$ and $R$ represent values of the left and right boundary conditions.

\begin{algorithm}
\caption{\small\em MOHL's algorithm for the weakly coupled heat and moisture transfer problem.}\label{alg:mohl}
\begin{algorithmic}[1]\bigskip
\State Initialization;
\State Define functions: ODE of $v$ and ODE of $u\,$; \Comment{from Equation~\eqref{eq:bvp_moist} and \eqref{eq:bvp_heat}}
\State Define functions: BC of $v$ and BC of $u\,$; \Comment{from Equation~\eqref{eq:bvp_moist_bc} and \eqref{eq:bvp_heat_bc}}
\State Define initial solution: $v_{\,0}$ and $u_{\,0}\,$;
\State Set relative and absolute tolerances ``$\mathsf{tol}$'' of the solver;
\While{$t\,<\,\tau$}
\State $[\,v^{\,n},\,v_{\,x}^{\,n}\,]\egal$ \texttt{bvp4c} (ODE of $v^{\,n}\,$, BC of $v^{\,n}\,$, $v_{\,0}$);
\State $[\,u^{\,n},\,u_{\,x}^{\,n}\,]\egal$ \texttt{bvp4c} (ODE of $u^{\,n}\,$, BC of $u^{\,n}\,$, $u_{\,0}$);
\State Compute refined solution;
\State Update: $v_{\,0}$ and $u_{\,0}\,$;
\State Increment: $t\, \coloneqq\, t\plus \Delta t $ and $n\, \coloneqq\, n \plus 1\,$;
\EndWhile\label{euclidendwhile}
\State \textbf{return} $v\,(x,\,t)$, $v_{\,x}\,(x,\,t)\,$, $u\,(x,\,t)$  and  $u_{\,x}\,(x,\,t)\,$.
\end{algorithmic}
\end{algorithm}

When dealing with highly coupled equations, the issue is to rewrite System~\eqref{eq:HAM_equation_dimless} in a vectorized way, such as Equation~\eqref{eq:HAM_dimlss_vecto}, and solve it at as only one equation. In the next section, the MOHL is tested in two cases of numerical application.


\section{Numerical benchmark}
\label{sec:numeric_app}

Simulations of one-dimensional coupled heat and moisture transport are carried out with the MOHL method, the \textsc{Euler} implicit method used in a numbers of works (see for instance \cite{Mendes2005, Janssen2014}) and a reference solution to verify the applicability and the advantages of the proposed method. The \textsc{Euler} implicit scheme approximates the continuous operator to the order $\O\,(\dx^{\,2}\, +\, \dt)\,$.  The reference solutions are computed using the \texttt{Matlab\texttrademark} open source toolbox \texttt{Chebfun} \cite{Janssen2014}.

To compare and validate the proposed method, the error between solution components $u^{\, \mathrm{num}}$ and $v^{\, \mathrm{num}}$, obtained by the MOHL method or the \textsc{Euler} implicit, and the reference solutions $u^{\, \mathrm{ref}}$ and $v^{\, \mathrm{ref}}$, are computed as functions of $x$ using the following Euclidean norms:
\begin{align*}
  \varepsilon_{\,2,\,u}\, (x)\ &\eqdef\ \sqrt{\,\frac{1}{N_{\,t}} \, \sum_{j\, =\, 1}^{N_{\,t}} \, \Bigl( \, u_{\, j}^{\, \mathrm{num}}\, ( x\,, t_{\,j}) \moins u_{\, j}^{\mathrm{\, ref}}\, (x\,, t_{\,j}) \, \Bigr)^{\,2}}\,, \\ 
  \varepsilon_{\,2,\,v}\, (x)\ &\eqdef\ \sqrt{\,\frac{1}{N_{\,t}} \, \sum_{j\, =\, 1}^{N_{\,t}} \, \Bigl( \, v_{\, j}^{\, \mathrm{num}}\, ( x\,, t_{\,j}) \moins v_{\, j}^{\mathrm{\,ref}}\, ( x\,, t_{\,j}) \, \Bigr)^{\,2}}\,,
\end{align*}
where $N_{\,t}$ is the number of temporal steps. The uniform norm errors $\varepsilon_{\,\infty,\,u}$ and $\varepsilon_{\,\infty,\,v}$ are given by the maximal values of $\varepsilon_{\,2,\,u}\,(x)$ and $\varepsilon_{\,2,\,v}\,(x)\,$: 
\begin{align*}
  \varepsilon_{\,\infty,\,u}\ \eqdef\ \sup_{x \ \in \ \bigl[\, 0 ,\, L \,\bigr]} \, \varepsilon_{\,2,\,u}\, (x) \,, & &
  \varepsilon_{\,\infty,\,v}\ \eqdef\ \sup_{x \ \in \ \bigl[\, 0 ,\, L \,\bigr]} \, \varepsilon_{\,2,\,v}\, (x) \,.
\end{align*}

In the aftermath, the numerical case studies are presented.


\subsection{Single layer case} 
\label{sec:validat}

In order to contemplate the applicability of the presented method, the first case study considers the combined effects of the heat and moisture transfer. A physical application is performed in a $10$-$\mathsf{cm}$ monolithic bearing material. To test the robustness of the scheme with strong nonlinearities, the properties of the material are gathered from \cite{Hagentoft2004} and reminded in Table~\ref{table:properties_mat1} of the Appendix~\ref{annexe:material_properties}. Initial conditions are considered uniform over the spatial domain, with an initial temperature of $\Ti \egalb 293.15\ \mathsf{K}$ and an initial vapour pressure of $\Pvi \egalb 1636.53\ \mathsf{Pa}\,$, referent to the relative humidity of $70\, \%\,$. 

\begin{figure}
\begin{center}
  \subfigure[][\label{fig_AN1:BC_T}]{\includegraphics[width=.48\textwidth]{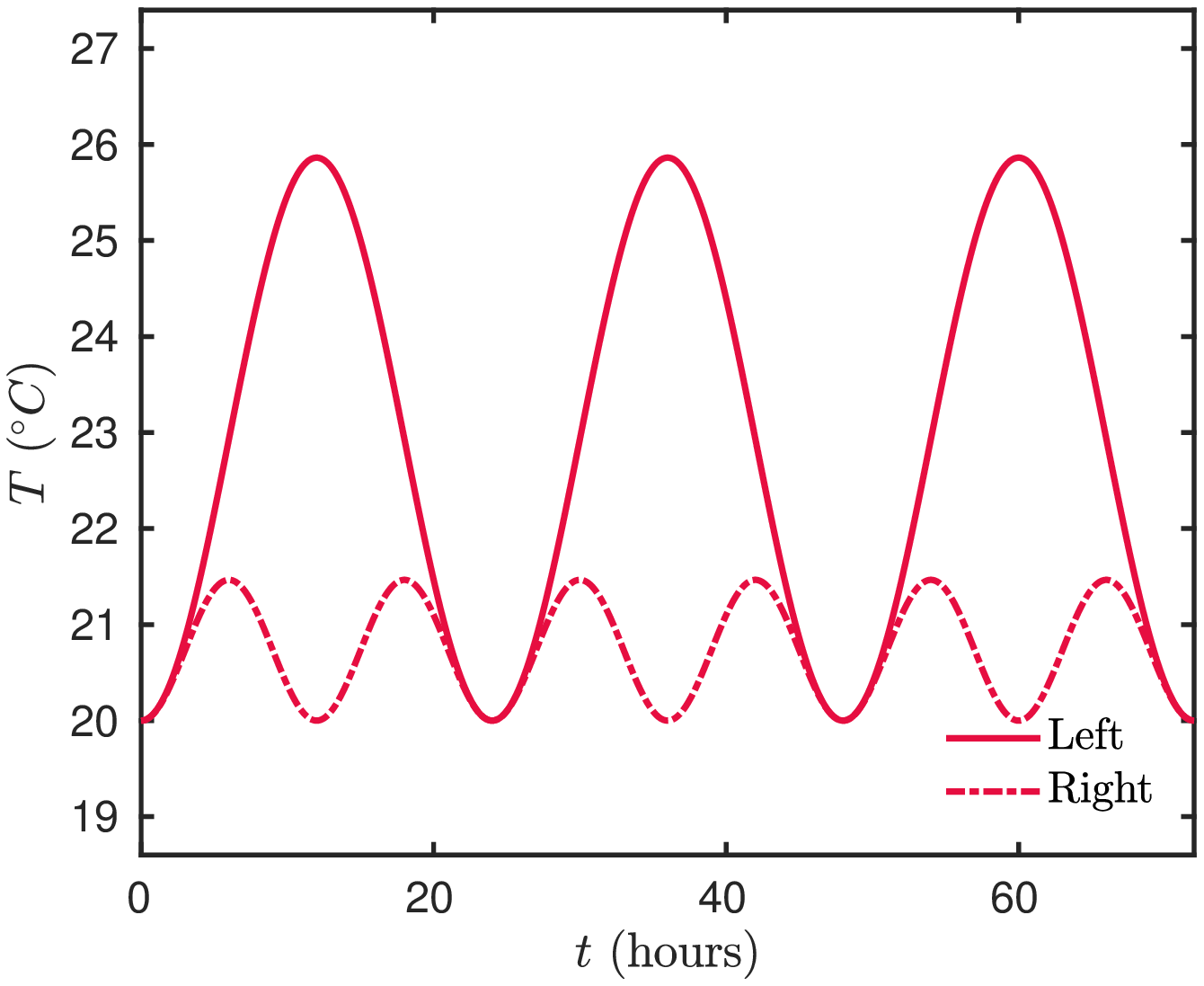}} \hspace{0.3cm}
  \subfigure[][\label{fig_AN1:BC_RH}]{\includegraphics[width=.485\textwidth]{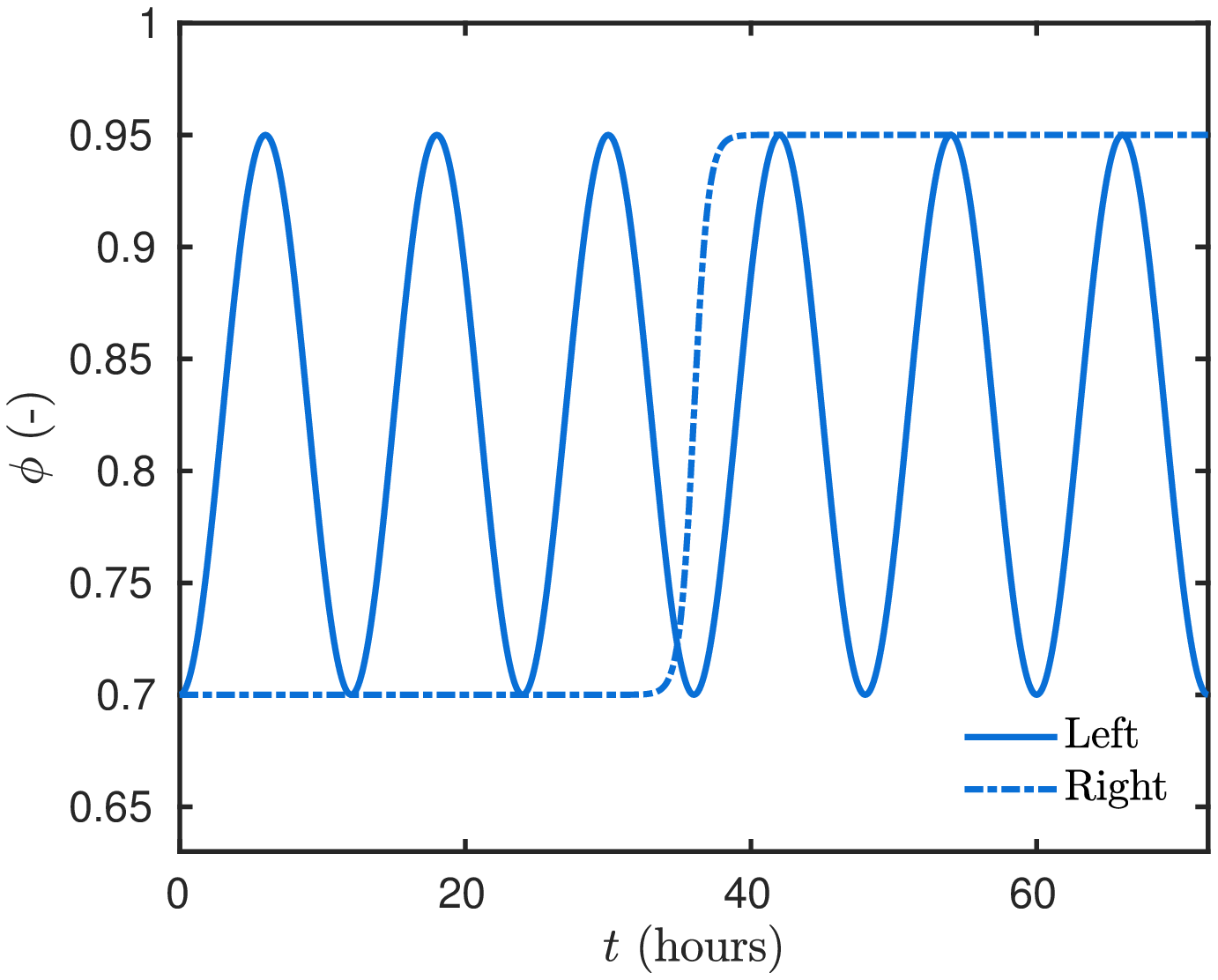}} 
  \caption{\small\em Temperature $T$ (a) and relative humidity $\phi$ (b) boundary conditions.}
  \label{fig_AN1:BC}
\end{center}
\end{figure}

The boundary conditions, represented by the relative humidity $\phi$ and temperature $T\,$, are shown in Figure~\ref{fig_AN1:BC}. They oscillate during $72$ hours of simulation. The convective mass and heat transfer coefficients are set to $\hM  \egalb 2 \cdot 10^{\,-7} \ \unitfrac{s}{m}$ and $\hT \egalb 25 \ \unitfrac{W}{(m^{\,2}\cdot K)}$ at the left boundary, and to $\hM \egalb 3 \cdot 10^{\,-8} \ \unitfrac{s}{m}$ and $\hT \egalb 8 \ \unitfrac{W}{(m^{\,2}\cdot K)}\,$, at the right boundary. No additional source term is considered for the moment.

Simulations with the MOHL method were performed using the solver \texttt{bvp4c} with the relative and absolute tolerances set to $\mathsf{tol} \egalb 10^{\,-5}$ for moisture and $\mathsf{tol} \egalb 10^{\,-6}$ for temperature. The time step discretization is of $\Delta \ts \egalb 10^{\,-1}$ and, for the space discretization an adaptive technique is employed, as explained in Section~\ref{sec:num}. The initial guess is composed by $10$ spatial nodes, for vapour pressure and temperature fields. For the \textsc{Euler} Implicit solution, simulations have been performed with $\Delta \xs \egalb 10^{\,-2}$ and $\Delta \ts \egalb 10^{\,-2}$.

\begin{figure}
\begin{center}
  \subfigure[][\label{fig_AN1:BVP_profile_T}]{\includegraphics[width=.48\textwidth]{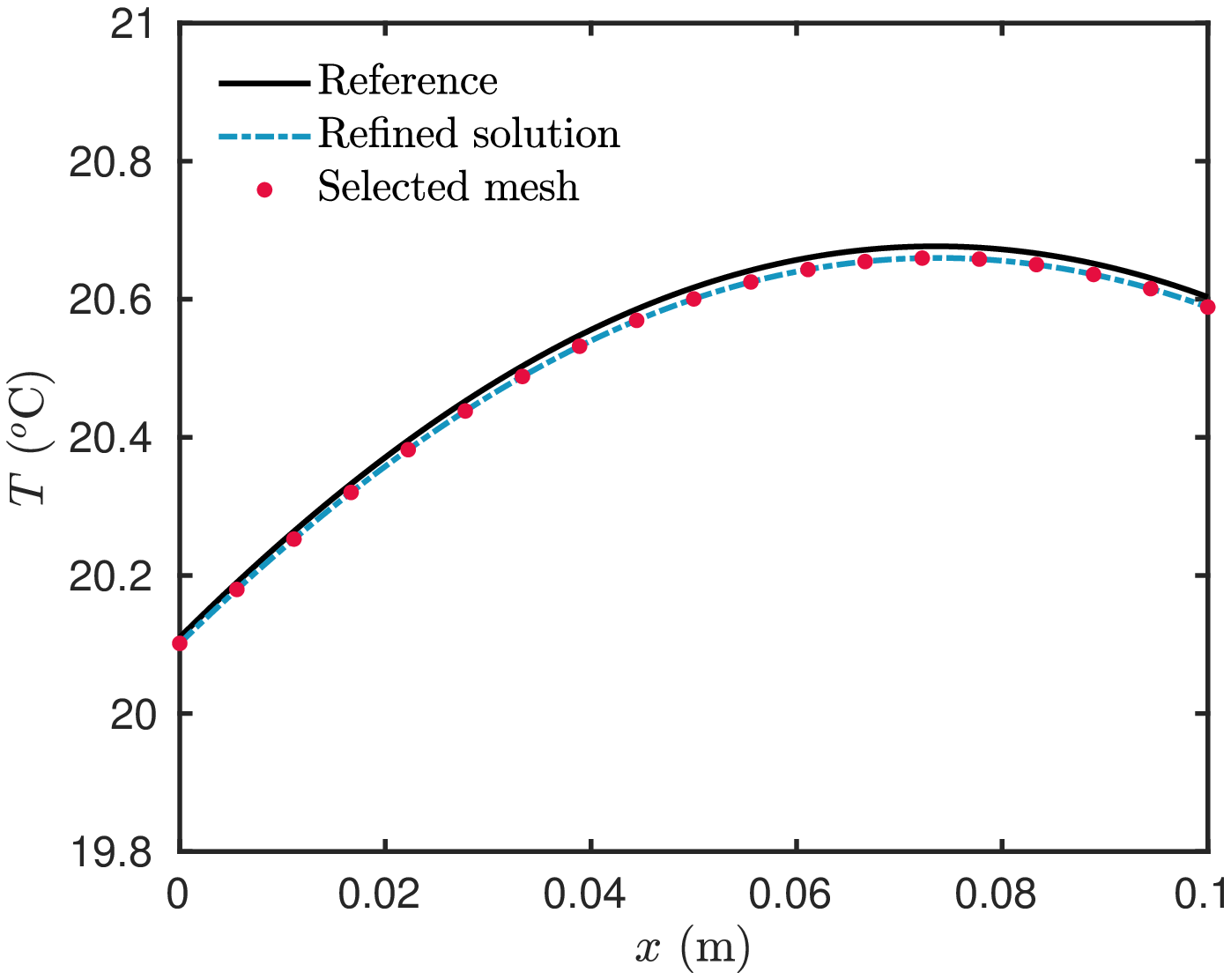}} \hspace{0.3cm}
  \subfigure[][\label{fig_AN1:BVP_profile_Pv}]{\includegraphics[width=.48\textwidth]{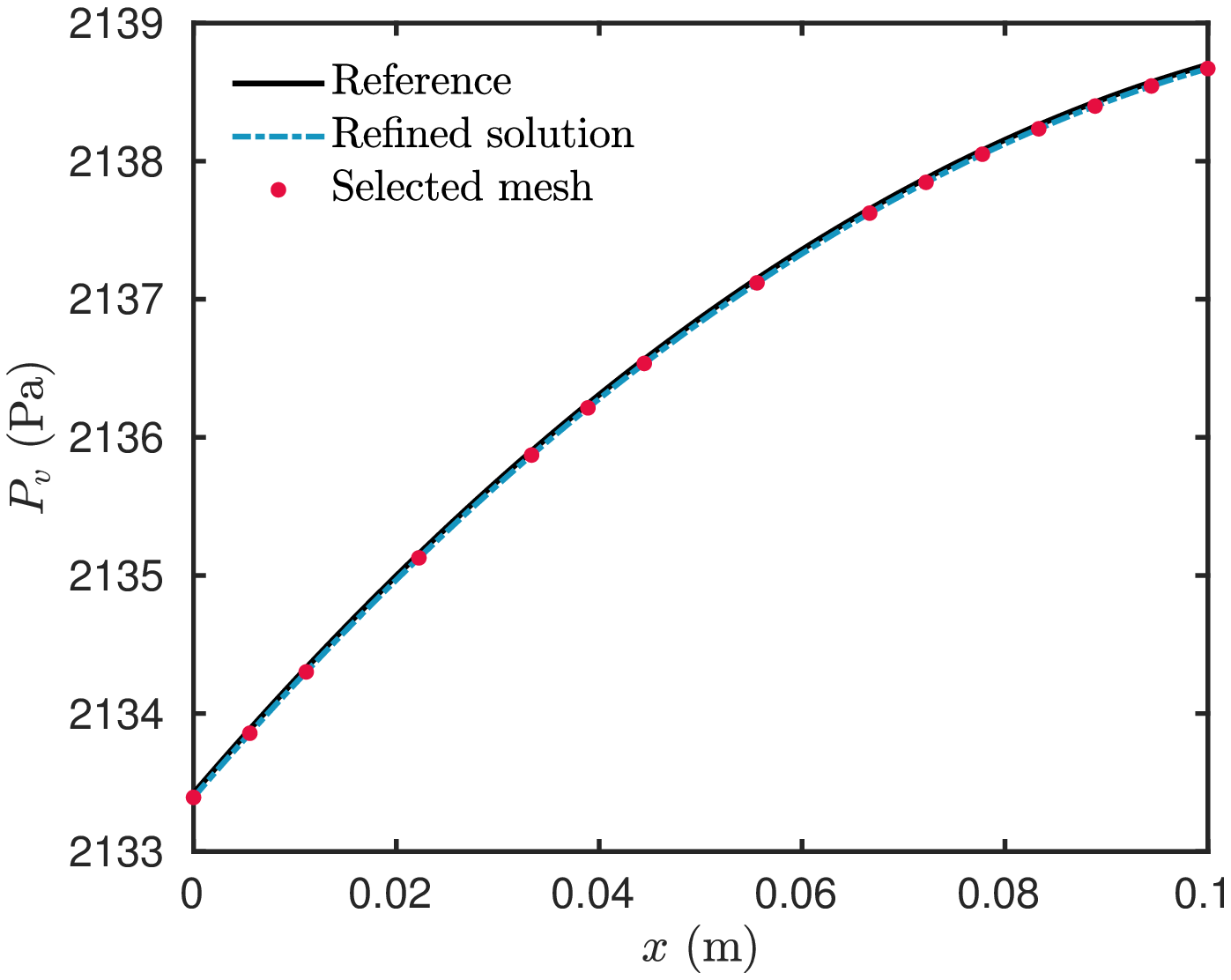}} 
  \caption{\small\em Selected mesh determined by \texttt{bvp4c} for the last profile (at $t\egalb 72\ \mathsf{h}$) of temperature (a) and vapour pressure (b) fields.}
  \label{fig_AN1:BVP_profile}
\end{center}
\end{figure}

Figure~\ref{fig_AN1:BVP_profile} presents the selected mesh determined by the solver for the last profile of temperature and vapour pressure fields, with the reference solution and with the refined solution. This last one is an evaluation of the solution that covers all the interval $[\,0,\,L\,]$ in more points than the selected mesh. Note that at each time layer the number of points of the selected mesh varies, with a maximum of $15$ points for the moisture field and $19$ points for the temperature field. Over the simulation time the number of nodes remained almost the same, which means it was sufficient for the set tolerance. If we decrease the tolerance, the number of nodes of the spatial grid would rise accordingly.

\begin{figure}
\begin{center}
  \subfigure[][\label{fig_AN1:evolution_T}]{\includegraphics[width=.47\textwidth]{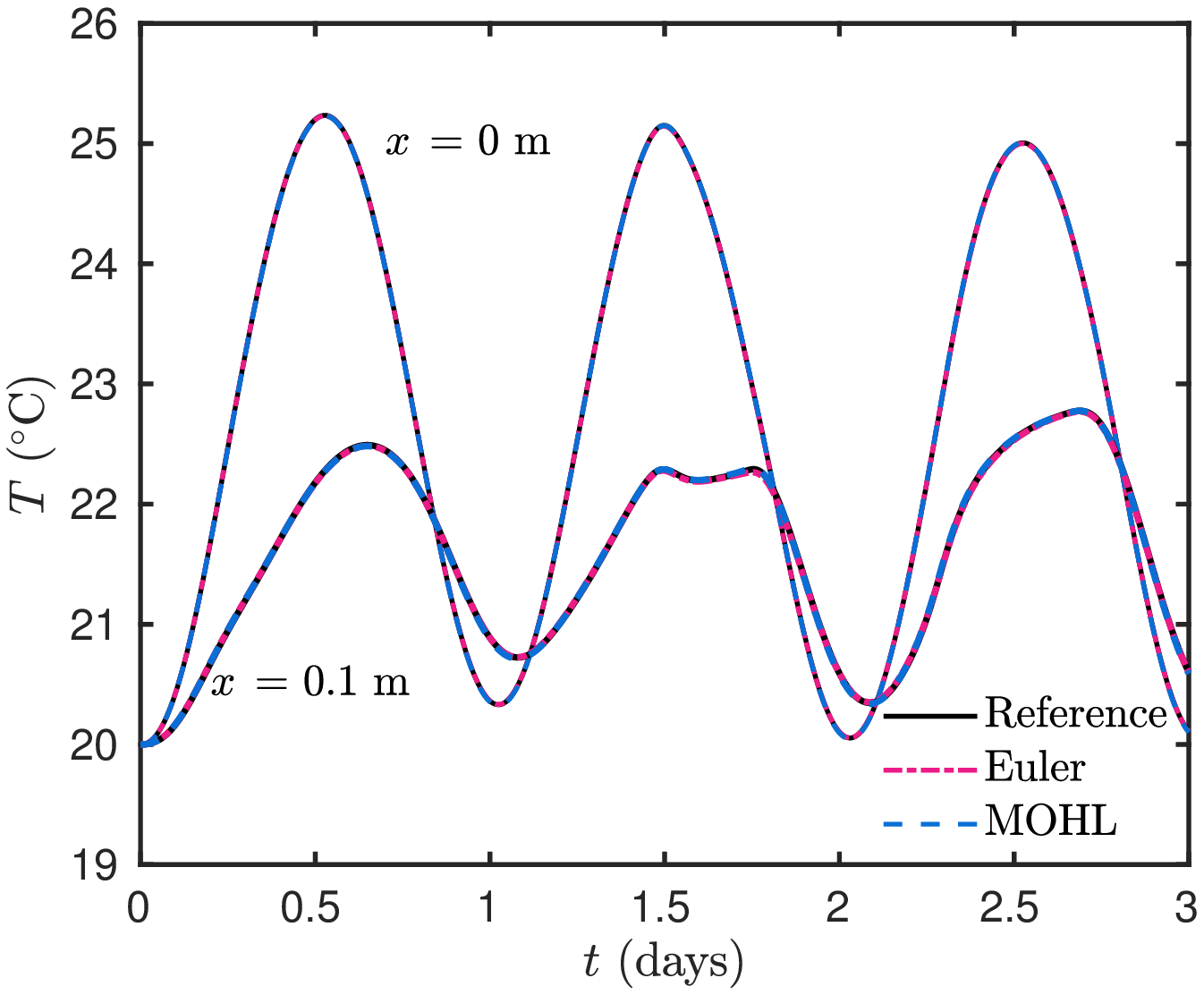}} \hspace{0.3cm}
  \subfigure[][\label{fig_AN1:evolution_Pv}]{\includegraphics[width=.48\textwidth]{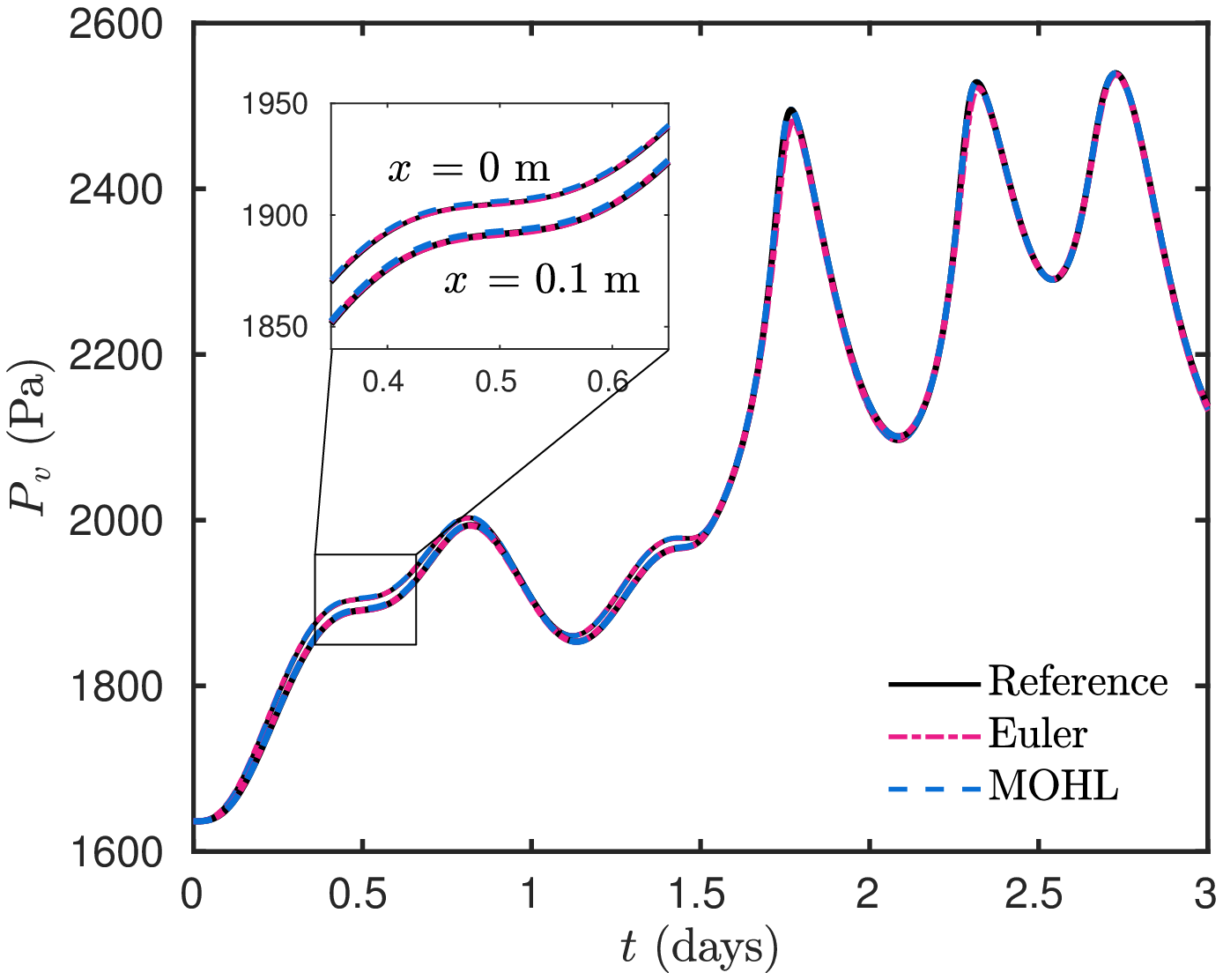}} \\
  \subfigure[][\label{fig_AN1:evolution_RH}]{\includegraphics[width=.48\textwidth]{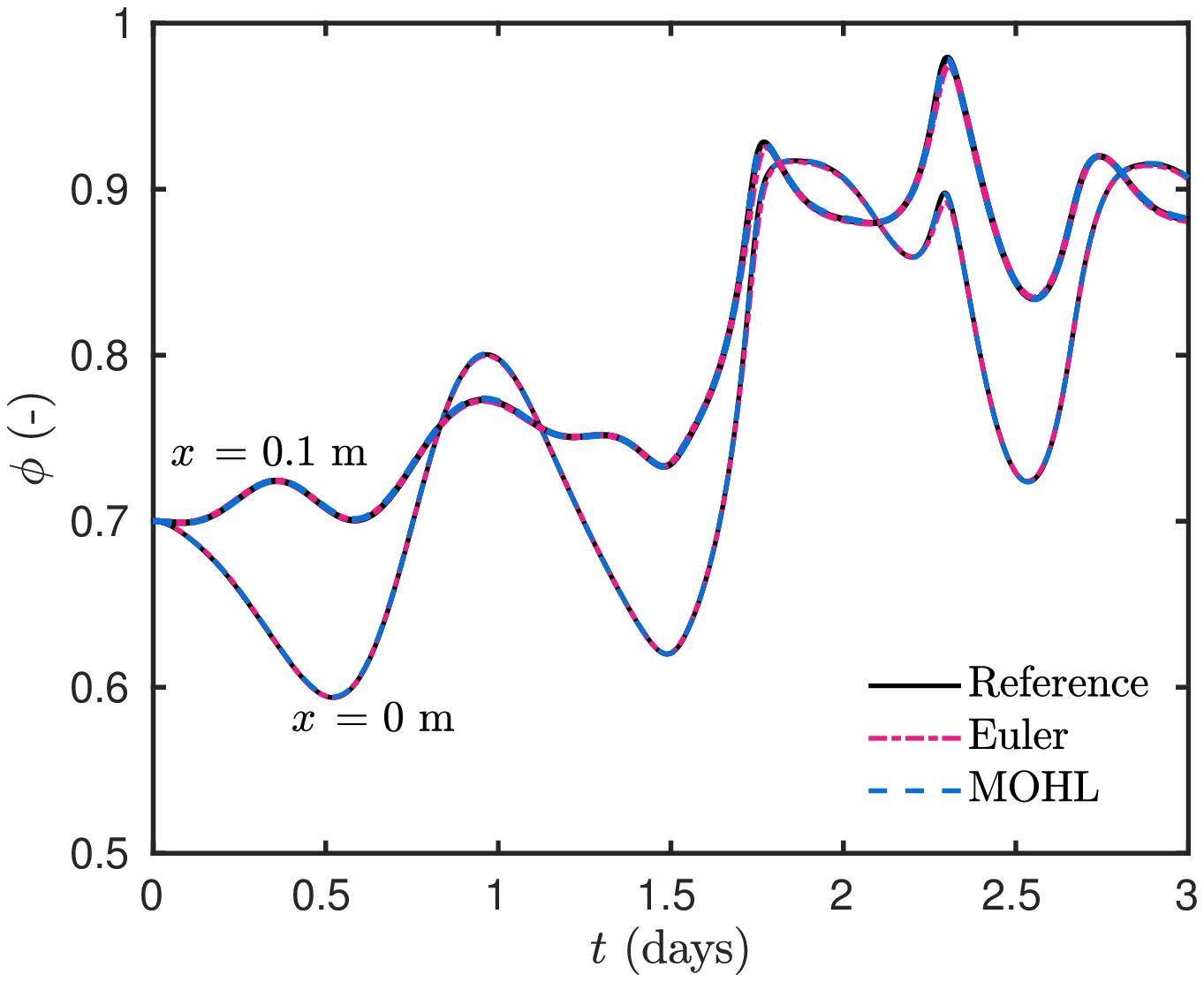}} 
  \caption{\small\em Evolution of the temperature (a), vapour pressure (b)  and of the relative humidity (c) on the edges of the material (at $x \egalb 0\ \mathsf{m}$ and $x \egalb 0.1\ \mathsf{m}\,$).}
  \label{fig_AN1:evolution}
\end{center}
\end{figure}

Simulation results of the temperature, vapour pressure and relative humidity evolution on the boundaries of the building component ($x \egalb 0\ \mathsf{m}$ and $x \egalb 0.1\ \mathsf{m}\,$) are presented in Figures~\ref{fig_AN1:evolution_T}, \ref{fig_AN1:evolution_Pv} and \ref{fig_AN1:evolution_RH}, respectively. The proposed method has been able to follow successfully the solicitations from the boundary conditions and represent the solution. The step on the relative humidity at the right boundary can be observed on the vapour pressure and relative humidity evolution. The vapour pressure diffuses over the material almost instantaneously with the step imposed. The level of moisture is high but the material is not saturated. It can be noted that for the simulation of such cases with really high levels of moisture, it is better to change the physical model using, for instance, the moisture content or the capillary pressure as potentials \cite{Mendes2005}.

\begin{figure}
\begin{center}
\subfigure[][\label{fig_AN1:profiles_T}]{\includegraphics[width=.47\textwidth]{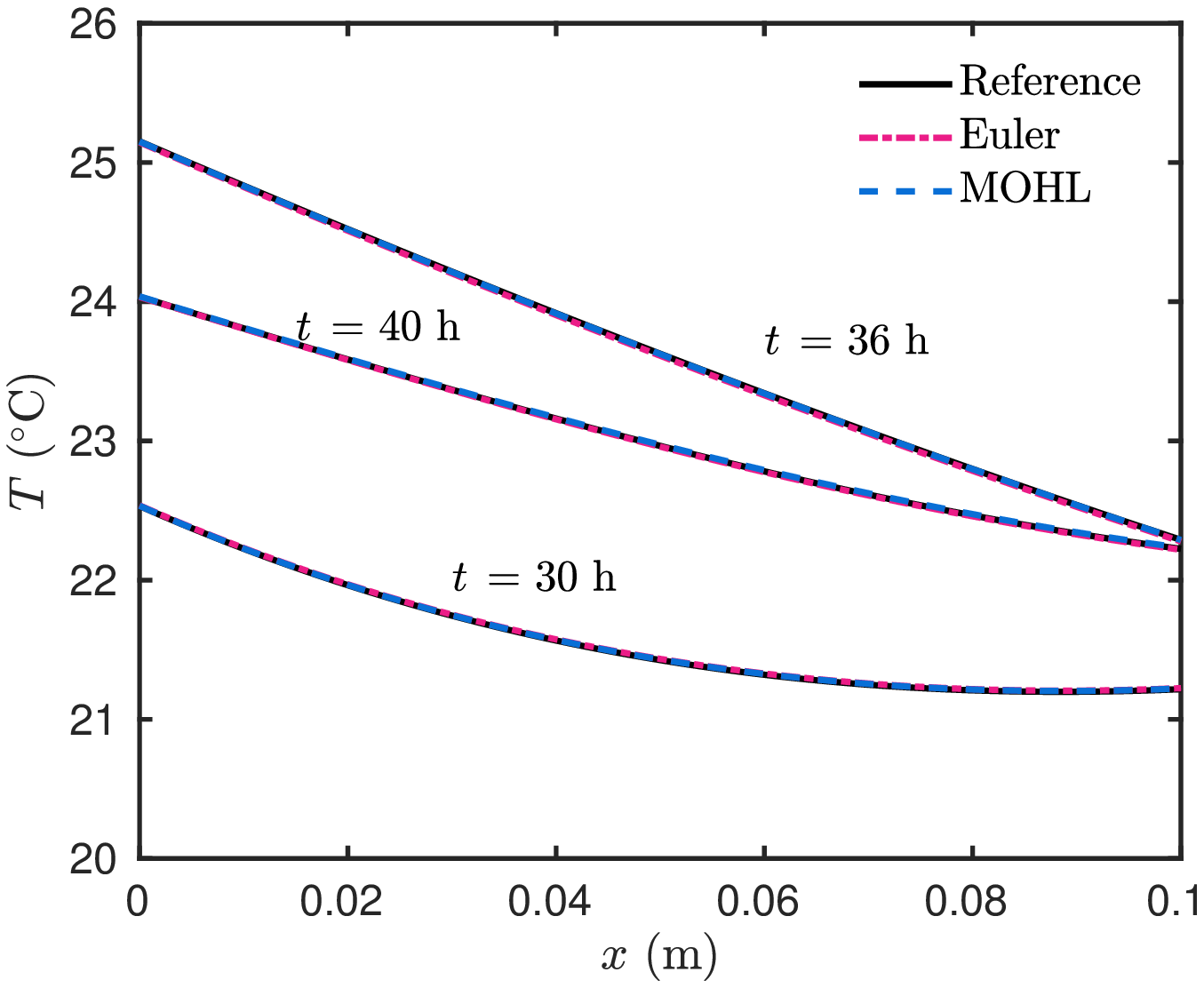}} \hspace{0.3cm}
\subfigure[][\label{fig_AN1:profiles_RH}]{\includegraphics[width=.48\textwidth]{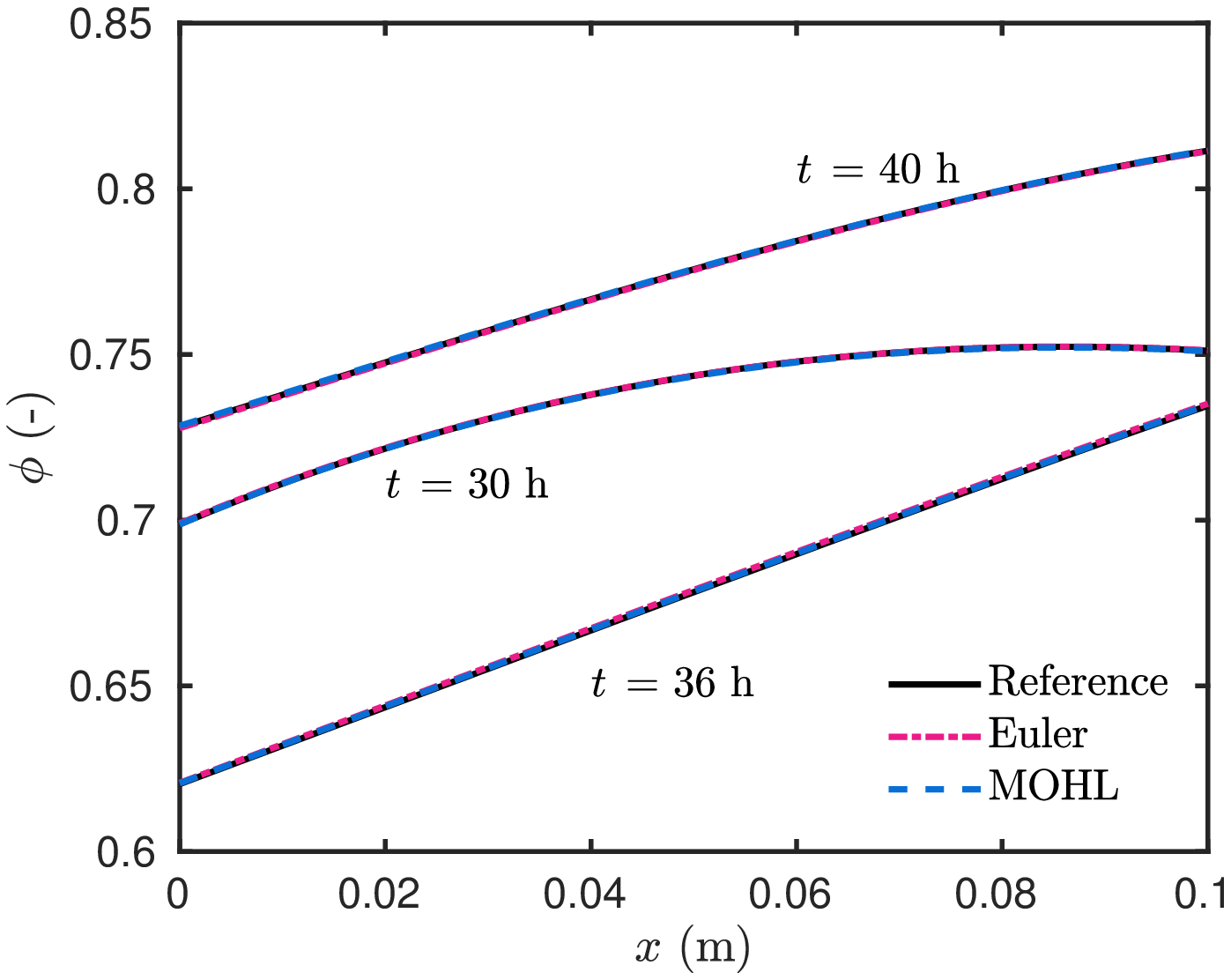}} 
\caption{\small\em Temperature (a) and relative humidity (b) profiles at different times.}
\label{fig_AN1:profile}
\end{center}
\end{figure}

The computed temperature and relative humidity distribution profiles for $t\egalb \{\,30,\,36,\,40\,\}\ \mathsf{h}$ are presented in Figures~\ref{fig_AN1:profiles_T} and \ref{fig_AN1:profiles_RH}. As this material has a fast liquid transfer, the profiles of relative humidity are almost straight lines. The material is highly permeable but with a low hygroscopicity, as it does not retain the moisture. For the temperature, it can also be observed that it diffuses quickly and it has the opposite effect of the relative humidity. Furthermore, it can be noted a good agreement between the MOHL method, the \textsc{Euler} implicit and the reference solution.

\begin{figure}
\begin{center}
  \includegraphics[width=.7\textwidth]{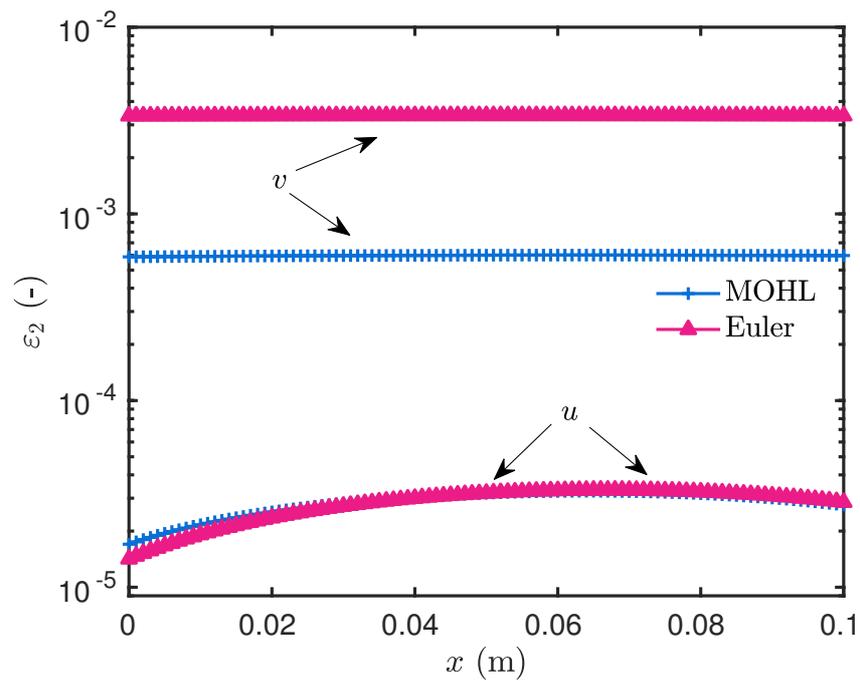}
  \caption{\small\em Error $\varepsilon_{\,2}$ computed over the time with the MOHL and the \textsc{Euler} approaches, for the dimensionless temperature -$u$- and vapour pressure -$v$- solutions.}
\label{fig_AN1:err_fx}
\end{center}
\end{figure}

The error $\varepsilon_{\,2}$ between the MOHL and the \textsc{Euler} implicit method with the reference solution are given in Figure~\ref{fig_AN1:err_fx}. For the temperature distribution, the error is to the order of $\O\,\bigl(10^{\,-5}\bigr)$ and, for the vapour pressure component is to the order of $\O\,\bigl(10^{\,-3}\bigr)\,$. The error is not of the same order for each solutions perhaps due to the following reasons: \textit{(i)} the properties are more nonlinear for the moisture balance equations; \textit{(ii)} or the residual effect of initial guess for the vapour pressure profile; \textit{(iii)} or moisture diffusive is much slower than heat, for the present case; \textit{(iv)} the order of the dimensionless fields $u$ and $v$ are not the same. Even with this difference, these results are consistent to the tolerance, that was set to $\mathsf{tol} \egalb 10^{\,-5}$ and to the discretization on time, that is second-order accurate $\O\,(\Delta t^{\,2})\,$. It is admissible that the actual error is slight larger than the prescribed tolerance. The solver estimates the residual and not the actual error \textit{strictu sensu} (since the exact solution is unavailable). It is required that the error decrease when the tolerance decrease as shown in Table~\ref{tab:tol_bvp}.

The accuracy of the solution depends on several factors, one of them is the absolute and relative tolerances of the solver. To compare different values of the tolerance with the solver \texttt{bvp4c}, Table~\ref{tab:tol_bvp} presents the error $\varepsilon_{\, \infty}$ for both solutions, the dimensionless temperature ($u$) and the dimensionless vapour pressure ($v$), as well as the number of mesh points used by the solver. As the tolerances get more restrictive, the solver adapts the solution making the number of mesh points to increase to provide a consistent solution with the specified accuracy. The initial number of points was $10\,$. For the tolerance $\mathsf{tol}\, \leqslant \,10^{\,-3}$, more than $10$ nodes are needed to compute the solution. For this case study, a tolerance of $\mathsf{tol} \egalb 10^{\,-3}$ would be enough to provide an accuracy to order of $\O (10^{\,-4})$.

\begin{table}
\centering
\small
\caption{\small\em Maximum error $\varepsilon_{\,\infty}$ computed with different tolerances with the solver \texttt{bvp4c}.}
\bigskip
\label{tab:tol_bvp}
\setlength{\extrarowheight}{.3em}
\begin{tabular}{cccccc}
\hline 
$\mathsf{tol}$ & $\varepsilon_{\, \infty,\,u}$ & \textit{mesh}  & $\varepsilon_{\, \infty,\, v}$ & \textit{mesh} & \textit{CPU time} \\
\hline
$10^{\,-6}$ & $3.5\cdot 10^{\,-5}$ & $19$  &  $2.1\cdot 10^{\,-4}$ & $21$ & $21.9\ \mathsf{s}$ \\
$10^{\,-5}$ & $1.5\cdot 10^{\,-4}$ & $10$  &  $4.5\cdot 10^{\,-4}$ & $15$ & $15.9\ \mathsf{s}$ \\
$10^{\,-4}$ & $1.5\cdot 10^{\,-4}$ & $10$  &  $8.9\cdot 10^{\,-4}$ & $11$ & $15.3\ \mathsf{s}$ \\
$10^{\,-3}$ & $1.5\cdot 10^{\,-4}$ & $10$  &  $5.4\cdot 10^{\,-4}$ & $10$ & $13.9\ \mathsf{s}$ \\
$10^{\,-2}$ & $1.5\cdot 10^{\,-4}$ & $10$  &  $2.3\cdot 10^{\,-4}$ & $10$ & $13.3\ \mathsf{s}$ \\
$10^{\,-1}$ & $1.5\cdot 10^{\,-4}$ & $10$  &  $7.2\cdot 10^{\,-3}$ & $10$ & $13.1\ \mathsf{s}$ \\
\hline
\end{tabular}
\end{table}

The CPU time required to compute the solution of this case study is provided in Table~\ref{tab:cpu_time}. The compuational time required for the \textsc{Euler} explicit approach is also included. It has been evaluated using \texttt{Matlab\texttrademark} platform on a computer with a processor Intel\textsuperscript{\textregistered} Core i$5$ CPU $2.80$ GHz and $15.6$ GB of RAM memory. The \textsc{Euler} explicit requires more time due to the CFL stability condition, since it is not possible to increase the time discretization. For the \textsc{Euler} implicit solution, it takes an average of $11$ iterations at each time step to converge to the solution which increases the global computational time of the method. The MOHL method is faster than the two \textsc{Euler} approaches implemented here. It can be remarked that for the same order of accuracy of the solution, the proposed method is two times faster. Moreover, as it can be seen in Table~\ref{tab:tol_bvp}, the accuracy of the solution can be increased with a relatively low increase of the spatial mesh and of the CPU time.

\begin{table}
\centering
\small
\caption{\small\em Features and CPU time of the methods implemented in the single layer case study.}
\label{tab:cpu_time}
\bigskip
\setlength{\extrarowheight}{.3em}
\begin{tabular}{cccc}
\hline 
  & \textit{BVP} & \textsc{Euler} \textit{implicit}  & \textsc{Euler} \textit{explicit} \\
\hline
$\Delta \ts$ & $10^{\,-1}$ & $10^{\,-2}$  &  $10^{\,-5}$  \\
$N_{\,x,\,u}$ & $10$ & $101$  &  $201$  \\
$N_{\,x,\,v}$ & $15$ & $101$  &  $201$  \\
$\varepsilon_{\, \infty,\,u}$ & $\O (10^{\,-4})$ & $\O (10^{\,-4})$  &  $\O (10^{\,-4})$  \\
$\varepsilon_{\, \infty,\,v}$ & $\O (10^{\,-4})$ & $\O (10^{\,-4})$  &  $\O (10^{\,-4})$  \\
\hline
\rowcolor[HTML]{C0C0C0} 
CPU time $\bigl[\,\mathsf{s}\,\bigr]$ & $18.5$ & $57.5$  &  $2572$  \\
\rowcolor[HTML]{C0C0C0} 
CPU time $\bigl[\,\mathsf{-}\,\bigr]$ & $0.49$ & $1$  &  $44$  \\
\hline
\end{tabular}
\end{table}

\begin{figure}
\begin{center}
\subfigure[][\label{fig_AN1:heat_flux_left}]{\includegraphics[width=.48\textwidth]{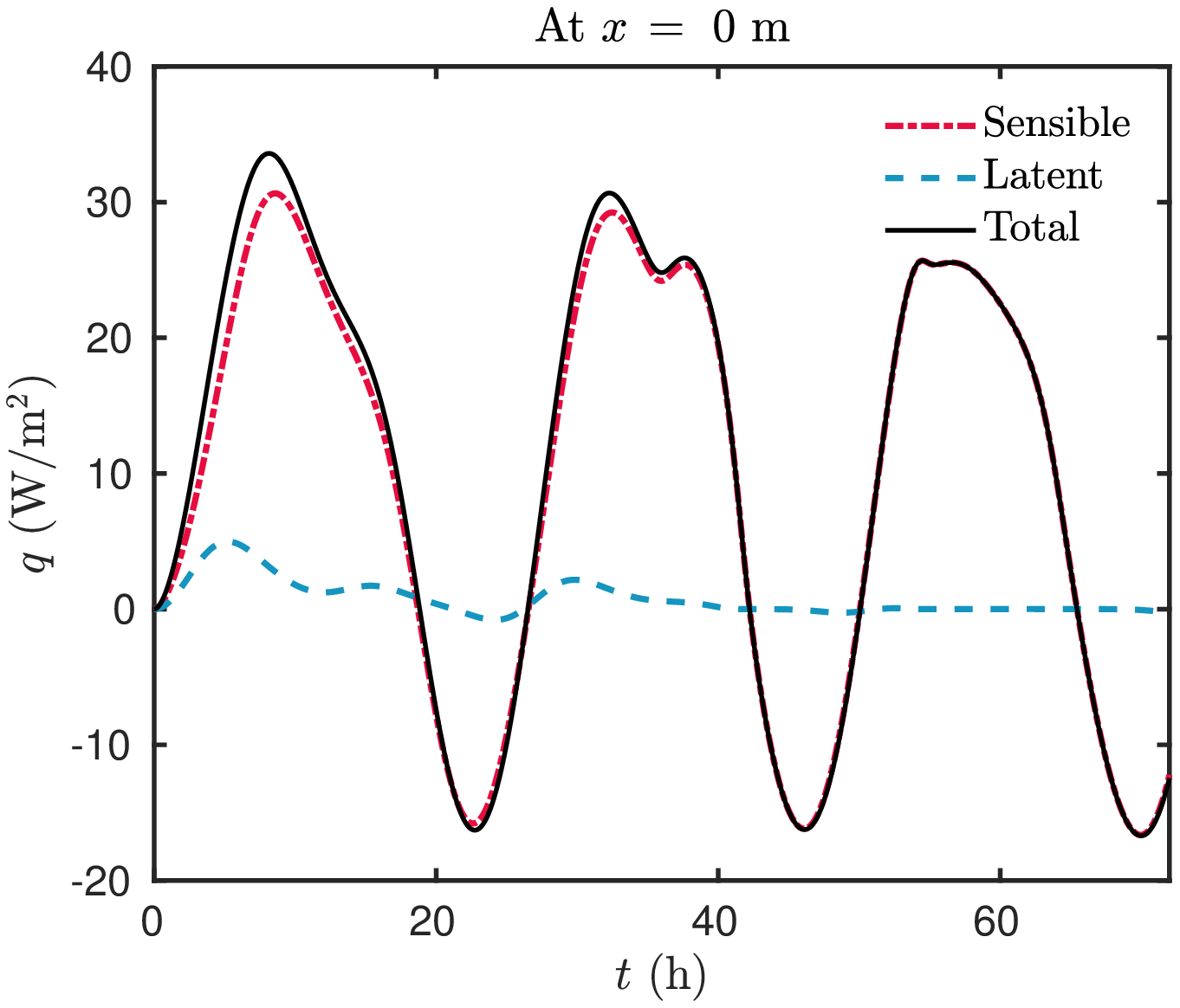}} \hspace{0.3cm}
\subfigure[][\label{fig_AN1:heat_flux_right}]{\includegraphics[width=.48\textwidth]{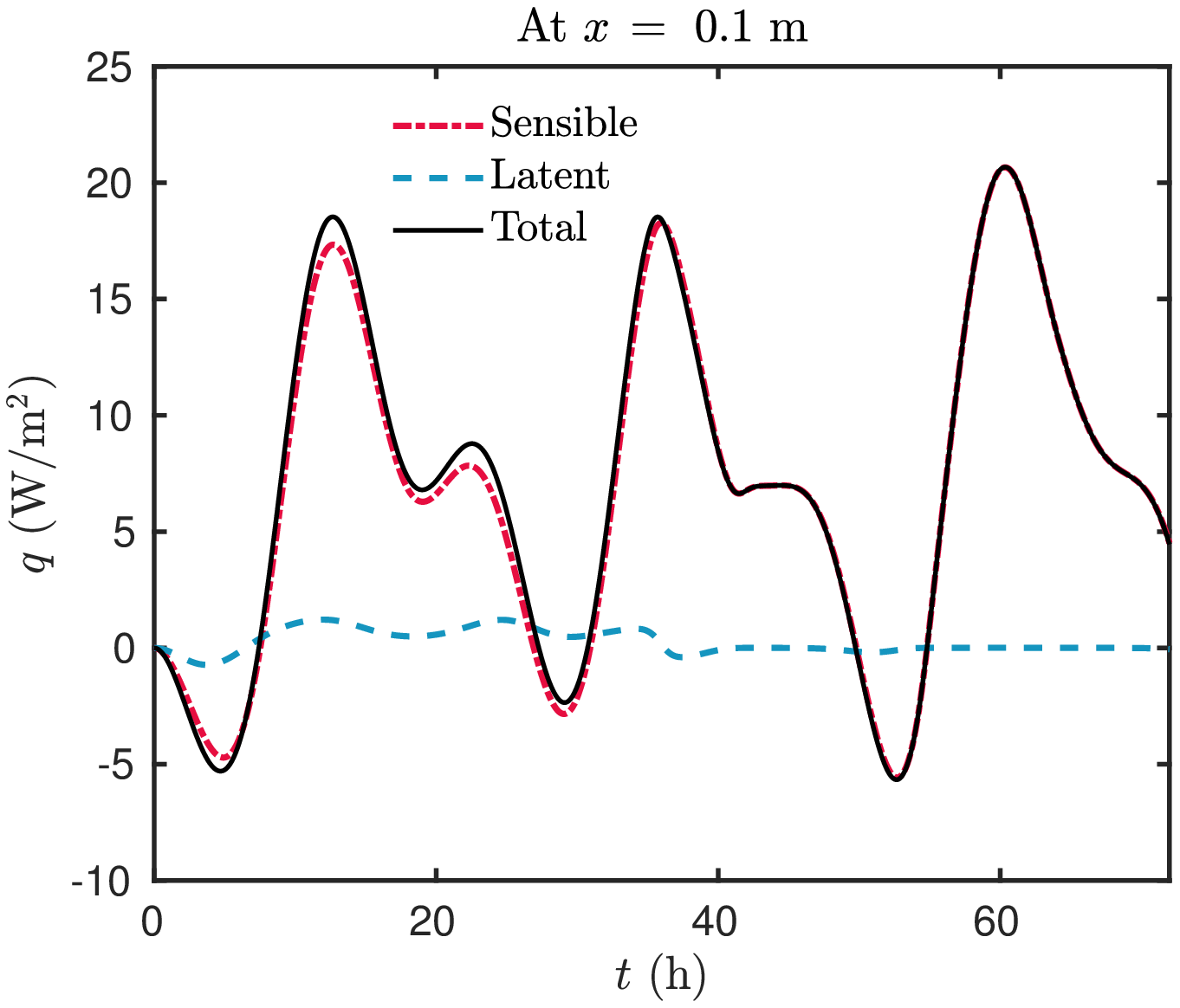}} 
\caption{\small\em Sensible, latent and total heat fluxes at the left boundary and (a) at the right boundary (b), for the one-layer case.}
\label{fig_AN1:heat_flux}
\end{center}
\end{figure}

One of the advantages of using the solver \texttt{bvp4c} is that it has as output the respective derivatives of the fields, making the computation of the fluxes straightforward. Thus, Figures~\ref{fig_AN1:heat_flux_left} and \ref{fig_AN1:heat_flux_right} present the sensible and latent heat fluxes at the left and right boundaries. As illustrated in those figures, the sensible heat flux is the main component of the surface energy balance. The latent heat flux at the right boundary slightly changes with the step on the relative humidity at $t \egalb 36 \ \mathsf{h}\,$, showing its little influence on the heat diffusion.

\begin{figure}
\begin{center}
\includegraphics[width=.699\textwidth]{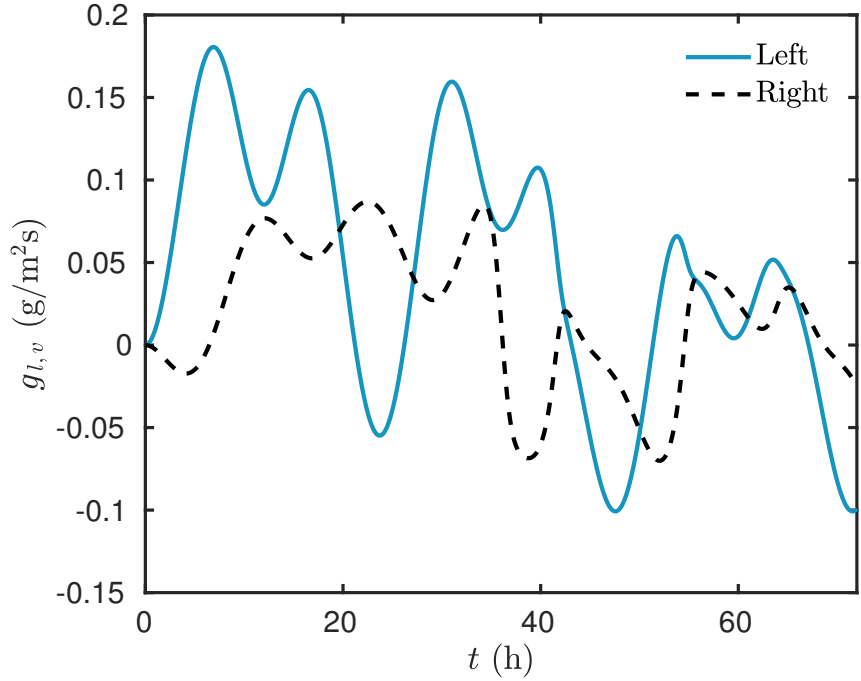}
\caption{\small\em Moisture flow at the boundaries, for the one-layer case.}
\label{fig_AN1:moisture_flow}
\end{center}
\end{figure}

The total moisture flow (liquid plus vapour) at both boundaries are shown in Figure~\ref{fig_AN1:moisture_flow}. As it can be observed, the flow follows variations of the boundary conditions. Furthermore, the left boundary has higher variations because its convective mass transfer coefficient is higher than the one at the other boundary. The step of relative humidity at the right boundary can also be observed on the moisture flow which suddenly changes.

\subsection{Multilayered domain}
\label{sec:layered_case}

In building constructions, multiple layers are commonly found. The configuration assumed at the interface between materials follows the hydraulic continuity ($P_{\,v,\,1} \egalb P_{\,v,\,2}$ and $T_{\,v,\,1} \egalb T_{\,v,\,2}\,$), which considers interpenetration of both porous structure layers \cite{DeFreitas1996}. Both materials are homogeneous and isotropic, and only heat and moisture transfer are simulated, through a perfectly airtight structure.

Consider Equation~\eqref{eq:heat_equation_dimless} over a multidomain in which $\xs\, \in\, [\, 0,\,x_{\,\text{int}}^{\,\star}\,)\, \cup\, [\, x_{\,\text{int}}^{\,\star},\,1\,]\,$. The coefficients are written in a general form as illustrated for the moisture diffusion:
\begin{align}\label{eq:properi_multidomain}
  \kMs (v,\, \xs) \egal \left\{ \begin{array}{cl}
  k_{M,\,1}^{\,\star}\, (v)\,, & \ \xs < x_{\,\text{int}}^{\,\star}\,, \\
  k_{M,\,2}^{\,\star}\, (v)\,, & \ \xs \geq x_{\,\text{int}}^{\,\star} \,,
  \end{array} \right.
\end{align}
where $x_{\,\text{int}}^{\,\star}$ is the location of the interface between materials and the indices $1$ and $2$ represent each material layer.

Thus, by using the MOHL approach, it is assumed that at the interface, the solution and the solution derivative of the problem are continuous. In this way, the method will search for the solution that can satisfy both conditions automatically. These interface assumptions are verified with an analytical solution in the Appendix \ref{annexe:analytical_sol_multi}.


\subsubsection{Application}

\begin{figure}
\begin{center}
\includegraphics[width=.75\textwidth]{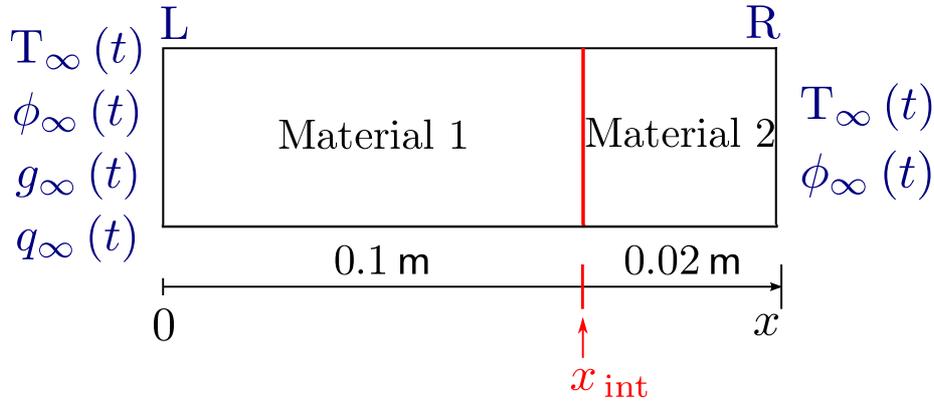}
\caption{\small\em Schematic representation of the two-layered wall.}
\label{fig_AN3:layred_configuration}
\end{center}
\end{figure}

This case study considers a porous wall formed by $2$ layers: $10$-$\mathsf{cm}$ of a load bearing material and $2$-$\mathsf{cm}$ of finishing material. Figure~\ref{fig_AN3:layred_configuration} shows schematically this physical situation. The selected materials further complicate the case, with the first layer having a faster liquid transfer while the second layer acts as a hygroscopic finish. The properties used for these materials were obtained from \cite{Hagentoft2004} which are given in Tables~\ref{table:properties_mat1} and \ref{table:properties_mat2}.

\begin{figure}
\begin{center}
  \subfigure[][\label{fig_AN3:BC_T}]{\includegraphics[width=.47\textwidth]{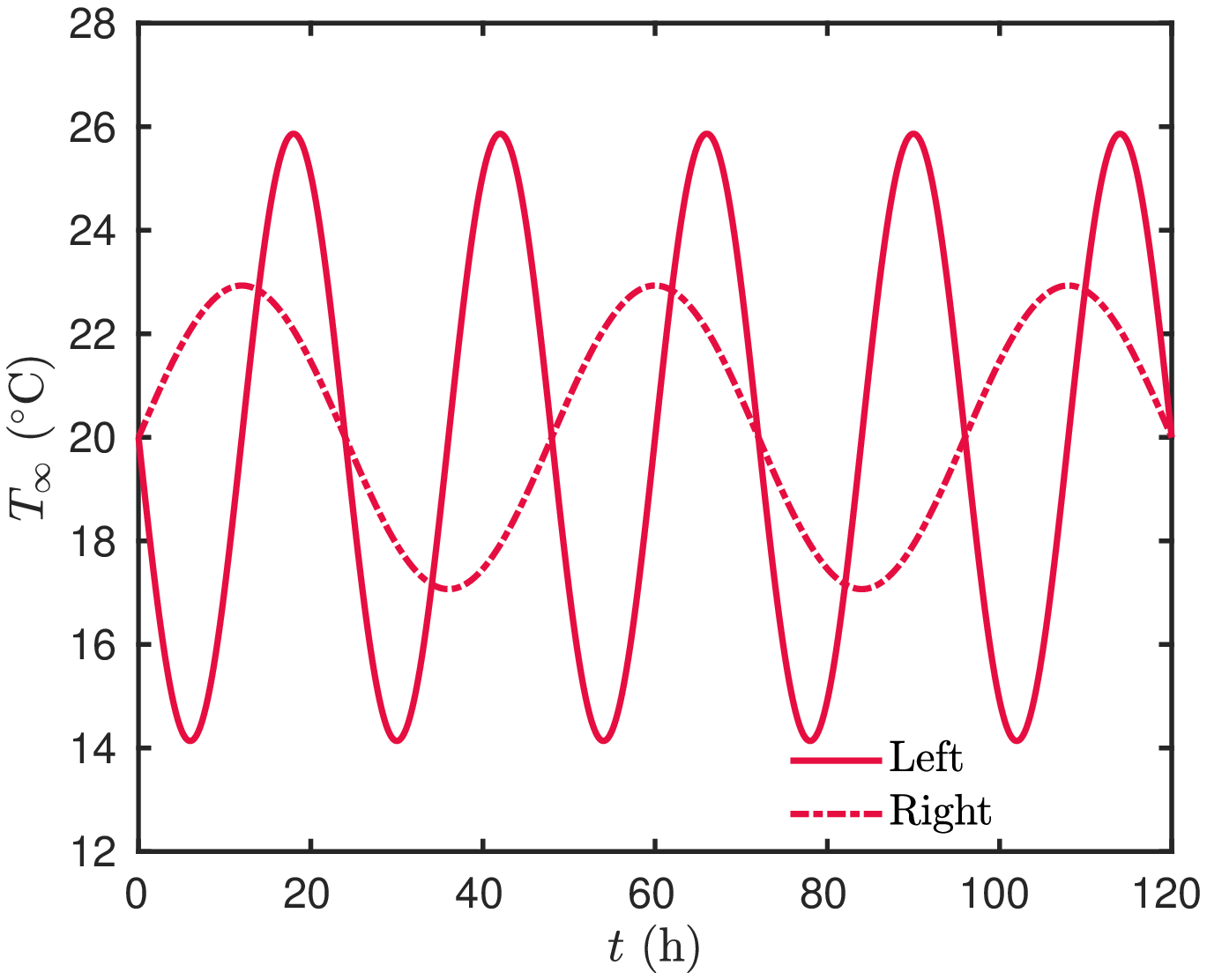}} \hspace{0.3cm}
  \subfigure[][\label{fig_AN3:BC_RH}]{\includegraphics[width=.48\textwidth]{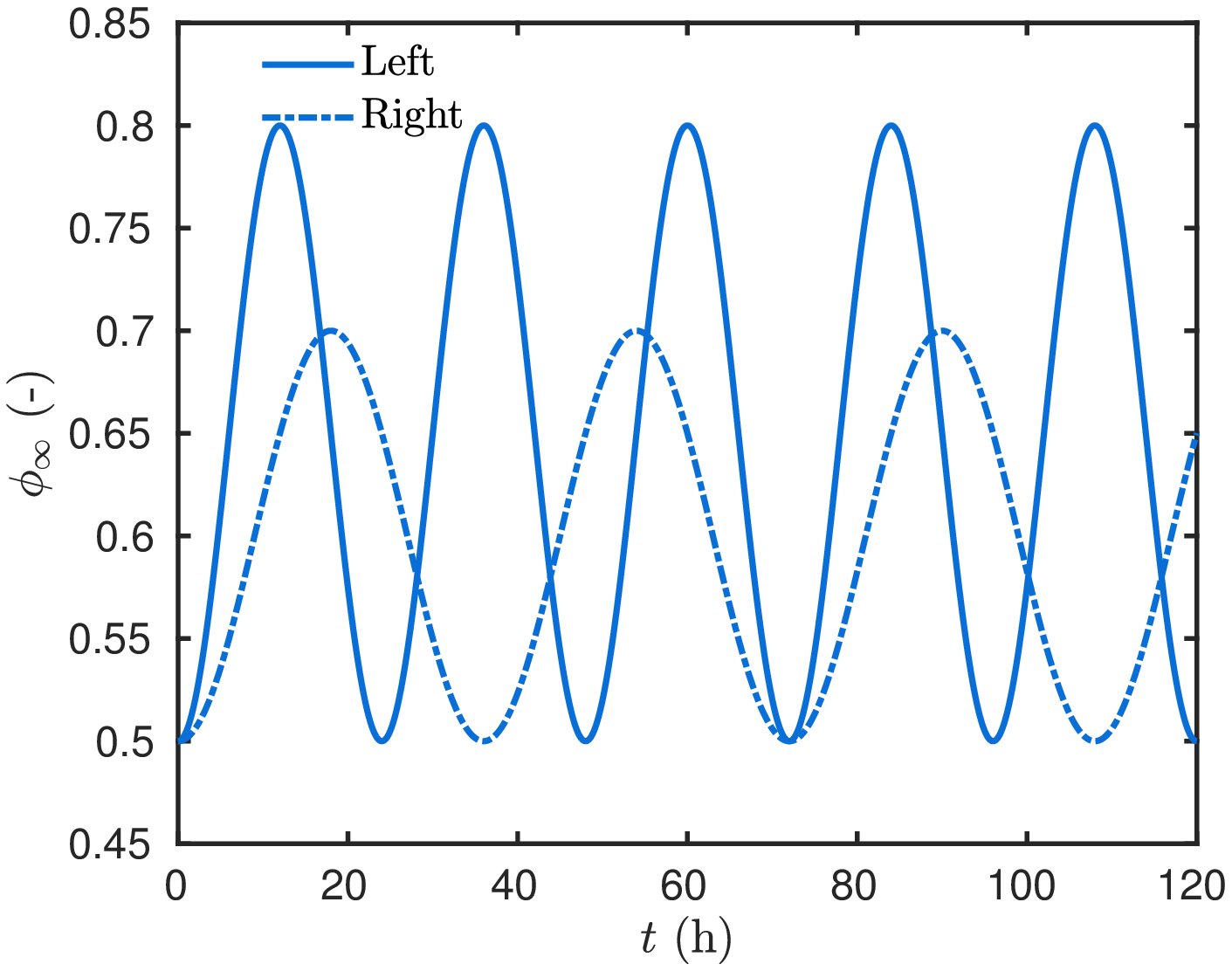}} 
  \caption{\small\em Boundary conditions of temperature $T$ (a) and relative humidity $\phi$ (b) for the multilayered case.}
  \label{fig_AN3:BC}
\end{center}
\end{figure}

Initial conditions are considered uniform over the spatial domain, with an initial temperature of $\Ti \egalb 293.15\ \mathsf{K}$ and an initial vapour pressure of $\Pvi \egalb 1.16\cdot 10^{\,3}\  \mathsf{Pa}\,$, referent to a relative humidity of $50\, \%\,$. The boundary conditions oscillate sinusoidally during $120$ hours of simulation, which are represented in Figure~\ref{fig_AN3:BC}. The convective mass and heat transfer coefficients are set with the same values as in the previous case study.

\begin{figure}
\begin{center}
  \subfigure[][\label{fig_AN3:Rain_flux}]{\includegraphics[width=.48\textwidth]{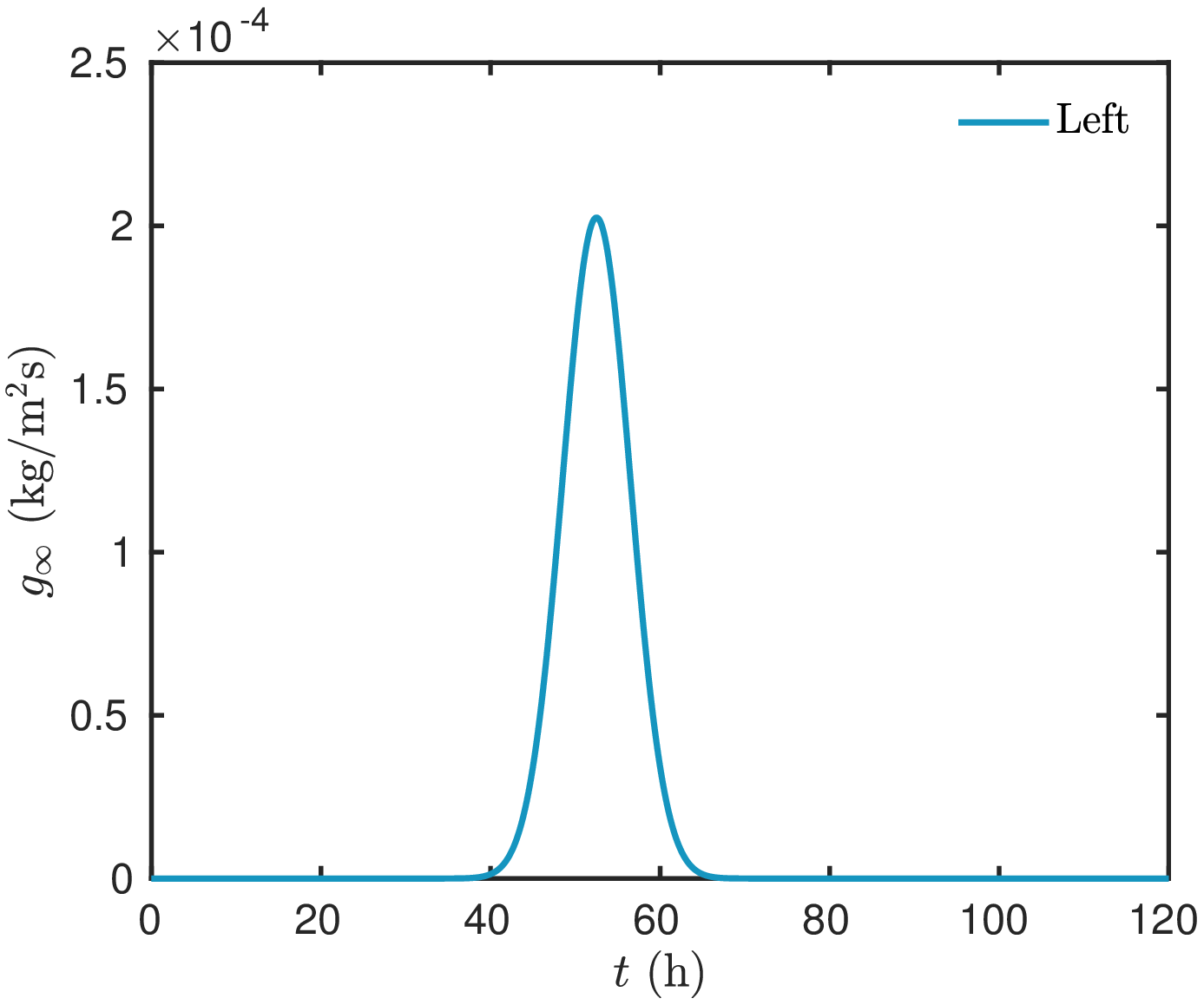}} \hspace{0.3cm}
  \subfigure[][\label{fig_AN3:sensible_heat_rain}]{\includegraphics[width=.47\textwidth]{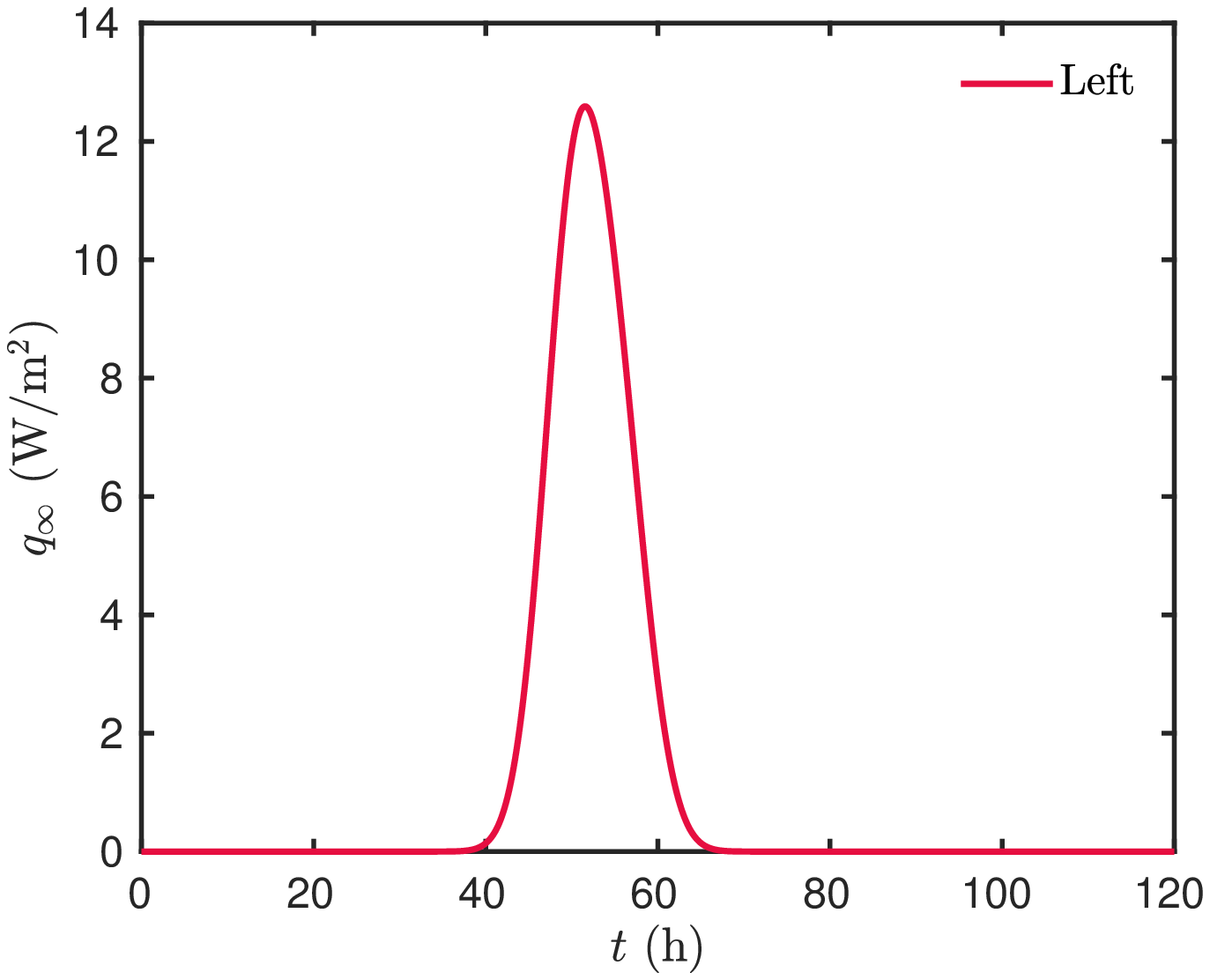}} 
  \caption{\small\em Rain flow imposed at the left boundary (a) and the sensible heat flux caused by the rain (b).}
  \label{fig_AN3:BC_flux}
\end{center}
\end{figure}

The rain is present in the simulation between hours $[\,40,\,65\,]\ \mathsf{h}\,$, reaching a maximum value of $2\cdot 10^{\,-4}\ \unitfrac{kg}{(m^2\cdot s)}$ at $52\ \mathsf{h}\,$, as presented in Figure~\ref{fig_AN3:Rain_flux}, which generates a sensible heat flux of $12\ \unitfrac{W}{m^2}$ at this boundary as shown in Figure~\ref{fig_AN3:sensible_heat_rain}. The rain flux is included at the left boundary, which causes a rapid increase of moisture within the material. The liquid flow from the rain can vary according to the wind speed \cite{Blocken2004}. In the literature \cite{Hagentoft2004, Rouchier2013, Gasparin2017, Janssen2014}, it is possible to find values within the range $[\,2\cdot 10^{\,-5},\,8\cdot 10^{\,-4}\,]\ \unitfrac{kg}{(m^2\cdot s)}\,$.

Simulations were performed using the \texttt{bvp4c}, with relative and absolute tolerances of the solver set to $10^{\,-5}\,$. The time is incremented with a discretization of $\Delta \ts \egalb 10^{\,-1}\,$, with an initial guess composed by $20$ spatial nodes, for both the vapour pressure and the temperature. The main advantage of using the approach of MOHL is that there is no need to perform sub-iterations at each time step to consider the nonlinearities of the material properties. Besides, no special treatment is needed at the interface. Everything is handled automatically. For the \textsc{Euler} implicit solution, simulations have been performed with $\Delta \xs \egalb 10^{\,-2}$ and $\Delta \ts \egalb 10^{\,-1}\,$.

\begin{figure}
\begin{center}
  \subfigure[][\label{fig_AN3:BVP_profile_T}]{\includegraphics[width=.47\textwidth]{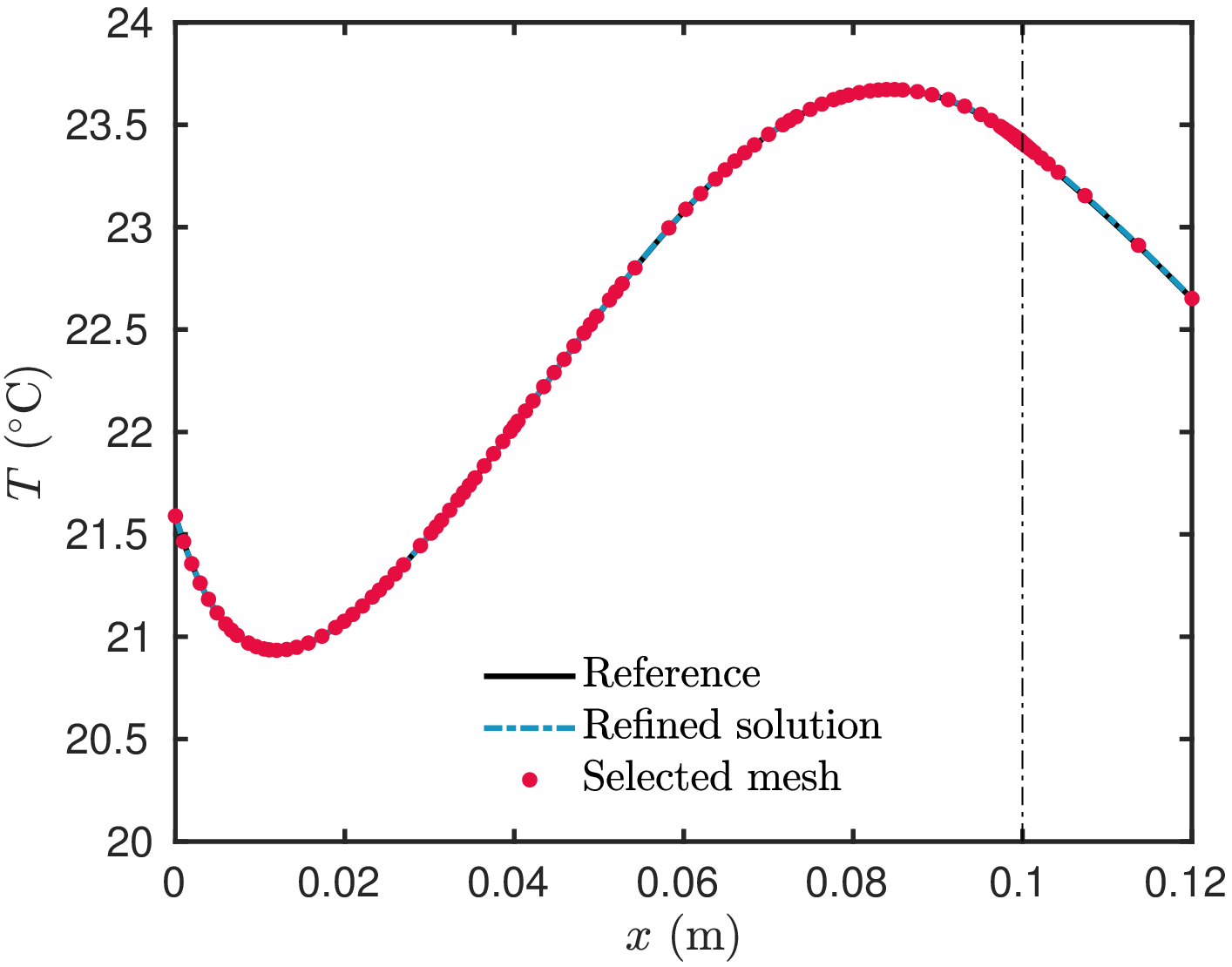}} \hspace{0.3cm}
  \subfigure[][\label{fig_AN3:BVP_profile_Pv}]{\includegraphics[width=.485\textwidth]{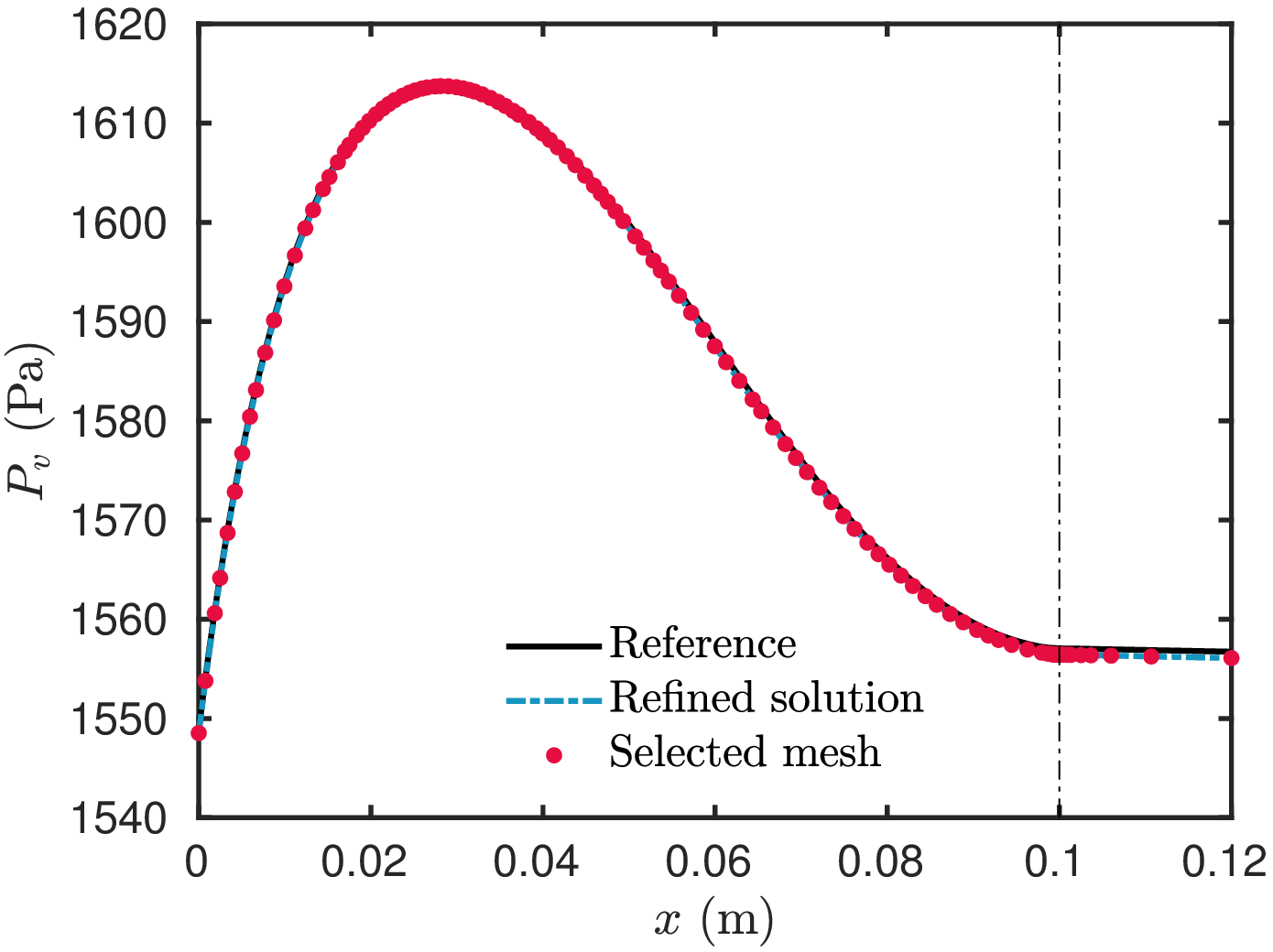}} 
  \caption{\small\em Selected mesh determined by \texttt{bvp4c} for the last profile (at $t\egalb 120\ \mathsf{h}\,$) of temperature (a) and vapour pressure (b).}
  \label{fig_AN3:BVP_profile}
\end{center}
\end{figure}

Figures~\ref{fig_AN3:BVP_profile_T} and \ref{fig_AN3:BVP_profile_Pv} present the selected mesh for the last profile of temperature and vapour pressure fields, respectively, and also, the refined and the reference solutions. At each time iteration, the number of mesh points varies with an average of $498$ mesh points for the moisture field and $235$ mesh points for the temperature field. Due to high nonlinearities, a concentration of nodes is noted at the interface between materials and where the gradients are higher. For the domain of the second layer, with the finishing material, only $33$ nodes were used to compute the vapour pressure field, where $20$ are concentrated closer to the interface between the two materials. It indicates that the method proposed can be adapted according to the physical phenomenon.

\begin{figure}
\begin{center}
  \includegraphics[width=.69\textwidth]{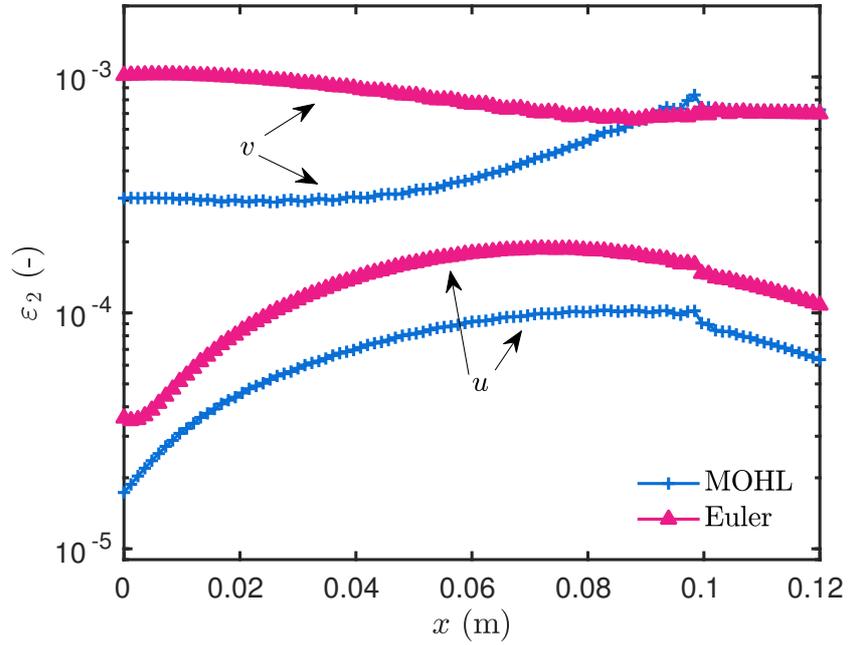}
  \caption{\small\em Error $\varepsilon_{\,2}$ computed for the MOHL method, for the temperature and vapour pressure solutions of the multilayered case.}
  \label{fig_AN3:Error_fx}
\end{center}
\end{figure}

The MOHL has demonstrated a good agreement with the reference solution given by \texttt{Chebfun} to reproduce the physical phenomenon. Distributions of the error $\varepsilon_{\,2}$ on function of $x$ are shown in Figure~\ref{fig_AN3:Error_fx}. The error of the MOHL is $\varepsilon_{\infty,\,v} \egalb 8.3\cdot 10^{\,-4}$ for the vapour pressure and $\varepsilon_{\infty,\,u} \egalb  10^{\,-4}$ for the temperature. These values depend on the order of the discretization in time and on the chosen tolerance of the solver.

\begin{figure}
\begin{center}
  \subfigure[][\label{fig_AN3:profiles_T}]{\includegraphics[width=.48\textwidth]{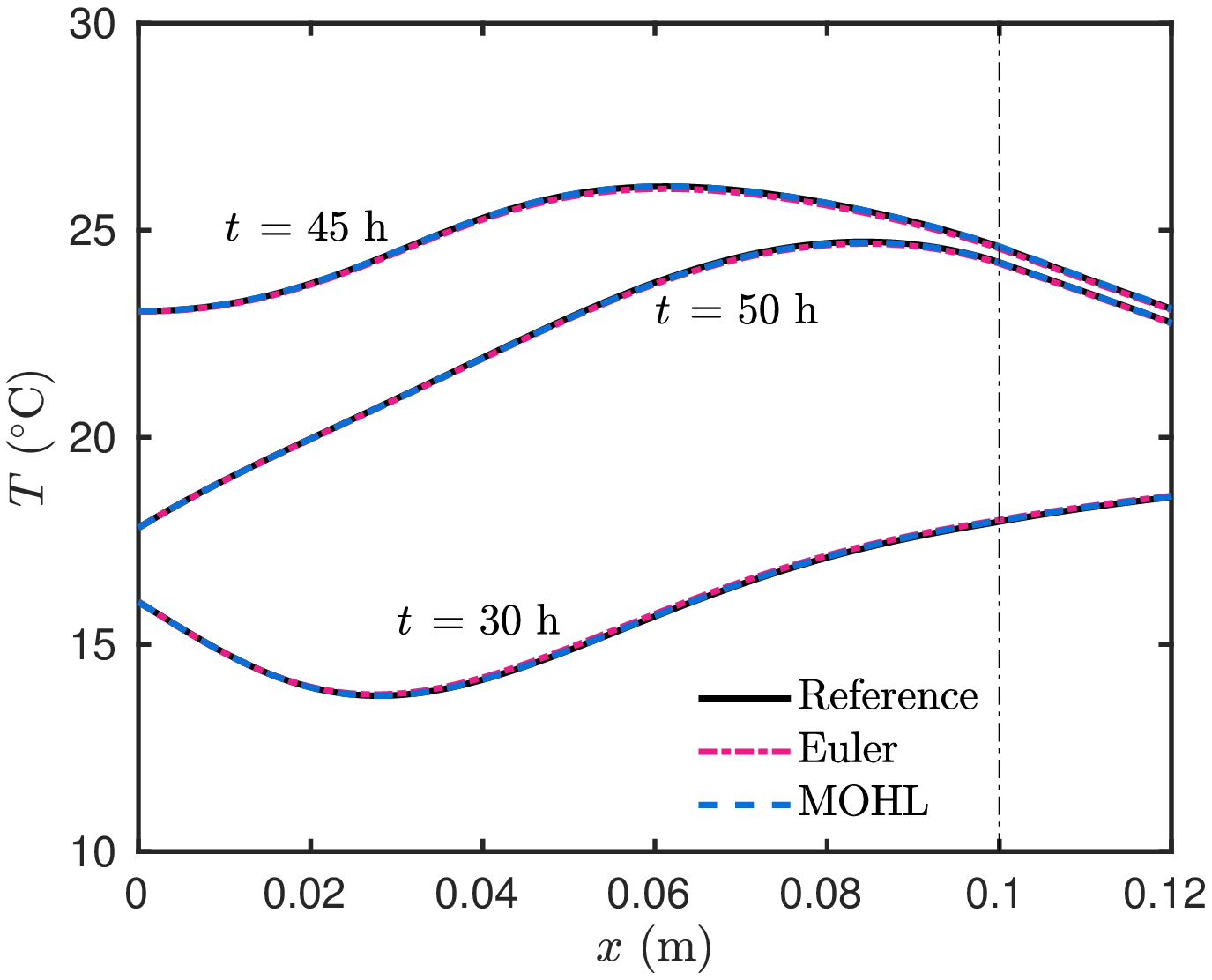}} \hspace{0.3cm}
  \subfigure[][\label{fig_AN3:profiles_RH}]{\includegraphics[width=.48\textwidth]{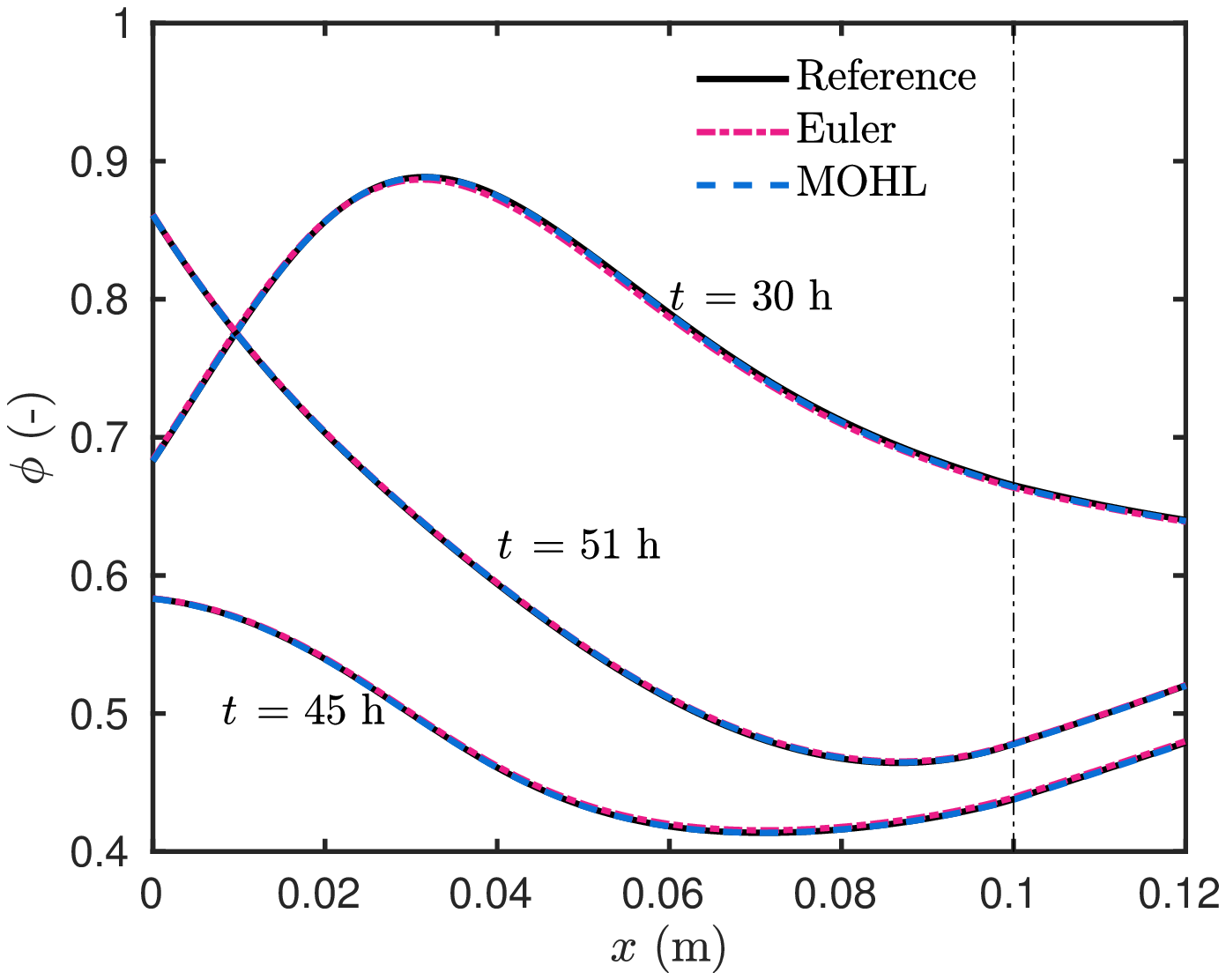}} 
  \caption{\small\em Temperature (a) and relative humidity (b) profiles at different times for the multilayered case.}
  \label{fig_AN3:profile}
\end{center}
\end{figure}

Figure~\ref{fig_AN3:profiles_RH} presents the relative humidity profiles when the rain flow occurs. It is possible to observe a liquid concentration on the left edge, and, a stable situation on the right edge. This behaviour is observed due to different material properties. As the second layer is less hygroscopic than the first layer, the rain flow crosses both materials and arrives at the right boundary with little intensity. Additionally, Figure~\ref{fig_AN3:profiles_T} presents the temperature profiles for the same time instants. The temperature increases in the middle of the material are due to the moisture content gradients on the storage coefficient.

\begin{figure}
\begin{center}
  \subfigure[][\label{fig_AN3:evolution_T_BC}]{\includegraphics[width=.475\textwidth]{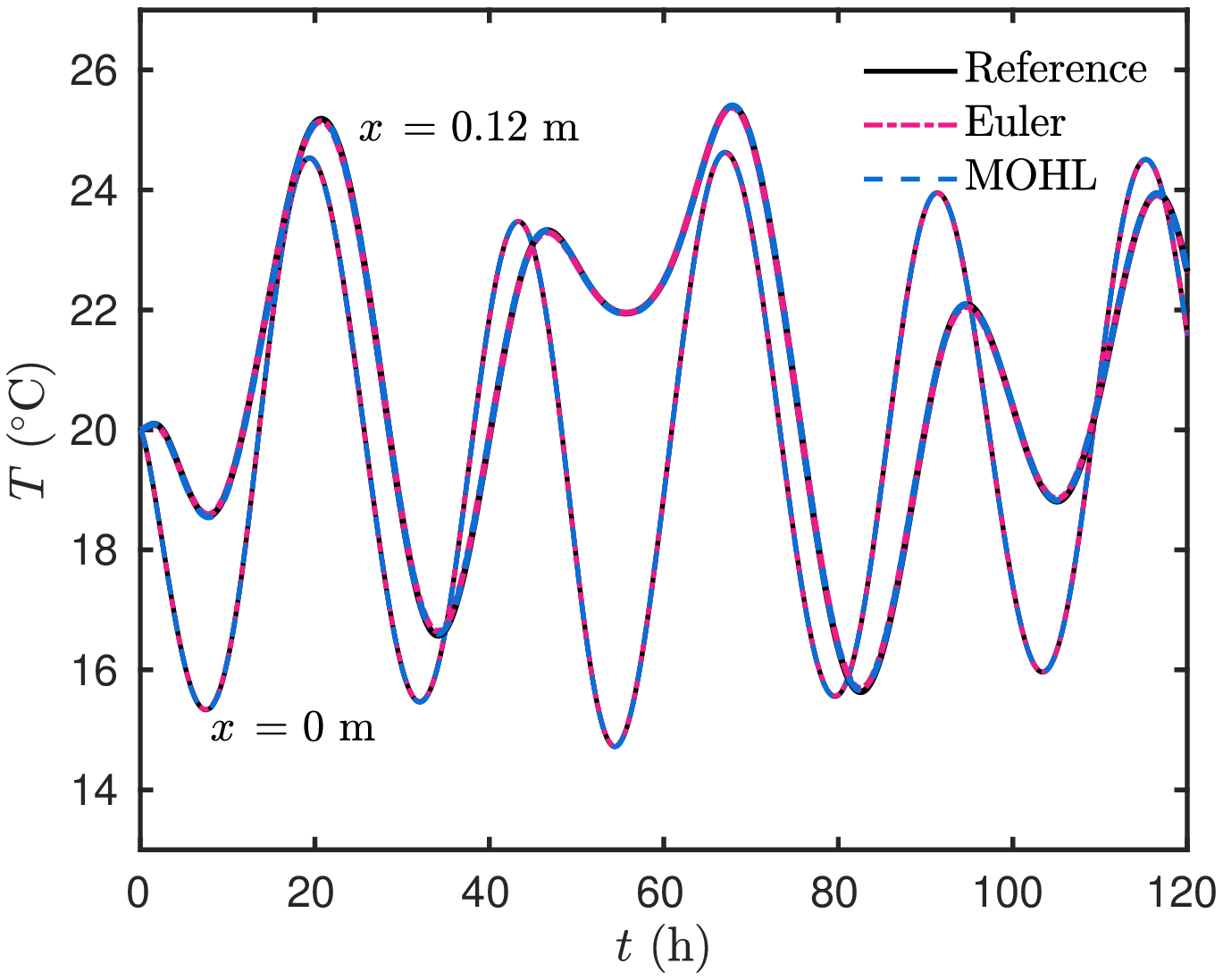}} \hspace{0.3cm}
  \subfigure[][\label{fig_AN3:evolution_Pv_BC}]{\includegraphics[width=.485\textwidth]{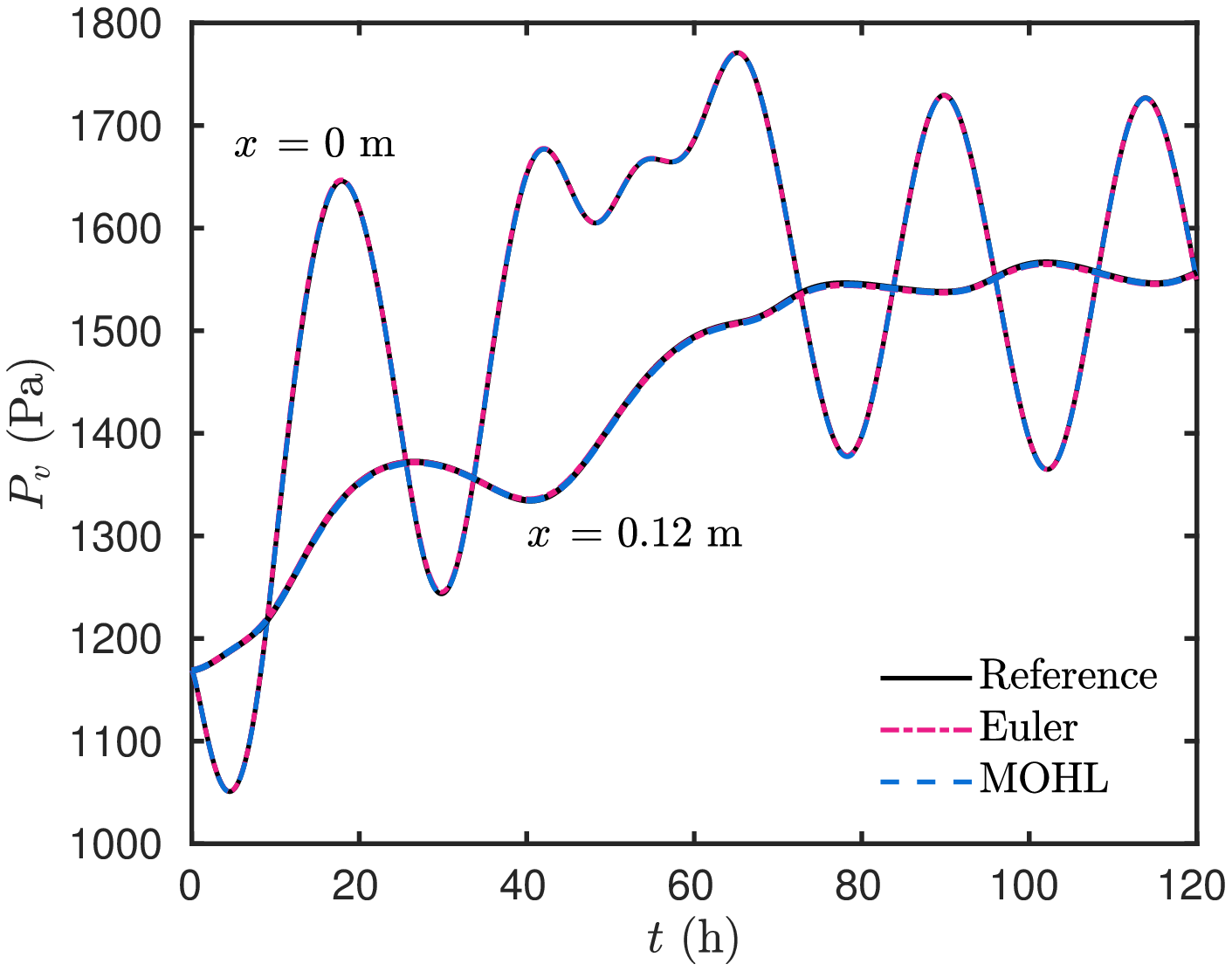}} \\
  \subfigure[][\label{fig_AN3:evolution_RH_BC}]{\includegraphics[width=.485\textwidth]{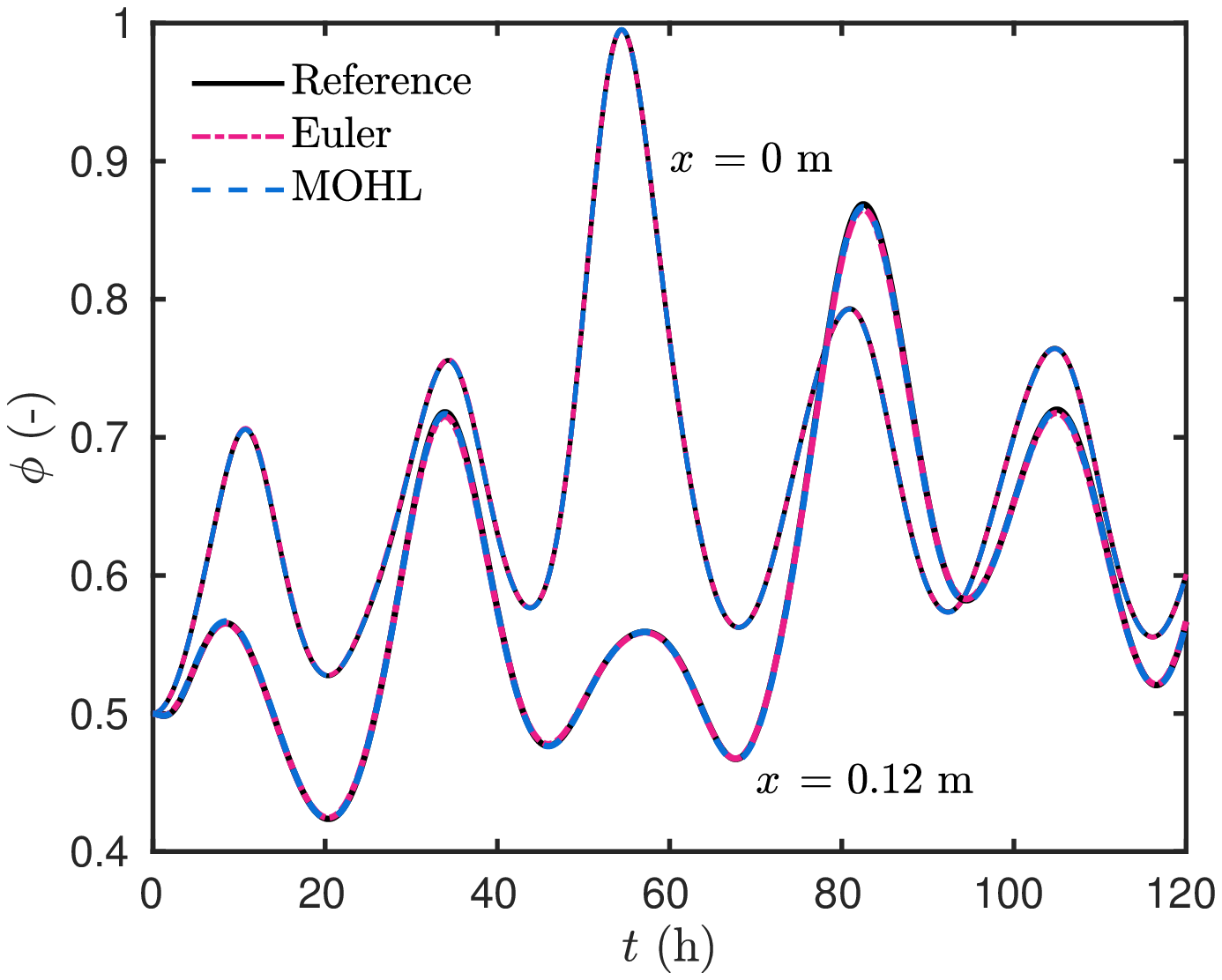}} 
  \caption{\small\em Evolution of the temperature (a), of vapour pressure (b) and of relative humidity (c) on the boundary surfaces for the multilayered case.}
  \label{fig_AN3:evolution}
\end{center}
\end{figure}

The evolution of temperature, vapour pressure and relative humidity at the boundary surfaces ($x \egalb 0\ \mathsf{m}$ and $x \egalb 0.12\ \mathsf{m}\,$) are shown in Figures~\ref{fig_AN3:evolution_T_BC}, \ref{fig_AN3:evolution_Pv_BC} and \ref{fig_AN3:evolution_RH_BC}, respectively. The vapour pressure varies according to the sinusoidal fluctuations of the boundary conditions affected by the rain flow. At $x \egalb 0\ \mathsf{m}\,$, the vapour pressure suddenly increases due to the rain flow imposed at the surface. The moisture due to the rain flow diffuses through both layers. Although, as the second layer is composed with a less hygroscopic material, the vapour pressure completely reaches this surface by $70\ \mathsf{h}\,$. In Figure~\ref{fig_AN3:evolution_T_BC}, it is possible to observe a peak of temperature on the right boundary, which is caused because the heat transfer coefficient under vapour pressure gradient $\kTMs$ is ten times higher in the second layer than in the first one.

\begin{figure}
\begin{center}
  \subfigure[][\label{fig_AN3:heat_flux_profile}]{\includegraphics[width=.485\textwidth]{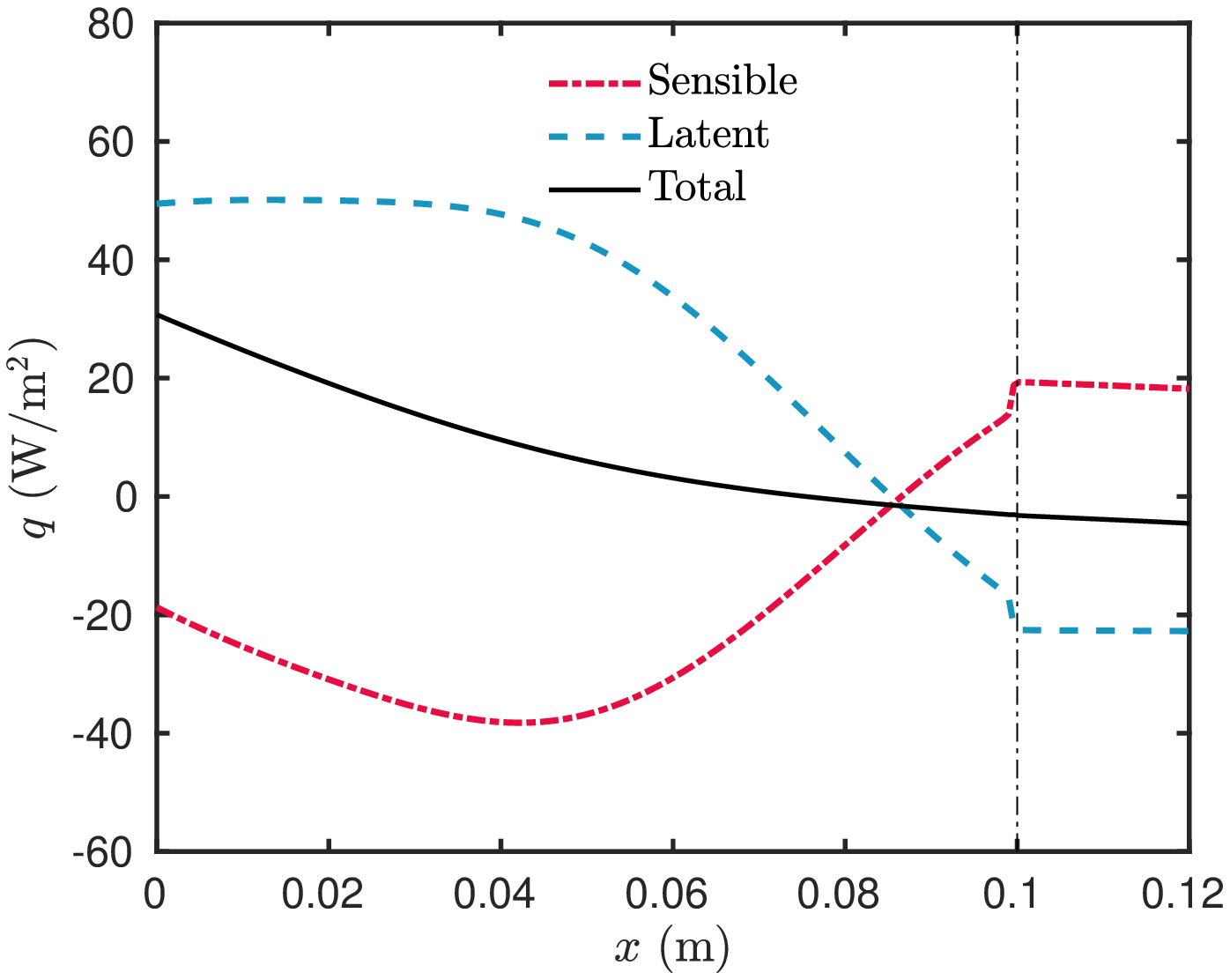}} \hspace{0.3cm}
  \subfigure[][\label{fig_AN3:moisture_flow_profile}]{\includegraphics[width=.48\textwidth]{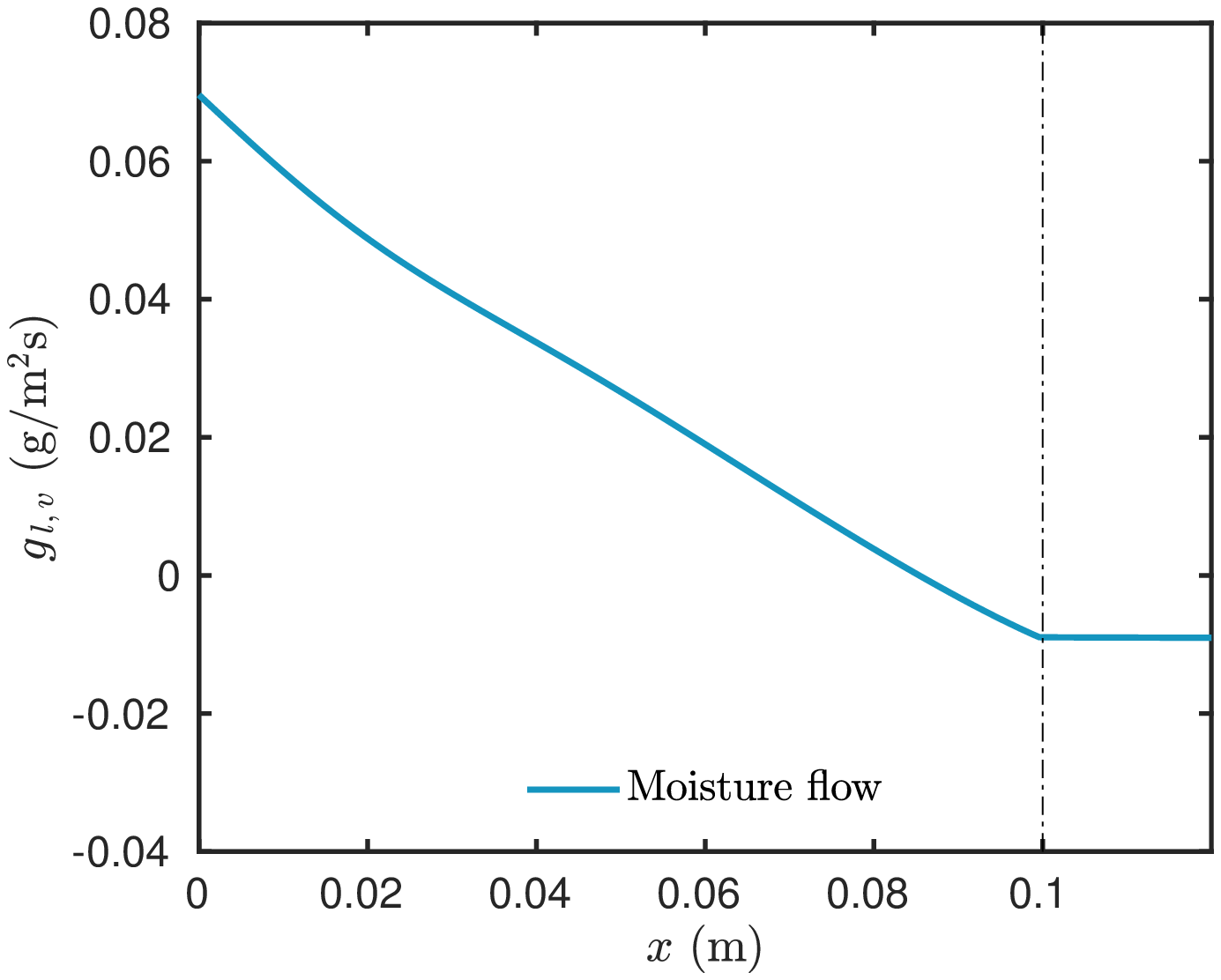}} 
  \caption{\small\em Profile of sensible, latent and total heat fluxes (a) and of moisture flow (b) at $t \egalb 60\  \mathsf{h}\,$.}
  \label{fig_AN3:flux_and_flow}
\end{center}
\end{figure}

Figure~\ref{fig_AN3:heat_flux_profile} indicates the sensible, the latent  and the total heat fluxes over both materials, at the instant $t \egalb 60\ \mathsf{h}\,$, a while after the peak of rain flow. Furthermore, Figure~\ref{fig_AN3:moisture_flow_profile} presents the total moisture flow for the same time. In Material $1\,$, the sensible and latent heat fluxes have high values but with opposite signs. As the latent heat flux is slight higher they do not cancel each other. Material $2$ has a higher diffusive coefficient than Material $1\,$, which explains the horizontal line for the fluxes and flow profiles. Another important point is that the total flux and flow across the interface between the two materials are continuous, which is consistent with the assumption of hydraulic continuity \cite{DeFreitas1996, Derluyn2011}.

\begin{figure}
\begin{center}
  \subfigure[][\label{fig_AN3:cM}]{\includegraphics[width=.48\textwidth]{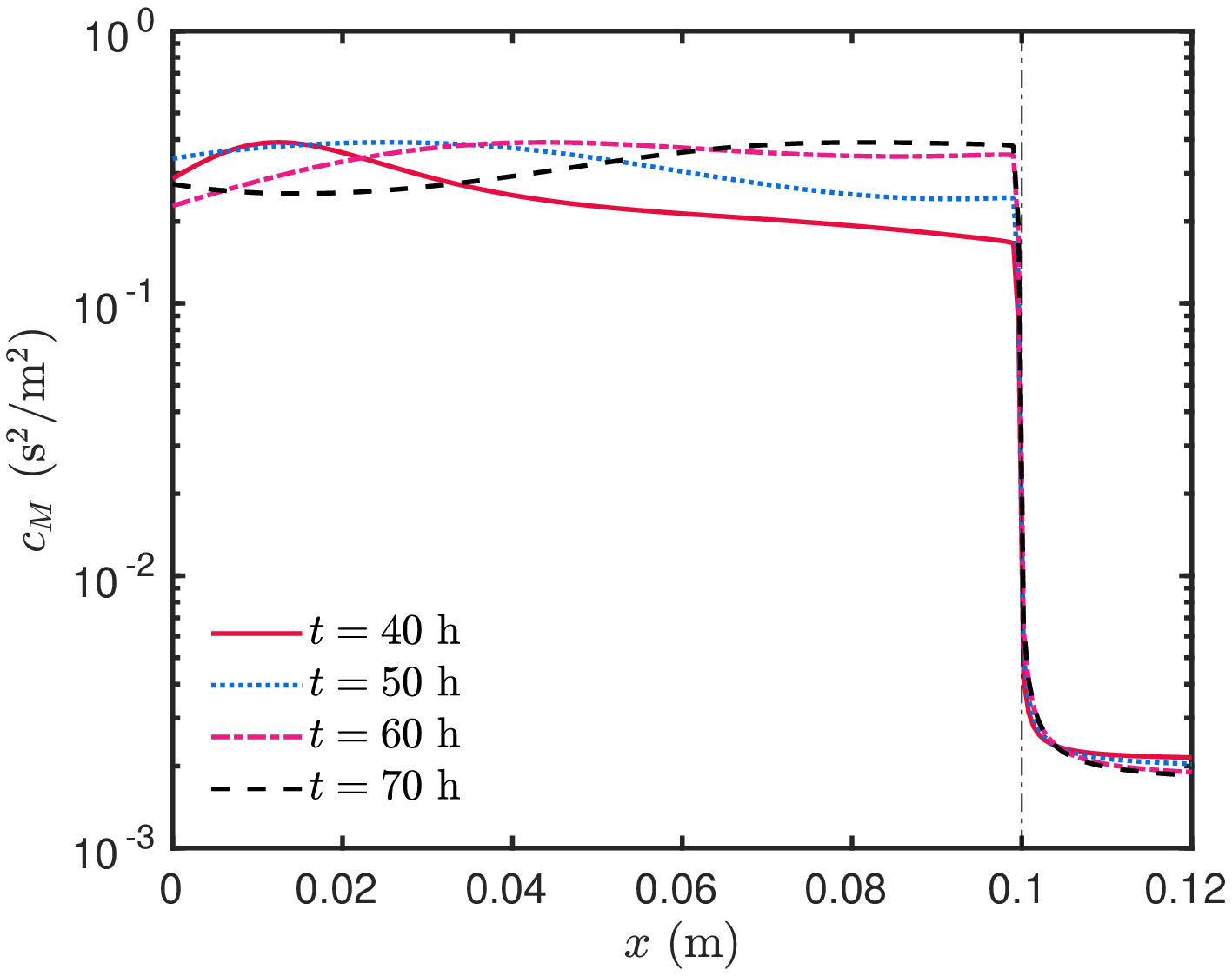}} \hspace{0.3cm}
  \subfigure[][\label{fig_AN3:kM}]{\includegraphics[width=.48\textwidth]{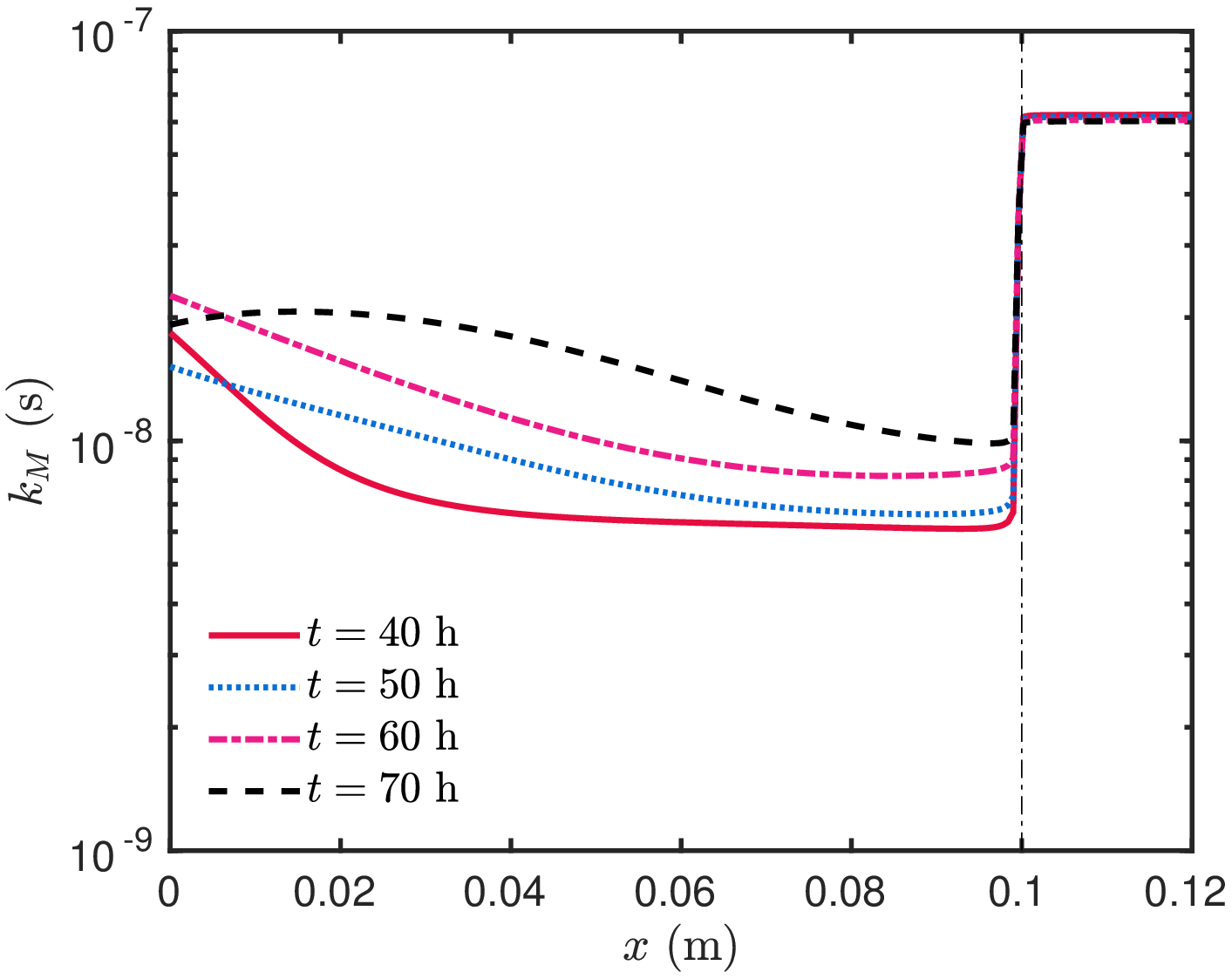}} 
  \caption{\small\em Moisture storage coefficient (a) and total moisture transfer coefficient (b) values over space domain for different times.}
  \label{fig_AN3:Properties}
\end{center}
\end{figure}

The moisture storage coefficient and the moisture diffusive coefficients are shown in Figures~\ref{fig_AN3:cM} and \ref{fig_AN3:kM}, respectively. They are presented on function of the space distribution for different time instant $t \egalb \{\,40,\,50,\,60,\,70\,\}\ \mathsf{h}$. The differences are high between the material properties and it is possible to observe the non-linearities due to the moisture content gradients.


\section{Comparison of the numerical predictions with experimental data}
\label{sec:validation}

In this section, the physical model with the MOHL numerical model is compared with experimental data gathered from the French project HYGRO-BAT \cite{HYGRO-BA2014}. The measurements were performed at the French laboratory LOCIE (Laboratory of Optimisation of the Conception and Engineering of the Environment). One dimensional coupled heat and moisture transfer through a single-layered wall is monitored by sensors placed within the material and at its surfaces. The liquid transfer $\kl$ in the moisture balance equation --- Equation~\eqref{eq:M_equation} --- is neglected in this study. Surface sensors provide boundary conditions for the coupled simulation, while the other sensors provide reference measurements for the model confirmation.

The wall considered in their study is composed of a $16$-$\mathsf{cm}$ layer of wood fibre material, which is subjected to variations on the temperature and relative humidity for a $2$-week period. The material properties are given in Table~\ref{table:properties_woodfibre} of the Appendix~\ref{annexe:material_properties}. The reference data for the evaluation is provided  by temperature and humidity sensors \texttt{SHT75 Sensirion}, located at $x\egalb \{\,4,\,8,\,12\,\}\ \mathsf{cm}$ inside the wall. They are located sparsely within the wall to avoid interferences among them.

The sensors have an uncertainty of measurement of $\sigma_{\,T}^{\,\text{meas}} \egalb 0.3\gC$ for temperature and of $\sigma_{\,\phi}^{\,\text{meas}} \egalb 0.018$ for relative humidity. In addition, the uncertainty regarding the position of the sensors is of $\sigma^{\,\text{pos}} \egalb 1\ \mathsf{cm} $ for the sensors located at $x\egalb \{\,4,\,12\,\}\ \mathsf{cm}$ and of $\sigma^{\,\text{pos}} \egalb 0.5\ \mathsf{cm} $ for the other sensors. The uncertainty on position is higher for the sensor at $x\egalb \{\,4,\,12\,\}\ \mathsf{cm}\,$, since according to the experiment design, they have been settled by perforating a whole in the wood fibre layer.

\begin{figure}
  \centering
  \subfigure[a][\label{fig_AN4:BC_T}]{\includegraphics[width=0.48\textwidth]{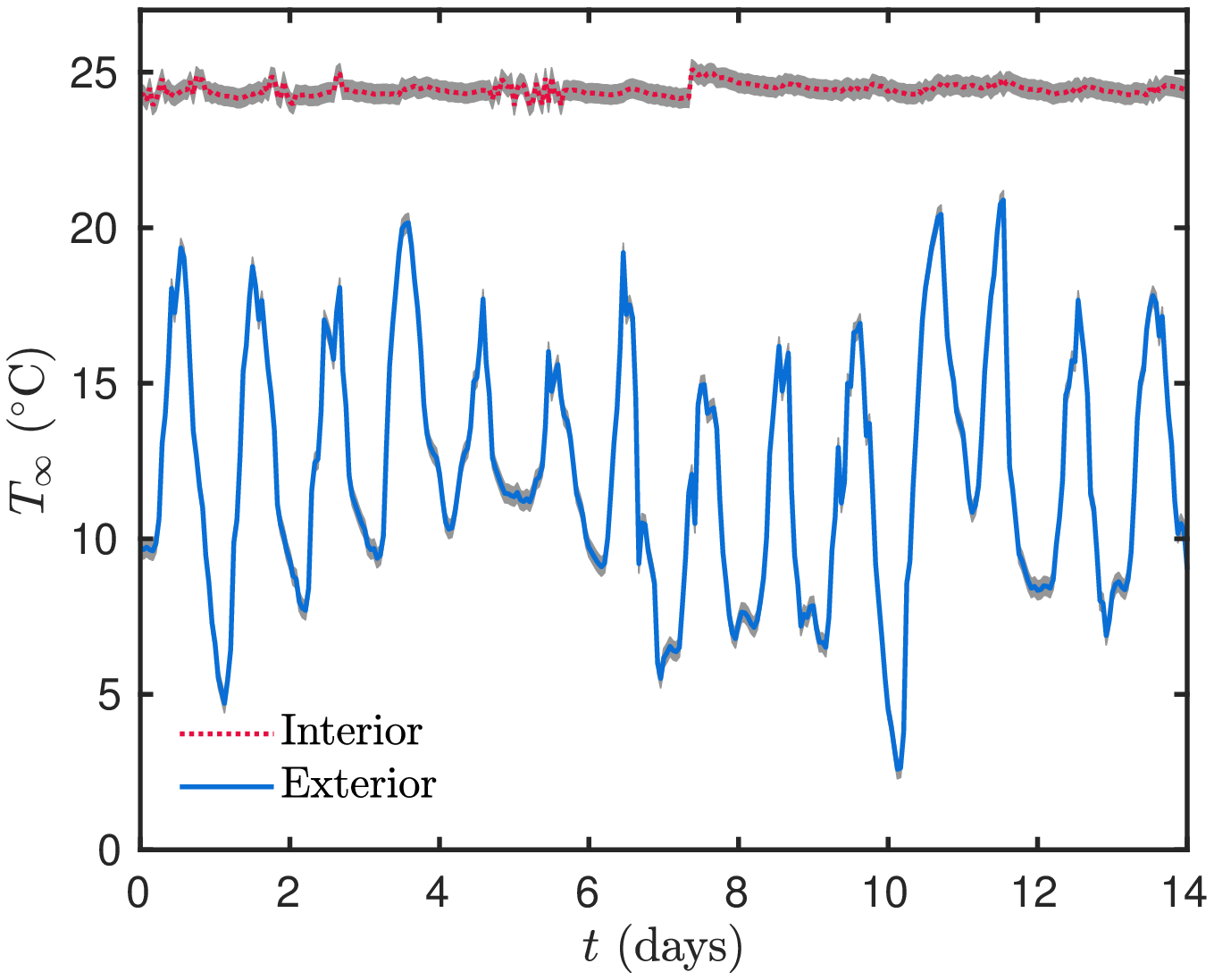}} \hspace{0.3cm}
  \subfigure[b][\label{fig_AN4:BC_RH}]{\includegraphics[width=0.48\textwidth]{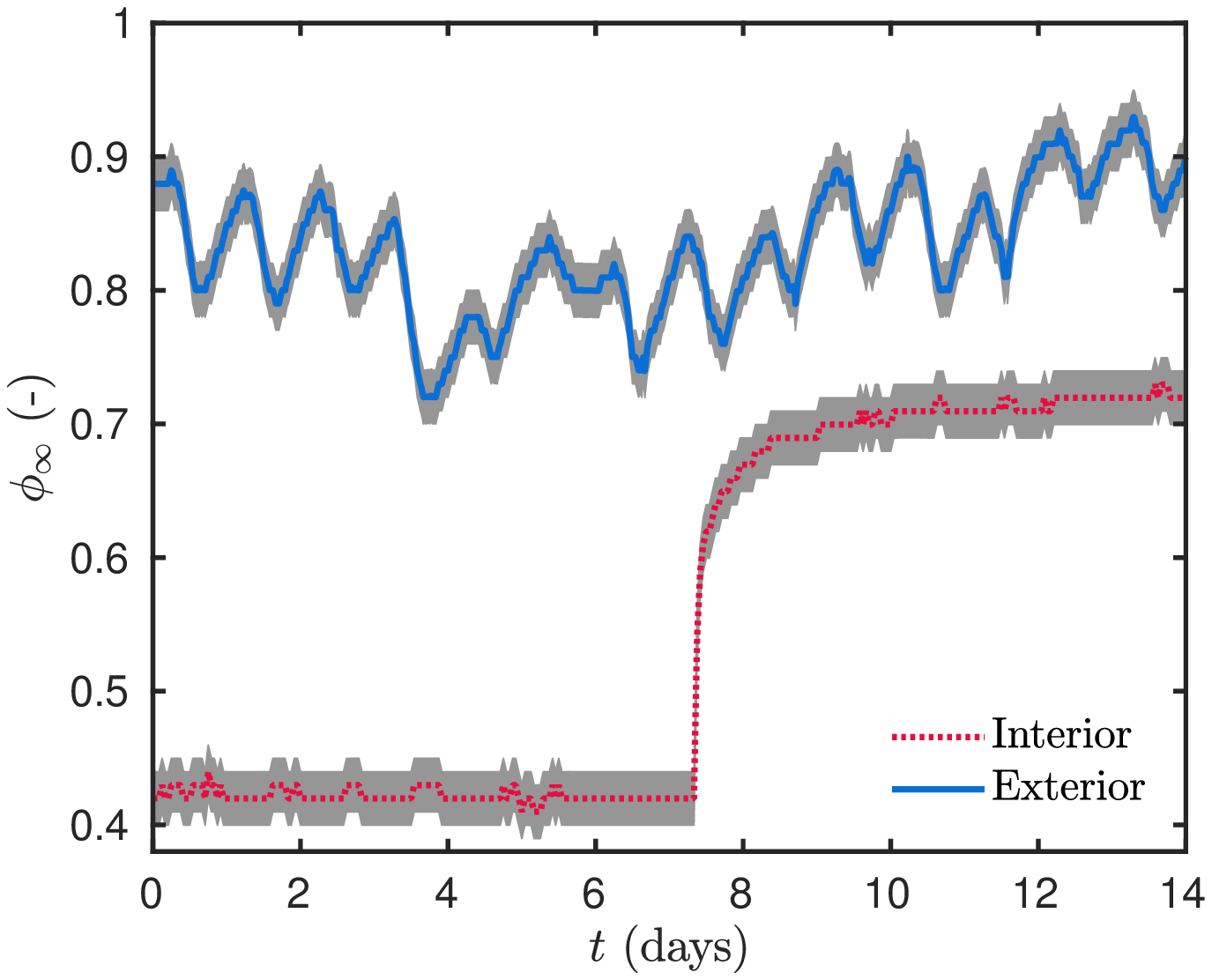}}
  \caption{\small\em Ambient temperature (a) and relative humidity (b) of the experimental investigation.}
\end{figure}

The temperature and relative humidity at the interior and exterior boundaries are measured and given in Figures~\ref{fig_AN4:BC_T} and \ref{fig_AN4:BC_RH}, respectively. In these figures, the uncertainties related to the sensors measurements are shown in gray color. At the interior boundary, the temperature is set to approximately $24\gC$ and the relative humidity set to $40\,\%\,$, at the first week, and to $70\,\%$ at the second week. The exterior temperature and relative humidity values are given by their measurement at the boundary, which is filtered by a $2$-$\mathsf{cm}$ coating layer, excluding solar radiation and driving rain phenomena. Thus, both boundaries are expressed as \textsc{Dirichlet} conditions for the model confirmation. For more details about the experiment one may refer to \cite{Rouchier2017} and \cite{Rouchier2016}. In this wok, the temporal derivative of $\rhow$ in Equation~\eqref{eq:H_equation} was also taken into account, differently from \cite{Rouchier2016}.

For confirming simulations, the relative error - $\epsilon$ - between the solutions and the data are computed as: 
\begin{align*}
\epsilon\, (t)\ &\eqdef\  \dfrac{\sqrt{\Bigl(Y_{\, k}^{\, \mathrm{num}}\, (x_{\,k},\, t) \moins Y_{\, k}^{\mathrm{\, meas}}\, ( x_{\,k},\, t)\Bigr)^{2}}}{Y_{\, k}^{\mathrm{\, meas}}( x_{\,k},\, t) }  \,, 
\end{align*}
where $Y_{\, k}^{\, \mathrm{num}}$ is the computed solution and $Y_{\, k}^{\mathrm{\, meas}}$ is the measured data.

Simulations are performed on a single-layered wall of a material that separates two ambiances. The initial condition for the temperature and vapour pressure profiles are given by an interpolation of the measurements at $x \egalb \{\,0,\,4,\,8,\,12,\,16\,\}\ \mathsf{cm}\,$, at the first time instant. The solver chosen to perform the MOHL method is the \texttt{bvp4c}, with relative and absolute tolerances set to $10^{\,-4}\,$. The solution has been computed with a time step of $\Delta \ts \egalb 0.1\,$, the equivalent of $6\ \mathsf{min}\,$.

\begin{figure}
  \centering
  \subfigure[a][\label{fig_AN4:T4}]{\includegraphics[width=0.48\textwidth]{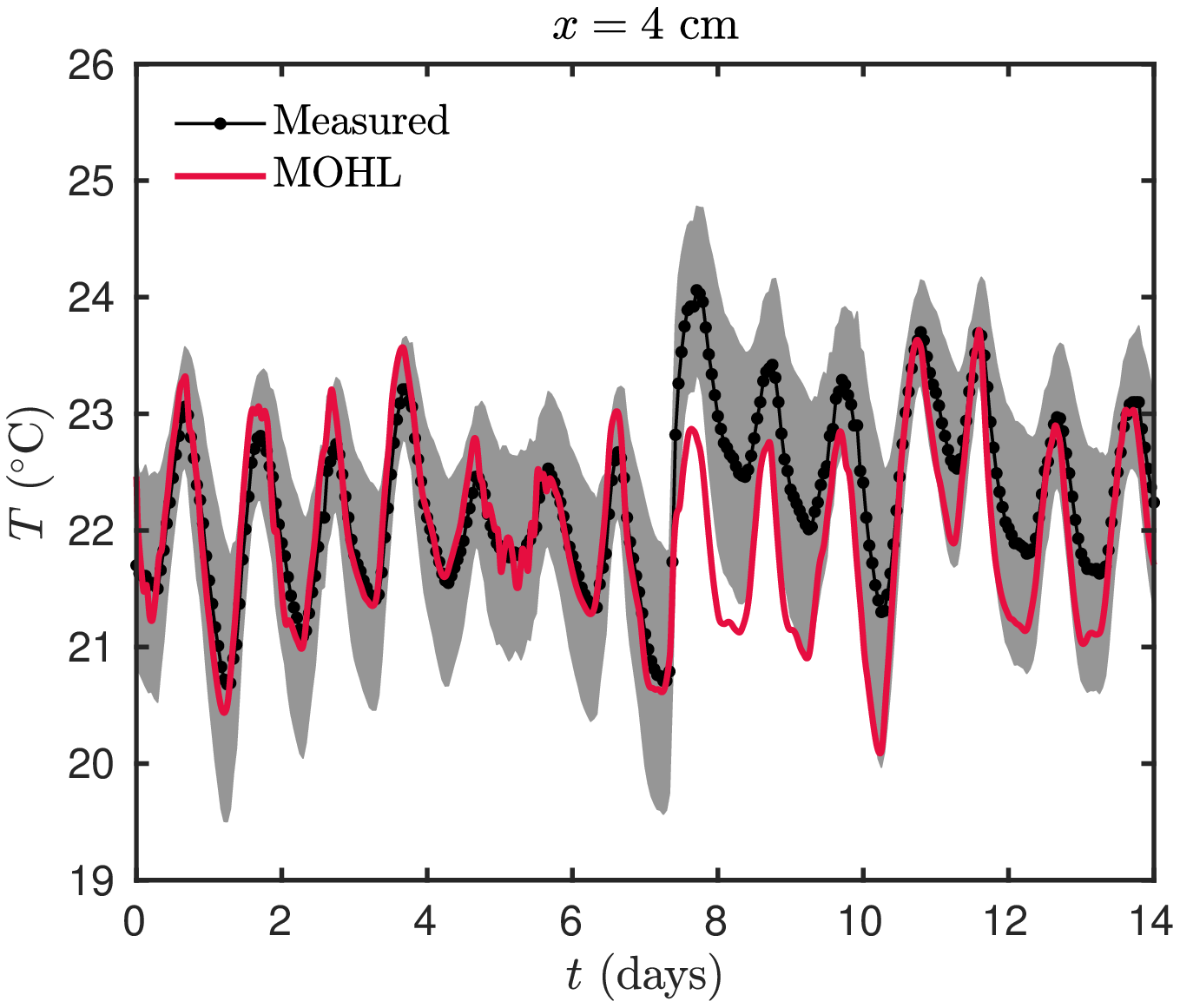}} \hspace{0.3cm}
  \subfigure[b][\label{fig_AN4:RH4}]{\includegraphics[width=0.485\textwidth]{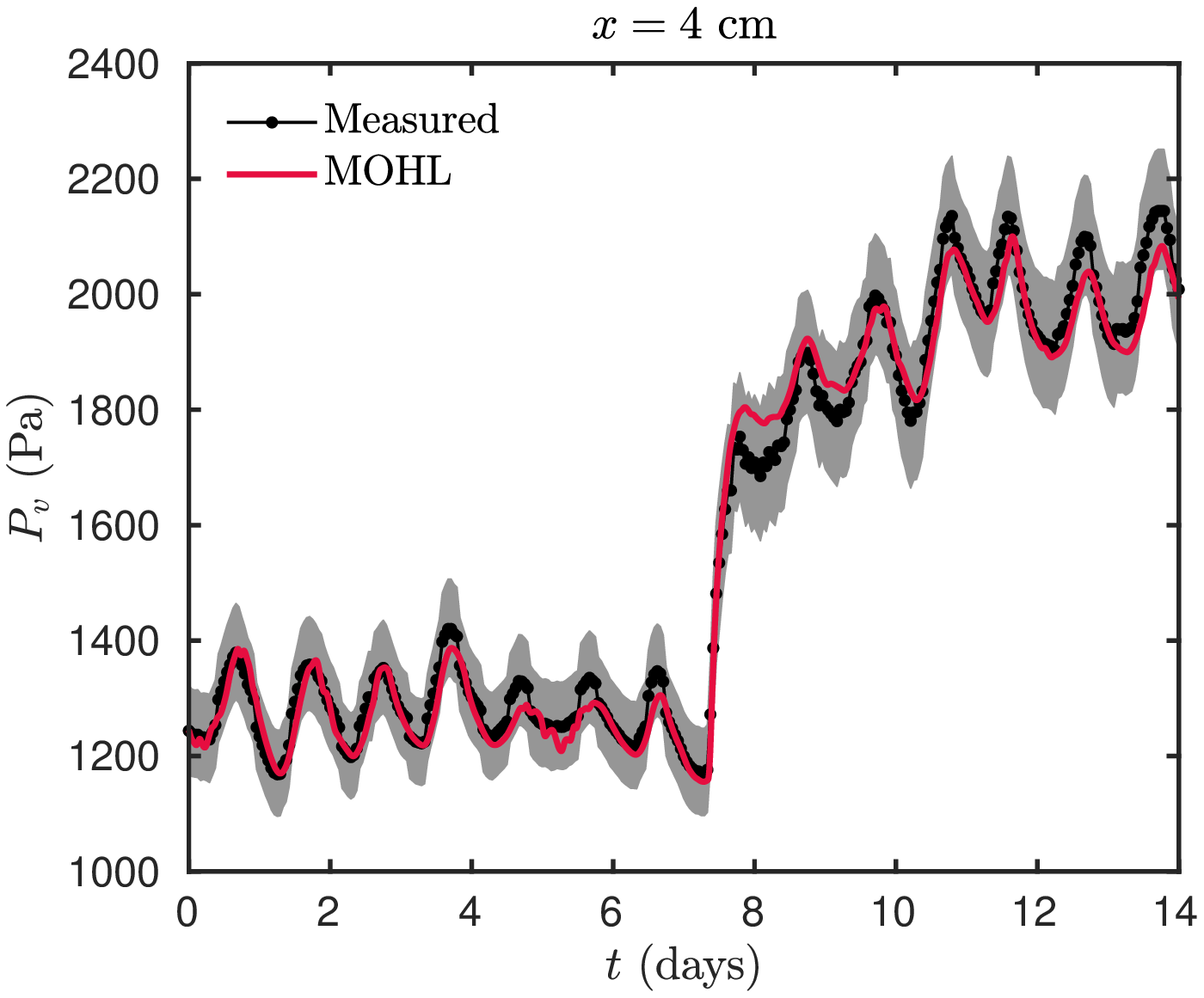}} \\
  \subfigure[c][\label{fig_AN4:T8}]{\includegraphics[width=0.48\textwidth]{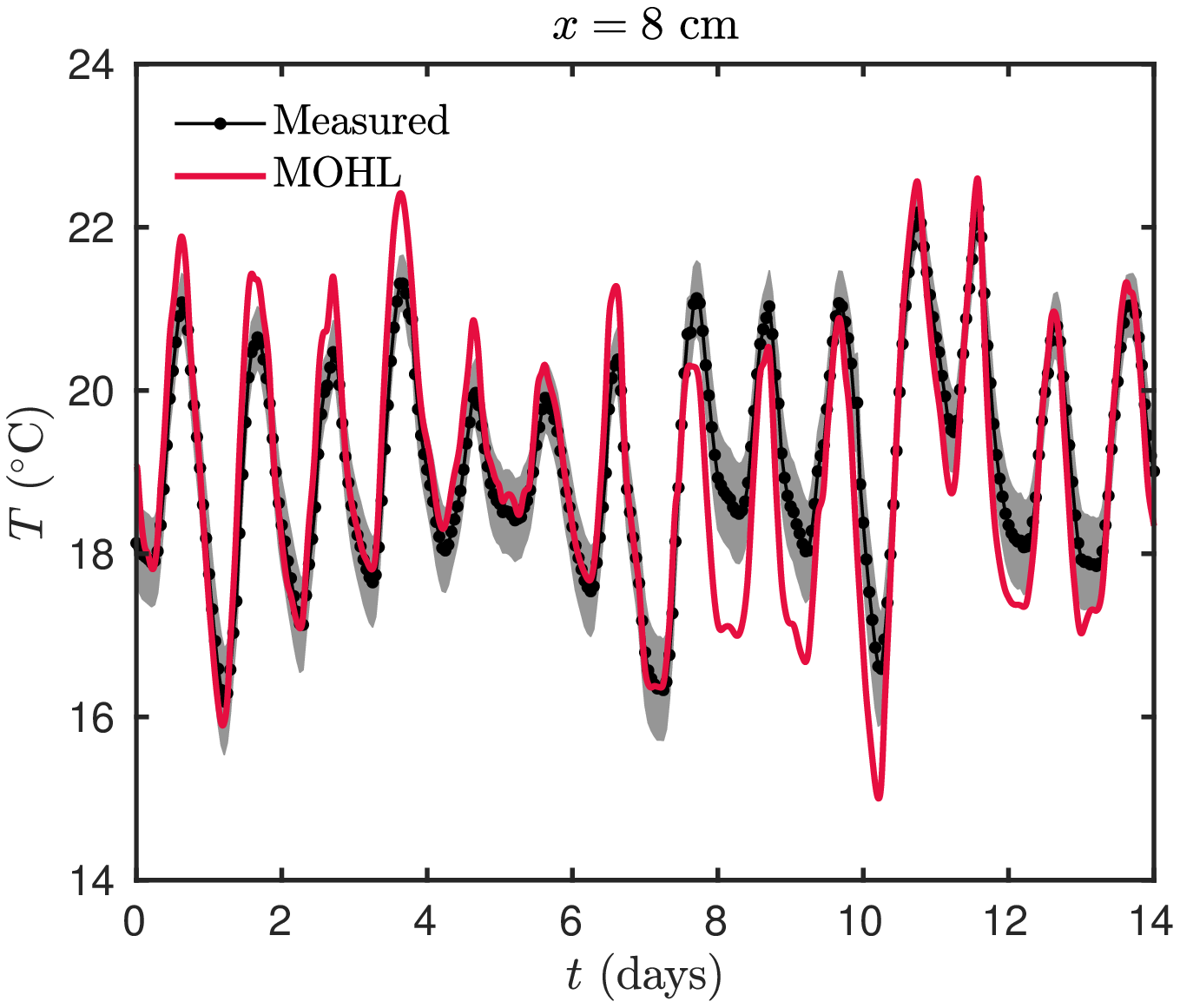}} \hspace{0.3cm}
  \subfigure[d][\label{fig_AN4:RH8}]{\includegraphics[width=0.485\textwidth]{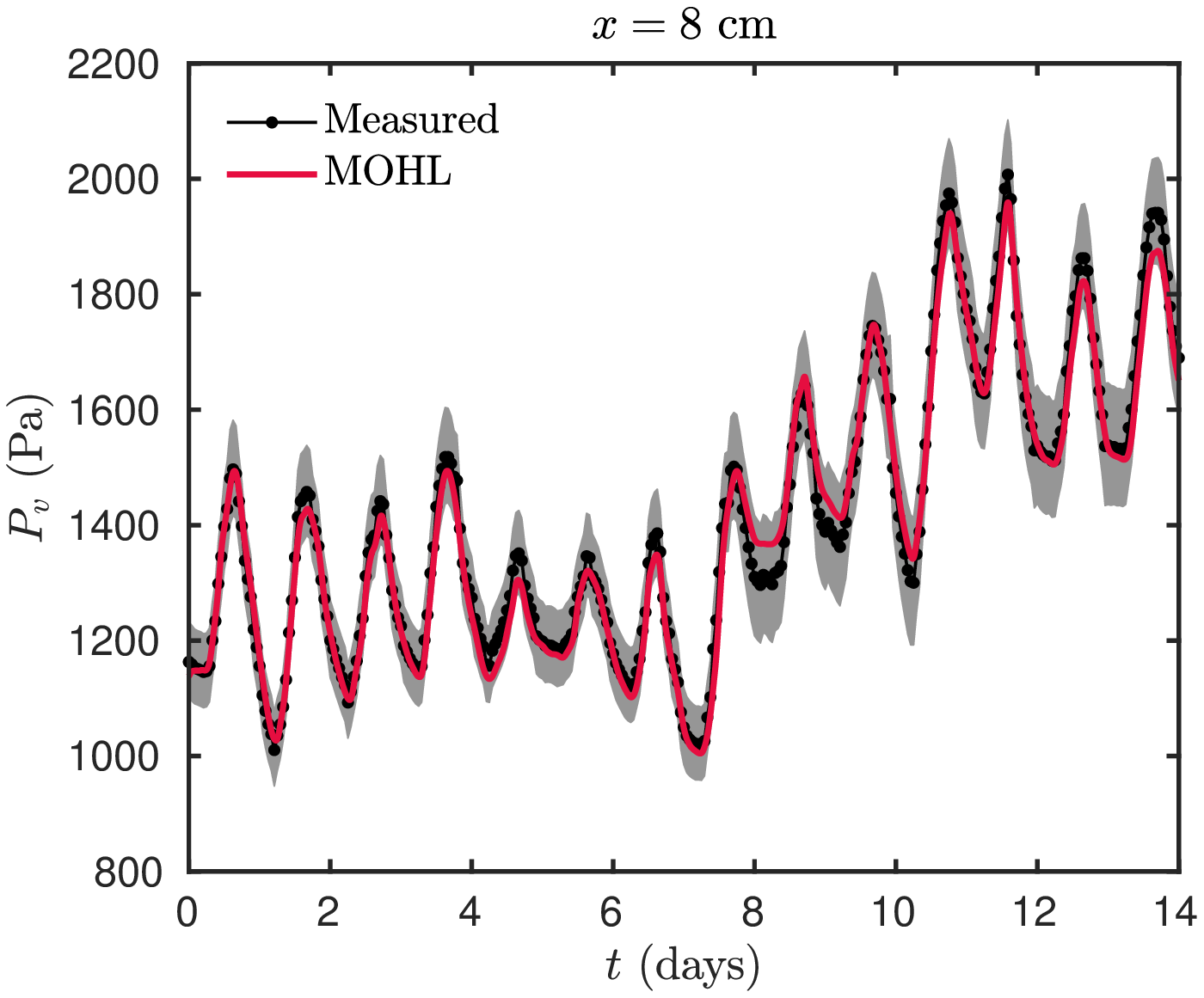}} \\
  \subfigure[e][\label{fig_AN4:T12}]{\includegraphics[width=0.48\textwidth]{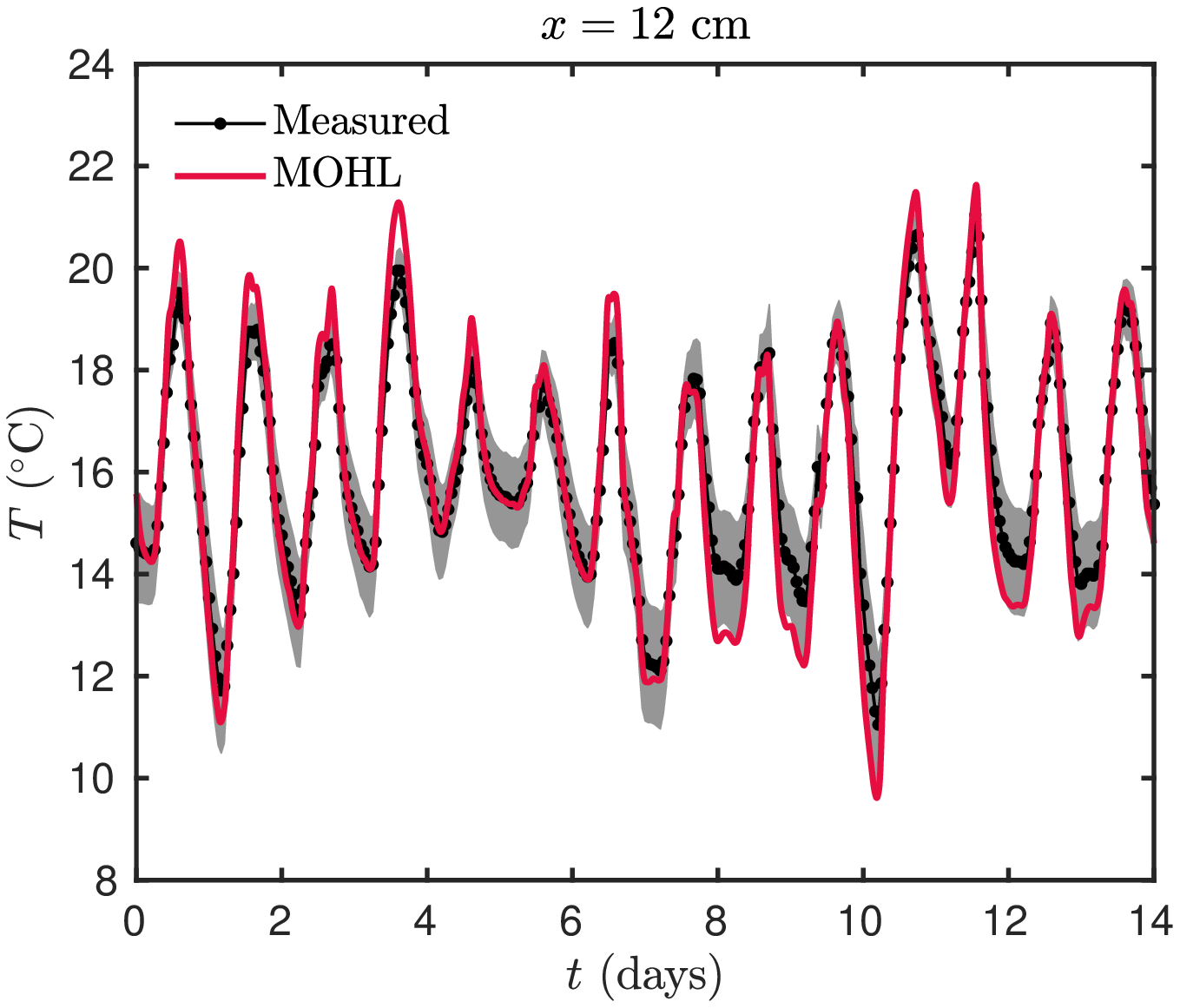}} \hspace{0.3cm}
  \subfigure[f][\label{fig_AN4:RH12}]{\includegraphics[width=0.485\textwidth]{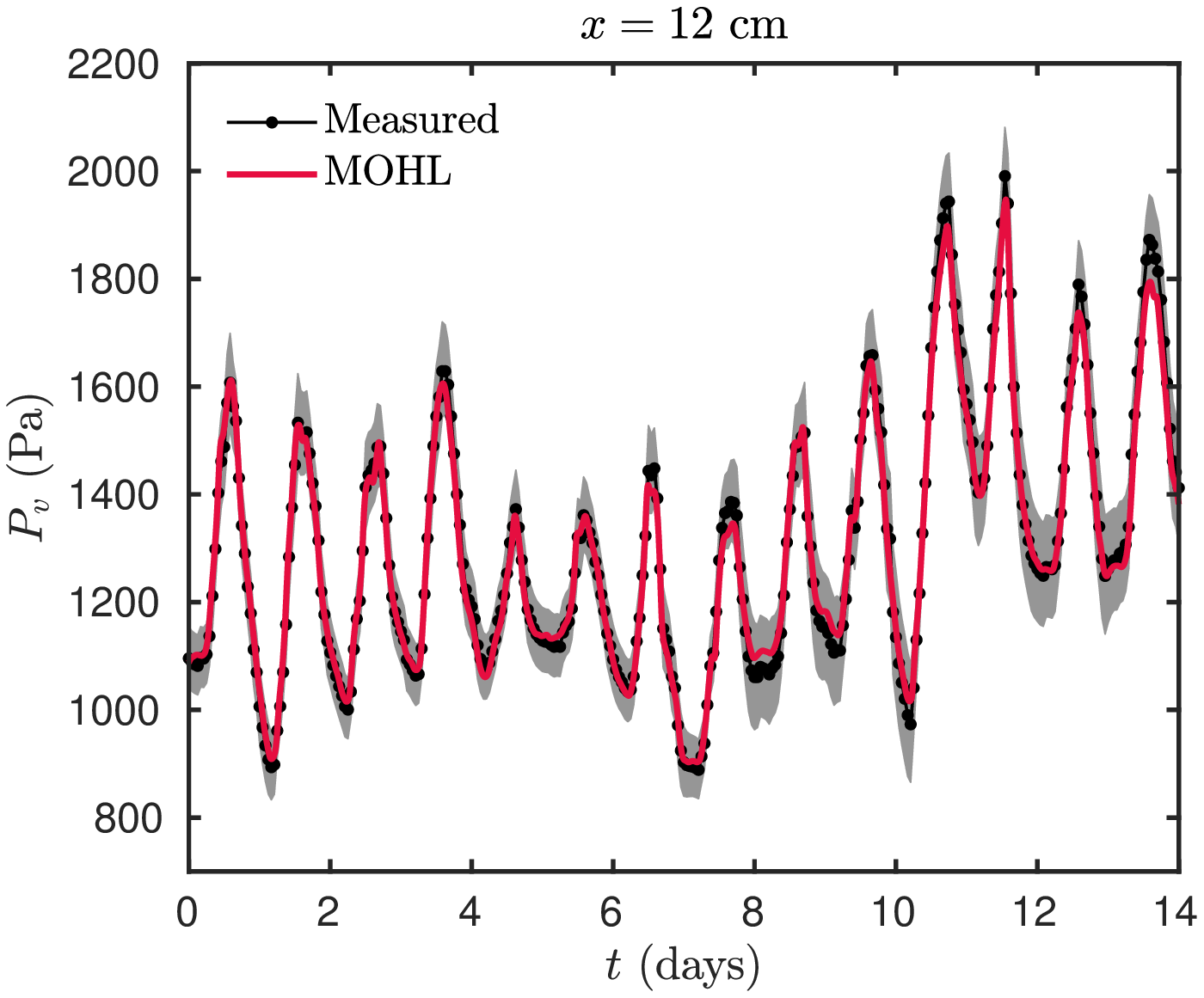}}
  \caption{\small\em Evolution of the temperature and vapour pressure at $x \egalb \{\,4,\,8,\,12\,\}\ \mathsf{cm}$ inside of the material of the measured data versus the simulation solution.}
  \label{fig_AN4:evolution}
\end{figure}

Results of the simulation are compared in terms of temperature and vapour pressure. Figure~\ref{fig_AN4:evolution} presents the measured data and the simulation results in each one of the locations of the sensors, at $x\egalb \{\,4,\,8,\,12\,\}\ \mathsf{cm}\,$. The error between the predicted solution and the experimental observations is between the uncertainties of the sensors during almost all simulation period. The uncertainties are displayed on gray in Figure~\ref{fig_AN4:evolution} and they are calculated according to the position of the sensor ($\sigma^{\,\text{posi}}\,$) and according to the sensor measurement error ($\sigma^{\,\text{meas}}\,$): 
\begin{align*}
  \sigma^{\,\text{total}} \egal \sqrt{(\sigma^{\,\text{meas}})^{\,2} \plus (\sigma^{\,\text{pos}})^{\,2} }
\end{align*}
where $\sigma^{\,\text{total}}$ are the total uncertainties regarding the temperature and the vapour pressure fields. The detailed computation of the measurement uncertainties can be verified in \cite{Berger2019}. For the temperature evolution, the highest difference is when the relative humidity at the left boundary change from $40\,\%$ to $70\,\%\,$. From the $7$\textsuperscript{th} day, the measured value of temperature rises more than on the simulated one as can be observed in Figure~\ref{fig_AN4:T4}. With the variations at the boundaries, the simulated temperature (v. Figures~\ref{fig_AN4:T8} and \ref{fig_AN4:T12}) varies more than the measured temperature. It seems that the total diffusion coefficient of the temperature used for the simulation is higher than the real one. The difference between the simulated and measured values of temperature reaches a maximum of $2\gC$ affected by the moisture flow.

A high influence of the step on the relative humidity can be observed at $x\egalb 4\ \mathsf{cm}$ in Figure~\ref{fig_AN4:RH4}, which decreases along the thickness. Simulations do not completely follow the variations of vapour pressure, but not because of the numerical method, but because of the physical model, which does not consider liquid transport neither hysteresis. In Figures~ \ref{fig_AN4:RH8} and \ref{fig_AN4:RH12}, the differences between simulations and measurements get higher, with the incoming moisture flow reaching a maximum of $77\ \mathsf{Pa}$ and $80\ \mathsf{Pa}\,$, respectively. The absolute difference is higher at $x\egal 4\ \mathsf{cm}\,$, as it is closer to the left boundary and consequently to the incoming flow.

\begin{figure}
  \centering
  \subfigure[a][\label{fig_AN4:error_T}]{\includegraphics[width=0.485\textwidth]{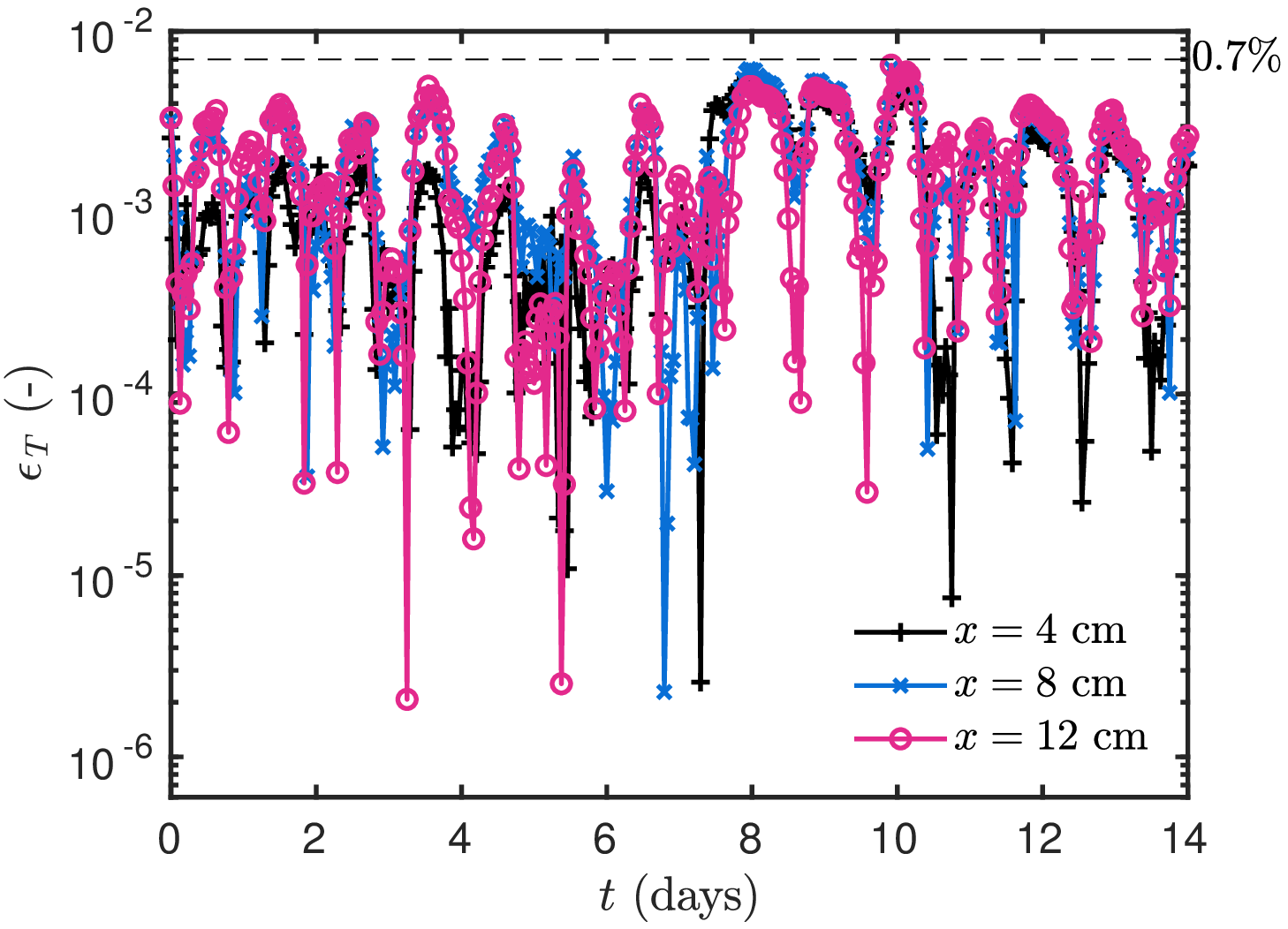}} \hspace{0.2cm}
  \subfigure[b][\label{fig_AN4:error_Pv}]{\includegraphics[width=0.485\textwidth]{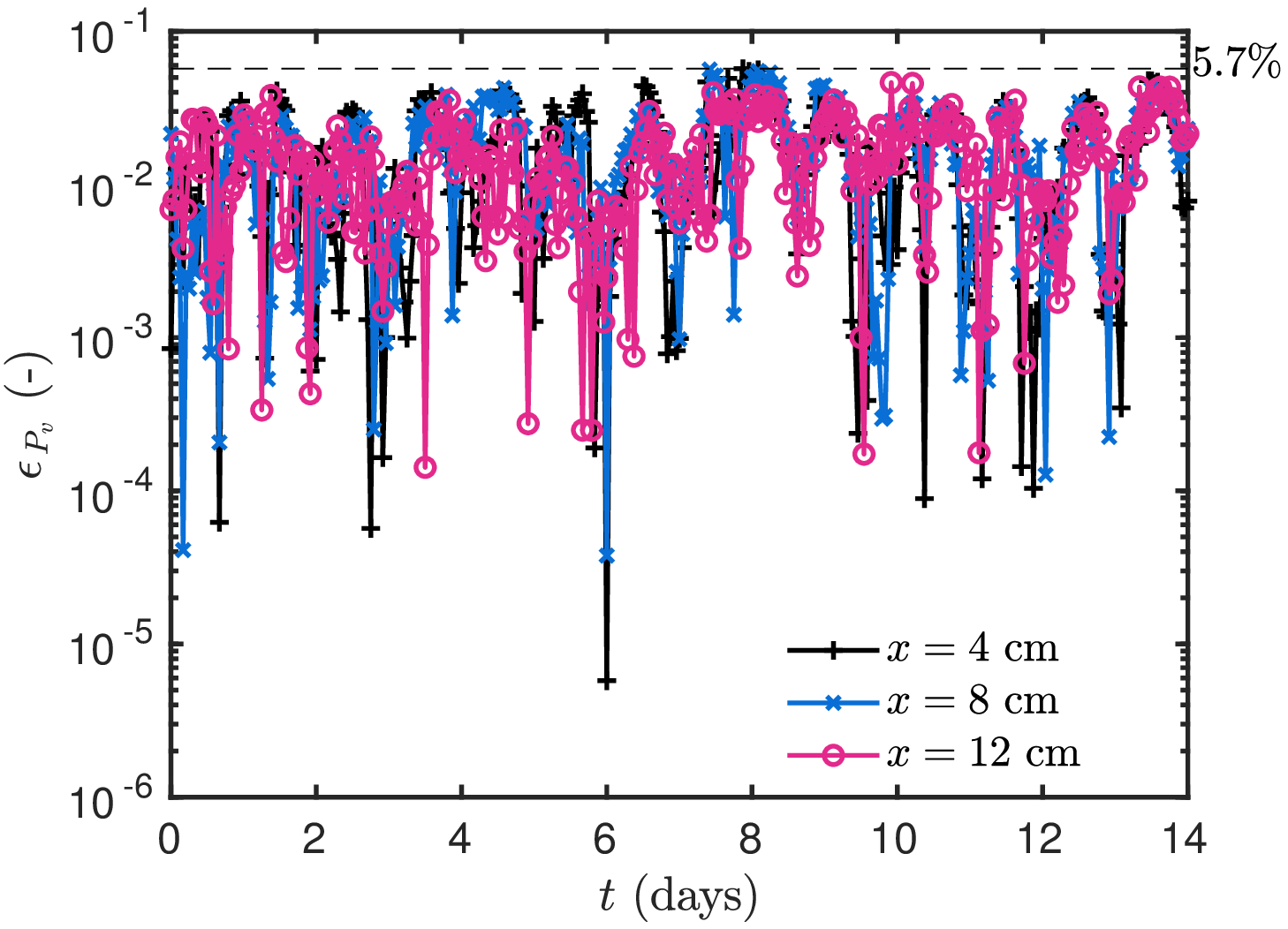}}
  \caption{\small\em Relative error of the simulated temperature (a) and vapour pressure (b) solutions computed at the interior of the material.}
  \label{fig_AN4:error}
\end{figure}

The relative error $\epsilon$ of the computed solutions are presented in Figure~\ref{fig_AN4:error}. Simulations with the MOHL method showed a good agreement with the reference data. The maximum relative error for temperature solution is $0.7\,\%$ and for vapour pressure is $5.7\,\%\,$. The average error on the dynamic profiles is $0.2\,\%$ for temperature and $1.7\,\%$ for vapour pressure, values which are close to the ones obtained by Rouchier \etal \cite{Rouchier2016}.

The MOHL method can provide a very accurate solution of the physical model. The difference that it is observed in Figure~\ref{fig_AN4:evolution} is attributed to the model that describes the phenomenon, as previously mentioned, and to the values of the estimated properties. Nevertheless, despite these discrepancies, simulated results are considered satisfactory for the prediction of heat and moisture transport.


\section{Conclusions}
\label{sec:conl}

Mesh refinement is a great issue in the field of numerical simulations of thermal sciences problems. Therefore, the present article shows that the unsteady problem of heat and moisture transfer can be solved from a different perspective. Namely, we interchange the order of discretization by proposing a numerical model accurate to the order $\O\,(\dx^{\,4}\thicksim \dx^{\,6})$ with a moderate decrease of computational efforts. Traditionally, the numerical analysts discretize the differential governing equations first in space and, then, the coupled system of ODEs is solved in time. This approach is usually referred to as the Method Of Lines (MOL) \cite{Hamdi2007}.

Here, we proposed another point of view on the same problem, the so-called Method Of Horizontal Lines (MOHL). It consists of a semi-discretization; first, in time and only then in space using appropriate methods. So, the Initial-Boundary Value Problem (IVP--BVP) is replaced by a sequence of BVPs. Moreover, we took advantage of the fact that today there are well-tested robust adaptive numerical methods to tackle BVPs in one space dimension \cite{Shampine2000, Hale2008}. The resulting discretization is fully implicit, thus, unconditionally stable. In other words, the semi-discretization in time is not subject to any kind of restriction regarding the time step \cite{Courant1928}. Thus, the time step can be chosen based on the accuracy considerations solely. The distribution of spatial nodes is adaptive and can change at every time step. The grid is further refined (or unrefined) to meet a prescribed error tolerance. The numerical solution error is estimated by computing the continuous residual (inside the domain and at the boundaries).

This new method has been evaluated on two numerical case studies of heat and moisture transfer in porous media. Each case aimed at exciting the non-linear properties of the material to induce sharp profiles of temperature and vapour pressure. The first case considered a single material layer with sinusoidal variations of boundary conditions, taking into account a sudden rain effect at one boundary. The second case studied the heat and moisture diffusive transfer through a multi-layered material. For both cases, the numerical method has shown a high accuracy and perfect agreement with the respective reference solutions. The error was of the order $\O\,(10^{\,-4})$ or less. The advantage of the proposed method is the adaptive spatial mesh grid according to the solicitations of the physical phenomena. Moreover, it has been demonstrated that for the same order of accuracy of the solution, the MOHL numerical model is twice faster than the classical implicit \textsc{Euler} with central finite-difference approach. In the multilayered domain case study, for example, the nodes were concentrated at the interface between materials and at higher gradients, which allowed a solution with high order of accuracy.

The last case study aimed at highlighting the use of the numerical model to compare the numerical predictions with experimental data. The configuration represents a one layer wall exposed to climatic boundary conditions. On the other side, the boundary conditions are controlled so as to impose a step in relative humidity from $40\,\%$ to $70\,\%$ with a constant temperature. The comparison revealed a satisfactory agreement between the numerical results and the experimental data. The MOHL numerical model can provide a very accurate solution of the physical model to predict the heat and moisture transfer in porous materials.

As a technical improvement of the present method, there is another possible direction. Namely, one could notice that we have used a higher-order (at least, fourth order) $C^{\,1}$ continuous interpolation in space and only second-order discretization in time. It might appear as an imbalanced discretization. Hence, this point might be improved in future works by choosing higher-order differentiating in time as well. Another improvement can be achieved if we generalize the method to non-uniform time steps as well.


\section*{Acknowledgements}

The authors acknowledge the Brazilian Agencies CAPES of the Ministry of Education for the financial support of this work, which was conducted during a scholarship supported by the International Cooperation Program CAPES/COFECUB (Grant $\#\, 774/13$). The authors also acknowledge the Junior Chair Research program ``Building performance assessment, evaluation and enhancement'' from the University of Savoie Mont Blanc in collaboration with The French Atomic and Alternative Energy Center (CEA) and Scientific and Technical Center for Buildings (CSTB).


\section*{Nomenclature}

\begin{tabular*}{0.7\textwidth}{@{\extracolsep{\fill}} |@{} >{\scriptsize} c >{\scriptsize} l >{\scriptsize} l| }
\hline 
\multicolumn{3}{|c|}{\emph{Latin letters}} \\
$\cz$ & material specific heat capacity & $[\unitfrac{J}{(kg\cdot K)}]$ \\
$\cw$ & liquid water specific heat capacity & $[\unitfrac{J}{(kg\cdot K)}]$ \\
$\cM$ & moisture storage coefficient & $[\unitfrac{s^2}{m^2}]$ \\
$\cTM$ & coupling storage coefficient & $[\unitfrac{W\cdot s^3}{(kg\cdot m^2)}]$ \\
$\cT$ & energy storage coefficient & $[\unitfrac{W\cdot s}{(m^3\cdot K)}]$ \\
$\hM$ & convective vapour transfer coefficient & $[\unitfrac{s}{m}]$ \\
$\hT$ & convective heat transfer coefficient & $[\unitfrac{W}{(m^2\cdot K)}]$ \\
$H_{\,l}$ & water enthalpy & $[\unitfrac{J}{kg}]$ \\
$g$ & moisture flow & $[\unitfrac{kg}{(m^2\cdot s)}]$ \\
$\kl$ & liquid permeability & $[\mathsf{s}]$ \\
$\kM$ & moisture transf. coeff. under vap. press. grad. & $[\mathsf{s}]$ \\
$\kTM$ & heat transf. coeff. under vap. press. grad.& $[\unitfrac{W\cdot s^2}{kg}]$ \\
$\kT$ & heat transf. coeff. under temp. grad. & $[\unitfrac{W}{(m\cdot K)}]$ \\
$L$ & length & $[\mathsf{m}]$ \\
$\Lv$ & latent heat of evaporation & $[\unitfrac{J}{kg}]$ \\
$\Pc$ & capillary pressure & $[\mathsf{Pa}]$ \\
$\Ps$ & saturation pressure & $[\mathsf{Pa}]$ \\
$\Pv$ & vapour pressure & $[\mathsf{Pa}]$ \\
$q$ &  heat flux & $[\unitfrac{W}{m^2}]$ \\
$\Rv$ & water gas constant & $[\unitfrac{J}{(kg\cdot K)}]$\\
$\mathrm{s}$ & view factor between surfaces & $[-]$\\
$t$ & time & $[\mathsf{s}]$ \\
$T$ & temperature & $[\mathsf{K}]$ \\
$w$ & moisture content & $[\unitfrac{kg}{m^3}]$ \\
$x$ & coordinate direction & $[\mathsf{m}]$ \\
\hline
\end{tabular*}

\vspace{0.3cm}

\begin{tabular*}{0.5\textwidth}{@{\extracolsep{\fill}} |@{} >{\scriptsize} c >{\scriptsize} l >{\scriptsize} l| }
\hline
\multicolumn{3}{|c|}{\emph{Greek letters}} \\
$\delta_{\,v}$ & permeability & $[\mathsf{s}]$ \\
$\xi$ & emissivity & $[-]$\\
$\lambda$ & thermal conductivity & $[\unitfrac{W}{(m\cdot K)}]$ \\
$\sigma$ & \textsc{Stefan}--\textsc{Boltzmann} constant & $[\unitfrac{W}{(m^2\cdot K^4)}]$ \\
$\phi$ & relative humidity & $[-]$ \\
$\rho$ & specific mass & $[\unitfrac{kg}{m^3}]$ \\

\hline
\end{tabular*}

\vspace{0.3cm}

\begin{tabular*}{0.7\textwidth}{@{\extracolsep{\fill}} |@{} >{\scriptsize} c >{\scriptsize} l >{\scriptsize} l| }
\hline
\multicolumn{3}{|c|}{\emph{Parameters involved in the dimensionless representation}} \\
$u$, $v$ & fields & $[-]$ \\
$\mathrm{Bi}$ & \textsc{Biot} number & $[-]$ \\
$\mathrm{Fo}$ & \textsc{Fourier} number & $[-]$ \\
$\nu$ & diffusion coefficient & $[-]$ \\
$c^{\,\star}$ & storage coefficient & $[-]$ \\
$k^{\,\star}$ & permeability coefficient & $[-]$ \\
$q^{\,\star}$, $g^{\,\star}$ & source terms & $[-]$ \\

\hline
\end{tabular*}


\begin{appendices}

\section{Material properties}
\label{annexe:material_properties}

\begin{table}[h!]
\centering
\setlength{\extrarowheight}{.3em}
\caption{\small\em Hygrothermal properties of the load bearing material \cite{Hagentoft2004}.}
\bigskip
\begin{tabular}{l p{9cm} l}
\hline
\textit{Property} & \textit{Value} & \textit{Unit} \\
\hline
Thermal capacity & $\rhoz \,\cz \egal 2005\cdot 840 $  &  $ [\unitfrac{J}{m^3\cdot K}]$  \\
Sorption isotherm &  \parbox[t]{9cm}{$w\,(\phi) \egal 47.1 \, \biggl[1 + \Bigl(-1692.94 \cdot \ln(\phi)\Bigr)^{1.65}\biggr]^{\,-0.39} + \\ 109.9\, \biggl[1 + \Bigl(-2437.83\cdot \ln(\phi)\Bigr)^{6}\,\biggr]^{\,-0.83} $}  &  $  [\unitfrac{kg}{m^3}]$    \\
Vapour permeability & $\delta_{\,v}\,(\phi) \egal 6.413\cdot 10^{\,-9}\cdot \dfrac{\Bigl(1-\frac{w\,(\phi)}{157}\Bigr)}{0.503\, \Bigl(1-\frac{w\,(\phi)}{157}\Bigr)^{2}+0.497} $  &  $  [\mathsf{s}] $\\
Liquid permeability & $\kl\,(\phi) \egal 2.52\dix{-4}\cdot \exp(-1.55\dix{6}\cdot \phi) $  &  $  [\mathsf{s}] $\\
Thermal conductivity & $\lambda\,(\phi) \egal 0.5 \plus 0.0045\cdot w\,(\phi)$  &  $  [\unitfrac{W}{(m\cdot K)}]$\\
\hline                
\end{tabular}
\label{table:properties_mat1}
\end{table}

\begin{table}[h!]
\centering
\caption{\small\em Hygrothermal properties of the finishing material \cite{Hagentoft2004}.}
\bigskip
\setlength{\extrarowheight}{.3em}
\begin{tabular}{l p{9cm} l}
\hline
\textit{Property} & \textit{Value} & \textit{Unit} \\
\hline
Thermal capacity & $\rhoz \,\cz \egal 790\cdot 870 $  &  $ [\unitfrac{J}{m^3\cdot K}]$  \\
Sorption isotherm &  $w\,(\phi) \egal 209 \, \biggl[1 + \Bigl(-2.7\dix{14} \cdot \ln(\phi)\Bigr)^{1.27}\biggr]^{\,-0.21} $  &  $  [\unitfrac{kg}{m^3}]$    \\
Vapour permeability & $\delta_{\,v}\,(\phi) \egal 6.413\cdot 10^{\,-9}\cdot \dfrac{\Bigl(1-\frac{w\,(\phi)}{209}\Bigr)}{0.503\, \Bigl(1-\frac{w\,(\phi)}{209}\Bigr)^{\,2}+0.497} $  &  $  [\mathsf{s}] $\\
Liquid permeability & \parbox[t]{9cm}{$\kl\,(\phi) \egal \exp[-33\plus 0.0704\cdot (w-120)\\-1.742\dix{-4}\cdot (w-120)^2 -2.795\dix{-6}\cdot (w-120)^3\\ -1.157\dix{-7}\cdot (w-120)^4 +2.597\dix{-9}\cdot (w-120)^5] $ } &  $  [\mathsf{s}] $\\
Thermal conductivity & $\lambda\,(\phi) \egal 0.2 \plus 0.0045\cdot w\,(\phi)$  &  $  [\unitfrac{W}{(m\cdot K)}]$\\
\hline                
\end{tabular}
\label{table:properties_mat2}
\end{table}

\begin{table}[h!]
\centering
\caption{\small\em Hygrothermal properties of the wood fibre \citep{Rouchier2016}.}
\bigskip
\setlength{\extrarowheight}{.3em}
\begin{tabular}{l p{9.5cm} l}
\hline
\textit{Property} & \textit{Value} & \textit{Unit} \\
\hline
Thermal capacity & $\rhoz\, \cz\egal 1103\cdot 146 $  &  $   [\unitfrac{J}{(m^3\cdot K)}] $  \\
Sorption isotherm &  $w \, (\phi) \egal 7.063\dix{-5} \cdot \phi^{\,3} - 0.00736 \cdot \phi^{\,2} + 0.4105\cdot \phi + 0.2688 $  &  $  [\unitfrac{kg}{m^3}] $     \\
Vapour permeability  & $\delta_{\,v} \, (\phi)  \egal 4.85\dix{-13} \cdot \phi + 3.28\dix{-11} $  &  $  [\mathsf{s}] $\\
Thermal conductivity  & $\lambda\, (\phi,\,T)  \egal 0.038 + 0.192\cdot \dfrac{w\, (\phi)}{\rhol} + 1.08\dix{-4}\cdot T  $  &  $   [\unitfrac{W}{(m\cdot K)}] $ \\[10pt]
\hline 
\end{tabular}
\label{table:properties_woodfibre}
\end{table}


\section{Dimensionless parameters}
\label{annexe:dimensionless}

\subsection{Case from Section~\ref{sec:validat}}

Problem \eqref{eq:HAM_equation_dimless} is considered with $\FoT \egalb 1.6\cdot 10^{\,-1}\,$, $\FoM \egalb 3.2\cdot 10^{-2}\,$, $\gamma_{\,1} \egalb 2.3\cdot 10^{-2}$ and $\gamma_{\,2} \egalb 1.58 \cdot 10^{-1}\,$. The reference time is $\tref \egalb  1\ \mathsf{h}$, thus, the final simulation time is fixed to $\tau^{\,\star} \egalb  72\,$. The reference temperature and vapour pressure were taken the same as the initial condition. At the boundaries, the \textsc{Biot} numbers assume the following values: $\BiML \egalb 3.65$, $\BiMR \egalb 0.55\,$, $\BiTL \egalb 6.45\,$, $\BiTR \egalb 2.06\,$, $\BiTML \egalb 0.13$ and $\BiTMR \egalb 0.02\,$. The temperature and vapour pressure vary sinusoidally over the time as the following expressions:
\begin{align*}
\uinfL \,(\ts) &\egal 1 \plus 0.02 \cdot \sin \left(\, 2\,\pi \, \ts /48 \,\right)^{\,2}  \,, \\
\uinfR \,(\ts) &\egal 1 \plus 0.005 \cdot \sin \left(\, 2\,\pi \, \ts /24 \,\right)^{\,2} \,, \\
\vinfL \,(\ts) &\egal \Bigl(0.7 \plus 0.25 \cdot \sin (\, 2\,\pi \, \ts /24 \,)^{\,2}\Bigr)\cdot \Bigl(\Ps (\uinfL \,(\ts)\cdot \Ti)/\Pvi \Bigr)  \,,  \\
\vinfR \,(\ts) &\egal \Bigl(0.825 \plus 0.125 \cdot \tanh (\ts\moins  36)\Bigr)\cdot \Bigl(\Ps (\uinfR \,(\ts)\cdot \Ti)/\Pvi \Bigr) \,.
\end{align*}
The properties are dimensionalised with the following reference values:
\begin{align*}
& \cMref \egal 0.061 \,, & & \cTref \egal 8.6125\cdot 10^{\,5} \,, & & \cTMref \egal 5.0963\cdot 10^{\,3} \,, \\
& \kMref \egal 5.4712\cdot 10^{\,-9} \,, & & \kTref \egal 0.3873 \,, & & \kTMref \egal  0.0154 \,.
\end{align*}
For the dimensionless properties of the material, they can be written as:
\begin{align*}
\cMs\,(v) \egal & \dfrac{169.5\,v^{\,5}\moins 814.2\,v^{\,4}\plus534.4\,v^{\,3}\plus2625\,v^{\,2}\moins 4642\,v\plus2217}{v^{\,5}\plus2182v^{\,4}\moins 12520v^{\,3}\plus 27210v^{\,2}\moins 26680\,v\plus 10050} \,,\\
\cTs\,(v) \egal & \dfrac{246.6\,v^{\,2}\moins 778.9\,v\plus 656.9}{v^{\,4}\moins 41.37\,v^{\,3}\plus 395.2\,v^{\,2}\moins 985.6\,v\plus 760.7} \,,\\
\cTMs\,(v)\egal & \dfrac{4207\,v^{\,4}\moins 24860\,v^{\,3}\plus 50920\,v^{\,2}\moins 43030\,v\plus 14570}{v^{\,3}\plus 8614\,v^2\moins 28190\,v\plus 23480} \,,\\
\kMs\,(v) \egal & 4.045\,v^{\,6.448}\plus 16.23 \,,\\
\kTs\,(v) \egal & \dfrac{15.3\,v^{\,2}\moins 46.53\,v\plus 38.04}{v^{\,4}\moins 10.46\,v^{\,3}\plus 46.24\,v^{\,2}\moins 85.34\,v\plus 56.1} \,,\\
\kTMs\,(v)\egal & \dfrac{1.644\,v^{\,2}\moins 7.013\,v\plus 7.505}{v^{\,4}\moins 3.133\,v^{\,3}\plus 4.859\,v^{\,2}\moins 8.003\,v\plus 7.408} \,.
\end{align*}


\subsection{Case from Section~\ref{sec:layered_case}}

Problem \eqref{eq:HAM_equation_dimless} is considered with $\FoT \egalb 0.02\,$, $\FoM \egalb 0.07\,$, $\gamma_{\,1} \egalb 0.01$ and $\gamma_{\,2} \egalb 0.13\,$. The reference time is $\tref \egalb 1\ \mathsf{h}\,$, and, the final simulation time is fixed to $\tau^{\,\star} \egalb  120\,$. The reference temperature and vapour pressure were taken the same as the initial condition. At the boundaries, the \textsc{Biot} numbers assume the following values: $\BiML \egalb 4.4\,$, $\BiMR \egalb 0.6\,$, $\BiTL \egalb 6\,$, $\BiTR \egalb 2\,$, $\BiTML \egalb 0.1$ and $\BiTMR \egalb 0.02\,$. The ambient temperature, vapour pressure and rain flow are written as follows:
\begin{align*}
\uinfL \,(\ts) &\egal 1 \moins 0.02 \cdot \sin\, (\, 2\,\pi \, \ts /24 \,)  \,, \\
\uinfR \,(\ts) &\egal 1 \plus 0.01 \cdot \sin\, (\, 2\,\pi \, \ts /48 \,) \,, \\
\vinfL \,(\ts) &\egal \Bigl(0.5 \plus 0.3 \cdot \sin\, (\, 2\,\pi \, \ts /48 \,)^{\,2}\Bigr)\cdot \Bigl(\Ps (\uinfL \,(\ts)\cdot \Tref)/\Pvref \Bigr)  \,,  \\
\vinfR \,(\ts) &\egal \Bigl(0.5 \plus 0.2 \cdot \sin\, (\, 2\,\pi \, \ts /72 \,)^{\,2}\Bigr)\cdot \Bigl(\Ps (\uinfR \,(\ts)\cdot \Tref)/\Pvref \Bigr) \,,\\
\gsinfL \,(\ts) &\egal 3.8\cdot \sin\,(\pi \, \ts/105)^{\,70}\,, \\
\qsinfL \,(\ts) &\egal 1.8\dix{-4} \cdot \gsinfL\,(\ts) \cdot (\uinfL \moins u_{\,\text{ref}})\,.
\end{align*}
The properties are scaled with the following reference values:
\begin{align*}
& \cMref \egal 0.061 \,, & & \cTref \egal 1.6862\cdot 10^{\,6} \,, & & \cTMref \egal 5.0963\cdot 10^{\,3} \,, \\
& \kMref \egal 5.4712\cdot 10^{\,-9} \,, & & \kTref \egal 0.5021 \,, & & \kTMref \egal  0.0161 \,.
\end{align*}
Thus, the dimensionless properties of the Material $1$ are written as:
\begin{align*}
\cMs\,(v) \egal & \dfrac{\moins 0.1244\,v^{\,4}\plus 0.4949\,v^{\,3}\moins 0.6025\,v^{\,2}\plus 0.1802\,v\plus 0.06364}{v^{\,5}\moins 5.101\,v^{\,4}\plus 9.802\,v^{\,3}\moins 8.408\,v^{\,2}\plus 2.713\,v\plus 0.0055} \,,\\
\cTs\,(v) \egal & \dfrac{16330\,v^{\,4}\moins 80020\,v^{\,3}\plus 127900\,v^{\,2}\moins 75310\,v\plus 17920}{v^{\,5}\plus 16340\,v^{\,4}\moins 80050\,v^{\,3}\plus 127900\,v^{\,2}\moins 75340\,v\plus 17930} \,,\\
\cTMs\,(v)\egal &  \dfrac{\moins 0.124\,v^{\,4}\plus 0.4937\,v^{\,3}\moins 0.6015\,v^{\,2}\plus 0.18\,v\plus 0.0638}{v^{\,5}\moins 5.103\,v^{\,4}\plus 9.812\,v^{\,3}\moins 8.42\,v^{\,2}\plus 2.719\,v\plus 0.0055}\,,\\
\kMs\,(v) \egal &  \dfrac{\moins 0.8682\,v^{\,4}\plus 6.371\,v^{\,3}\moins 10.02\,v^{\,2}\plus 1.842\,v\plus 3.542}{v^{\,5}\moins 7.075\,v^{\,4}\plus 18.65\,v^{\,3}\moins 20.95\,v^{\,2}\plus 6.635\,v\plus 2.601}\,,\\
\kTs\,(v) \egal &  \dfrac{4692\,v^{\,4}\moins 17720\,v^{\,3}\plus 16330\,v^{\,2}\moins 3385\,v\plus 8137}{v^{\,5}\plus 4678\,v^{\,4}\moins 17650\,v^{\,3}\plus 16210\,v^{\,2}\moins 3336\,v\plus 8152}\,,\\
\kTMs\,(v)\egal &  \dfrac{\moins 1002\,v^{\,4}\plus 1727\,v^{\,3}\moins 94.85\,v^{\,2}\plus 253.6\,v\plus 2091}{v^{\,5}\moins 1002\,v^{\,4}\plus 1714\,v^{\,3}\moins 70.93\,v^{\,2}\plus 240.7\,v\plus 2092}\,,
\end{align*}
and, the dimensionless properties of the Material $2$ are written as:
\begin{align*}
\cMs\,(v) \egal & \dfrac{\moins 11870\,v^{\,4}\plus 36160\,v^{\,3}\moins 27730\,v^{\,2}\plus 11480\,v\plus 878}{v^{\,5}\plus 223500\,v^{\,4}\moins 212400\,v^{\,3}\plus 172100\,v^{\,2}\plus 41290\,v\plus 14.34}\,,\\
\cTs\,(v) \egal & \dfrac{\moins 3.72.3\,v^{\,4}\plus 203.7\,v^{\,3}\plus 344.9\,v^{\,2}\plus 956\,v\plus 1440}{v^{\,5}\moins 320\,v^{\,4}\moins 975.6\,v^{\,3}\plus 2050\,v^{\,2}\plus 1061\,v\plus 3012}\,,\\
\cTMs\,(v)\egal & \dfrac{\moins 65880\,v^{\,4}\plus 196800\,v^{\,3}\moins 163100\,v^{\,2}\plus 49850\,v\plus 7210}{v^{\,5}\moins 105800\,v^{\,4}\plus 583000\,v^{\,3}\moins 146000\,v^{\,2}\plus 336500\,v\plus 119.1}\,,\\
\kMs\,(v) \egal & \dfrac{\moins 7049\,v^{\,4}\moins 8193\,v^{\,3}\plus 36200\,v^{\,2}\moins 10330\,v\plus 55820}{v^{\,5}\moins 2888\,v^{\,4}\plus 7947\,v^{\,3}\moins 7103\,v^{\,2}\plus 3269\,v\plus 4471}\,,\\
\kTs\,(v) \egal & \dfrac{139.5\,v^{\,4}\moins 668\,v^{\,3}\moins 28.81\,v^{\,2}\plus 1191\,v\plus 974.1}{v^{\,5}\plus 333.6\,v^{\,4}\moins 1308\,v^{\,3}\plus 448.4\,v^{\,2}\plus 943.6\,v\plus 1468}\,,\\
\kTMs\,(v)\egal & \dfrac{\moins 77400\,v^{\,4}\plus 202700\,v^{\,3}\moins 245700\,v^{\,2}\plus 239400\,v\plus 120600}{v^{\,5}\moins 8436\,v^{\,4}\plus 22580\,v^{\,3}\moins 26780\,v^{\,2}\plus 25320\,v\plus 12160}\,.
\end{align*}

\subsection{Case from Section~\ref{sec:validation}}

Problem \eqref{eq:HAM_equation_dimless} is considered with $\FoT \egalb 0.06\,$, $\FoM \egalb 0.02\,$, $\gamma_{\,1} \egalb 5.17\dix{-3}$ and $\gamma_{\,2} \egalb 7.72\dix{-3}\,$. The reference time is $\tref \egalb 1\ \mathsf{h}\,$, and, the final simulation time is fixed to $\tau^{\,\star} \egalb  336\,$. The reference temperature is $\Tref \egalb 293.15\ \mathsf{K}$ and the reference vapour pressure is $\Pvref \egalb 1166.91\ \mathsf{Pa}\,$.

The boundary conditions are gathered from the experimental data and just adimensionalized. The dimensionless form of the initial condition are:
\begin{align*}
u_{\,0} \, (\xs) &\egal \moins 0.08806 \cdot (\xs)^{\,4} \plus 0.1688\cdot (\xs)^{\,3} \moins 0.1143\cdot (\xs)^{\,2} \moins  0.01621\cdot \xs \plus  1.015 \,,\\
v_{\,0} \, (\xs) &\egal \moins 0.408\cdot (\xs)^{\,4} \plus 1.188\cdot (\xs)^{\,3} \moins 1.053\cdot (\xs)^{\,2} \plus  0.08969\cdot \xs \plus  1.092 \,.
\end{align*}

The dimensionless properties of the material can be written as:
\begin{align*}
& \cMs\,(v)  \egal  \moins 0.01799\, v \plus  1.018 \,, & &
\kTMs\,(v)\egal  0.007343\, v \plus  0.9927 \,,\\
& \cTs\,(v)  \egal  0.005168\, v \plus  0.9948 \,, & &
\kMs\,(v)\egal  0.007343\, v \plus  0.9927  \,,\\
& \cTMs\,(v) \egal  \moins 0.01799\, v \plus  1.018 \,, & &
\kTs\,(v)\egal  7.343\dix{-4}\, v \plus  0.9994 \,,
\end{align*}
with the following reference values:
\begin{align*}
& \kMref \egalb 3.31\dix{-11}\, \mathsf{s}\,, & & \kTref \egalb  6.98\dix{-2}\, \mathsf{W/(m\cdot K)}\,, & & \kTMref \egalb  8.27\dix{-5}\, \mathsf{m^2/s)}\,, \\ 
& \cMref \egalb  1.72\dix{-4}\, \mathsf{s^2/m^2}\,, & & \cTref \egalb  163073.8\, \mathsf{W\cdot s/(m^3\cdot K)}\,, & & \cTMref \egalb  211.7\, \mathsf{W\,s^3/(kg\cdot m^2)}\,.
\end{align*}


\section{Analytical solution for the interface assumptions}
\label{annexe:analytical_sol_multi}

Consider the simplified boundary value problem:
\begin{align*}
\left\{ \begin{array}{l}
\moins \left( k\,(x) \cdot u^{\,\prime}\, (x) \right)^{\,\prime} \egal f(x) \egal \sin\, (\pi\,x) \,, \quad x \in [\,-1\,,1\,] \,,\\  [8pt]
u \,(-\ 1) \egal -\ 1 \,;\quad  u\, (1) \egal 1 \,.
\end{array} \right.
\end{align*}
with 
\begin{align*}
 k\,(x) \egal \left\{ \begin{array}{l}
k_{\,1}\,, \quad x \ < \ 0\,, \\ 
k_{\,2}\,, \quad x\ \geqslant\ 0\,.
\end{array} \right.
\end{align*}

The problem at the left can be written as:
\begin{align} \label{eq_AN:left_prob}
\left\{ \begin{array}{l}
-k_{\,1} \cdot u^{\prime \prime}\, (x) \egal f(x) \,,\\ 
u \,(-1) \egal -1 \,.
\end{array} \right.
\end{align}

And, the problem at right as:
\begin{align}\label{eq_AN:right_prob}
\left\{ \begin{array}{l}
-k_{\,2} \cdot u^{\prime \prime}\, (x) \egal f(x) \,,\\  
u\, (1) \egal 1 \,.
\end{array} \right.
\end{align}

The analytical solution of Problems \eqref{eq_AN:left_prob} and \eqref{eq_AN:right_prob} are, respectively:
\begin{align*}
u_{\,1}\, (x) \egal \dfrac{\sin\,(\pi\, x)}{k_{\,1}\,\pi^{\,2}} \plus x \plus C_{\,1}\, (1 \plus  x)\,, \\ 
u_{\,2}\, (x) \egal \dfrac{\sin\,(\pi\, x)}{k_{\,2}\,\pi^{\,2}} \plus x \plus C_{\,2}\, (1 \moins x)\,, 
\end{align*}
whit the following derivatives:
\begin{align*}
u_{\,1}^{\,\prime}\, (x) \egal \dfrac{\cos\,(\pi\, x)}{k_{\,1}\,\pi} \plus 1 \plus C_{\,1}\,, \\ 
u_{\,2}^{\,\prime}\, (x) \egal \dfrac{\cos\,(\pi\, x)}{k_{\,2}\,\pi} \plus 1 \moins C_{\,2}\,. 
\end{align*}

By assuming that the solution is continuous, $u_{\,1}(0)\ \equiv\ u_{\,2}(0)\,$, we can have the relation between the constants: $C_{\,1} \egal C_{\,2} \egal C$.

If we assumed that the \textit{flux is continuous} at the interface $x \egalb 0$ we have:
\begin{align*}
k_{\,2} \, u_{\,2}^{\, \prime}\, (0) \egal k_{\,1} \, u_{\,1}^{\, \prime}\, (0) \,, 
\end{align*}
which leads to the following relation:
\begin{align*}
C \egal\dfrac{k_{\,2} \moins k_{\,1}}{k_{\,2} \plus k_{\,1}} \egal C_{\,A}\,. 
\end{align*}

Although, if we assumed that the \textit{derivative solution is continuous} at the interface $x \egalb 0$ we have:
\begin{align*}
u_{\,2}^{\, \prime}\, (0) \egal u_{\,1}^{\, \prime}\,(0) \,,
\end{align*}
which leads to the following relation:
\begin{align*}
C \egal\dfrac{k_{\,1} \moins k_{\,2}}{2\, \pi\, k_{\,1}\, k_{\,2}} \egal C_{\,B}\,.
\end{align*}

\begin{figure}
\begin{center}
\includegraphics[width=.75\textwidth]{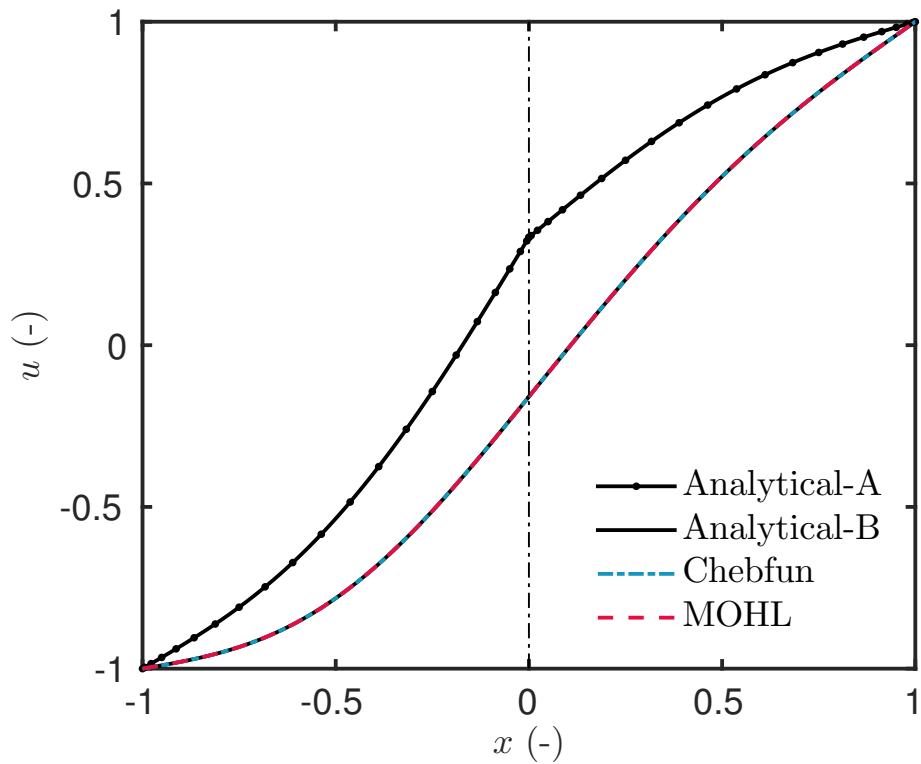}
\caption{\small\em Comparison of the solutions computed for a two-layers domain problem.}
\label{fig_annex:analytic}
\end{center}
\end{figure}

Now, we have two analytical solutions, one that considers the \textit{flux continuous}, called here as Analytical-A, and the other analytical solution that consider the \textit{derivative solution continuous}, called here as Analytical-B. Figure~\ref{fig_annex:analytic} presents both analytical solutions, with the \texttt{Chebfun} solution and the one computed with MOHL. It is shown that the \texttt{Chebfun} and MOHL solutions are in perfect agreement with the solution Analytical-B. This means that both numerical solutions compute the more regular solution and the flux is not continuous at the interface.

\end{appendices}


\bigskip
\addcontentsline{toc}{section}{References}
\bibliographystyle{abbrv}
\bibliography{biblio}
\bigskip

\end{document}